\pdfoutput=1
\documentclass[aps,prd,amsmath,floats,floatfix, twocolumn,
superscriptaddress,nofootinbib,showpacs]{revtex4-1}

\usepackage{txfonts}
\usepackage{tikz}
\usepackage[T1]{fontenc}
\usepackage[utf8]{inputenc}
\usepackage{tensor}
\usepackage{multirow}
\usepackage{textcomp}
\usepackage{enumitem}
\usepackage[hypertexnames=false, unicode, colorlinks=true, linkcolor=linkcolor,citecolor=linkcolor, filecolor=linkcolor,urlcolor=linkcolor, pdfusetitle,pdfencoding=auto, psdextra]{hyperref}
\usepackage{orcidlink}

\usepackage{verbatim}
\usepackage[utf8]{inputenc}
\usepackage{xcolor}
\definecolor{linkcolor}{rgb}{0.0,0.3,0.5}

\usepackage[all]{hypcap}
\usepackage{graphicx}
\usepackage{xspace}
\usepackage{amsmath,amssymb}
\usepackage[normalem]{ulem} %for \sout
\usepackage{bm} % boldmath
\usepackage{mathrsfs}
\usepackage[caption=false]{subfig}
\usepackage{pifont}
\usepackage{microtype}

\graphicspath{%
  {figs/}%
}

\DeclareMathAlphabet{\mathpzc}{OT1}{pzc}{m}{it}

\newcommand{\be}{\begin{equation}}
\newcommand{\ee}{\end{equation}}
\newcommand{\bs}{\begin{split}}
\newcommand{\es}{\end{split}}

\newcommand{\etal}{\textit{et al.\ }}
\newcommand{\ind}{{SXS:BBH:0207}}
\newcommand{\indnew}{{SXS:BBH:1936}}

\newlist{todolist}{itemize}{2}
\setlist[todolist]{label=$\square$}

\newcommand{\scrip}{$\mathscr{I}^+$}
\newcommand{\scrim}{$\mathscr{I}^-$}
\newcommand{\Hp}{$\mathscr{H}^+$}
\newcommand{\Hm}{$\mathscr{H}^-$}
\newcommand{\mycomment}[1]{}
\newcommand{\lmw}{{lm\omega}}
\begin{document}
\title{Gravitational-wave echoes from numerical-relativity waveforms via space-time construction near merging compact objects}
\newcommand\caltech{\affiliation{Theoretical Astrophysics 350-17, California
 Institute of Technology, Pasadena, CA 91125, USA}}
\newcommand\Cornell{\affiliation{Cornell Center for Astrophysics and Planetary
     Science, Cornell University, Ithaca, New York 14853, USA}}
\newcommand\MaxPlanck{\affiliation{Max Planck Institute for Gravitational
     Physics (Albert Einstein Institute), Am M{\"u}hlenberg 1, D-14476 Potsdam,
     Germany}}

\author{Sizheng Ma  \orcidlink{0000-0002-4645-453X}}
\email{sma@caltech.edu}
\caltech

\author{Qingwen Wang}
\affiliation{Perimeter Institute and University of Waterloo, Canada}

 \author{Nils Deppe \orcidlink{0000-0003-4557-4115}} \caltech
 \author{Fran\c{c}ois H\'{e}bert \orcidlink{0000-0001-9009-6955}} \caltech
 \author{Lawrence E.~Kidder \orcidlink{0000-0001-5392-7342}} \Cornell
 \author{Jordan Moxon \orcidlink{0000-0001-9891-8677}} \caltech
 \author{William Throwe \orcidlink{0000-0001-5059-4378}} \Cornell
 \author{Nils L.~Vu \orcidlink{0000-0002-5767-3949}} \MaxPlanck
\author{Mark A. Scheel}
\caltech

\author{Yanbei Chen \orcidlink{0000-0002-9730-9463} }
\email{yanbei@caltech.edu}
\caltech

% Because hyperref only gets the *last* author, we need to be explicit.
\hypersetup{pdfauthor={Ma et al.}}

\date{\today}

%==========================================================================

\begin{abstract}
%Recently, it has been argued that near-horizon modifications of the standard (classical) black hole
%spacetime could lead to observable alterations of the gravitational waveform generated by a binary black
%hole coalescence.
We propose a new approach toward reconstructing the late-time near-horizon geometry of merging binary black holes, and toward computing gravitational-wave echoes from exotic compact objects.  A binary black-hole merger spacetime can be divided by a time-like hypersurface into a Black-Hole Perturbation (BHP) region, in which the space-time geometry can be approximated by homogeneous linear perturbations of the final Kerr black hole, and a nonlinear region.  At late times, the boundary between the two regions is an infalling shell.  The BHP region contains late-time gravitational-waves emitted toward the future horizon, as well as those emitted toward future null infinity. In this region, by imposing no-ingoing wave conditions at past null infinity, and matching out-going waves at future null infinity with waveforms computed from numerical relativity, we can obtain waves that travel toward the future horizon.  In particular, the Newman-Penrose $\psi_0$ associated with the in-going wave on the horizon is related to tidal deformations measured by fiducial observers floating above the horizon.  
%We use the hybrid method to compute waveforms of gravitational-wave echoes from comparable-mass merging binaries, based on inspiral-merger-ringdown waveforms obtained from Cauchy-characteristic extraction (a numerical-relativity technique). 
%
%In constructing this echo model, the space-time is split into an inner post-Newtonian region and an outer black-hole perturbation (BHP) region by a 3-dimensional time-like worldtube. In the BHP region, we take the Weyl scalar $\psi_0$ at future null infinity  from Cauchy-characteristic extraction and evolve it backward into the bulk. This process allows us to obtain $\psi_0$ at future horizon. 
%The portion of horizon $\psi_0$ that lies outside the worldtube serves as a representation for the actual gravitational wave that falls down the horizon.
We further determine the boundary of the BHP region on the future horizon by imposing that   $\psi_0$ inside the BHP region  can be faithfully represented by quasi-normal modes.  Using a  physically-motivated way to impose boundary conditions near the horizon, and applying the so-called Boltzmann reflectivity, we compute the quasi-normal modes of non-rotating ECOs, as well as gravitational-wave echoes. We also investigate the detectability of these echoes in current and future detectors, and prospects for parameter estimation. 
\end{abstract}
\maketitle

%==========================================================================
\section{Introduction} 
\label{sec:introduction}

Delayed and repeating gravitational wave echoes emitted by compact-binary mergers \cite{Cardoso:2016rao,Cardoso:2016oxy,Cardoso:2017cqb}, following the main gravitational waves (GWs), can be  signatures of: (i) deviations of laws of gravity from general relativity~\cite{Zhang:2017jze,Dong:2020odp}, (ii) near-horizon quantum structures surrounding black holes (BHs) \cite{Almheiri:2012rt,Giddings:2016tla,Oshita:2018fqu,Cardoso:2019apo,Wang:2019rcf,Oshita:2019sat,Abedi:2021tti,Chakraborty:2022zlq,Chakravarti:2021jbv,Chakravarti:2021clm}, and (iii) the absence of event horizon, namely the existence of horizonless Exotic Compact Objects (ECOs)~\cite{Mazur:2004fk,Visser:2003ge,Damour:2007ap,Holdom:2016nek,Mathur:2005zp}. 
We must emphasize that strong arguments (within the context of general relativity and standard model of matter) exist against the existence of echoes and ECOs, including: (i)   the ergoregion instability \cite{Cardoso:2007az,Vicente:2018mxl,Maggio:2017ivp,Maggio:2018ivz}, (ii) the formation of a trapped surface due to the pileup of energy near the stable photon orbit \cite{Cunha:2017qtt,Keir:2014oka,Cardoso:2014sna,Ghosh:2021txu}, (iii) the collapse of ECO due to the gravity of incident GWs \cite{Chen:2019hfg,Addazi:2019bjz}, and (iv) other nonlinear effects \cite{Cardoso:2019rvt}. Nevertheless, if GW echoes do exist, their detection will serve as an important tool to study the physics of BHs or ECOs. A lot of efforts have been made to search for echoes in observed data (see Ref.~\cite{Abedi:2020ujo} for a thorough review).  As a result, constructing accurate waveform models for GW echoes is necessary
and timely \cite{Conklin:2021cbc,Mukherjee:2022wws}.

% =================nonspinning=======================%
%\begin{itemize}
%\item Zach's paper \cite{Mark:2017dnq} scalar echoes were studied on a nonrotating ECO background using an infalling particle as an initial excitation
%\item \cite{Maselli:2017tfq} a toy model, not RWZ formalism nonrotating, Fisher matrix
%\item SongMing's paper \cite{Du:2018cmp}: stochastic GW background from echo
%\item \cite{Testa:2018bzd} Analytic model for Schwarzschild:Poschl-Teller potential, Fisher matrix
%\item \cite{Wang:2018mlp}  if the post-merger object is a white hole, which is slowly pinching off (and eventually will collapse into a black hole), the late-time ringdown waveform will exhibit a series of interval-increasing echoes.
%\item \cite{Maggio:2020jml} nonspinning use the membrane diagram to relate relfectivity to viscosity etc.
%\item echo from a three-body system \cite{Fang:2021iyf}
%\item Manu's paper 
%\item scalar field propagating around fuzzballs \cite{Ikeda:2021uvc}
%\end{itemize}
If we restrict deviation from general relativity (GR) to be localized near the would-be horizon, then due to Birkhoff's theorem,  the region outside a spherically symmetric ECO can still be described by a Schwarzschild geometry. 
Consequently, studies of echoes from non-spinning ECOs were mostly based on the black hole perturbation (BHP) theory and the Zerilli-Regge-Wheeler equations \cite{Regge:1957td,1969PhDT........13Z}. For instance, Cardoso \etal \cite{Cardoso:2016rao,Cardoso:2016oxy} showed that the initial ringdown signal of different ECO models has an universal feature,  and is identical to that of a Schwarzschild BH, even though the quasinormal mode (QNM) spectra of ECOs are completely different from the ones of the Schwarzschild BH.  This implies that the initial pulse of the ringdown is more related to space-time geometry near the light ring, rather than the formal spectra of QNMs. The following echoes do depend on the structure of the QNM spectra \cite{Hui:2019aox}, which is characterized by modes trapped between the ECO surface and the peak of BH potential barrier \cite{Cheung:2021bol}. Mark \etal  \cite{Mark:2017dnq} developed a framework to systematically compute scalar echoes from non-spinning ECOs, in terms of GWs propagating toward the would-be horizon, and transfer functions that convert this horizon-going wave into echoes toward infinity.
%where echoes are written in terms of a transfer function.
%In their studies, a reflecting boundary condition was imposed at the ECO surface, and it depends on two free parameters: the compactness and the reflectivity. 
%
Testa \etal \cite{Testa:2018bzd} used a Poschl-Teller potential to approximate the BH potential for perturbations, and and derived an analytical echo template. Meanwhile, Ref.~\cite{Du:2018cmp} estimated the contribution of GW echoes to stochastic background. In terms of the membrane diagram, Maggio \etal \cite{Maggio:2020jml} and Chakraborty \etal \cite{Chakraborty:2022zlq} treated the ECO surface as a dissipative fluid, and related the reflectivity to the bulk and the shear viscosity. Cardoso \etal \cite{Cardoso:2019nis} studied resonant excitation of the modes of non-spinning ECOs during an extreme-mass-ratio inspiral. More recently, the echoes of fuzzballs \cite{PhysRevLett.125.221601,PhysRevLett.125.221602} were computed numerically in Ref.~\cite{Ikeda:2021uvc}, and the GW echo from a three-body system was studied in Ref.~\cite{Fang:2021iyf}. 
% ===================Kerr==========================%
% \begin{itemize}
% \item \cite{Nakano:2017fvh} Kerr black hole with a complete reflecting boundary condition, boundary condition was imposed on SN. No source
% \item \cite{Bueno:2017hyj} Echoes of Kerr-like wormholes, Scalar echoes from rotating wormholes 
% \item \cite{Wang:2018gin} echo for Kerr, the outside prescription; SN ( Kerr An initial pulse is sent in from null infinity and compute echoes.)
% \item \cite{Conklin:2019fcs} SN Kerr An initial pulse is sent in from null infinity and compute echoes.
% \item \cite{Wang:2019rcf} qingwen's boltzmann reflectivity, Kerr. inside vs outside, no source; SN; to match with BBH
% \item \cite{Maggio:2019zyv}: extends \cite{Testa:2018bzd} to Kerr; Still SN; use sum of qnm as the main pulse; ignore the source term (essentially inside),  analytical form
% \item initially at rest at infinity and falls into a Kerr black hole \cite{Sago:2020avw} geodesic SN
% \item \cite{PhysRevD.101.084010} scalar perturbation of Kerr, but the trajectory w as treated as geodesic.
% \item using a new boundary condition on the ECO surface that is connected to tidal tensors measured locally by zeroangular-momentum observers floating right above the horizon (fiducial observers used in the membrane paradigm) \cite{Chen:2020htz}
% \item Manu's paper \cite{Srivastava:2021uku} with source 
% \end{itemize}

In astrophysical situations, merger remnants usually have non-negligible spins~\cite{LIGOScientific:2020ibl}, hence it is of great practical interest to model echoes from spinning ECOs.  Even if GR is valid away from ECOs, the space-time geometry there can deviate significantly from Kerr, having a general multipole structure \cite{Geroch:1970cd,Hansen:1974zz}.  Nevertheless, we shall restrict ourselves to Kerr geometry, whose linear perturbation is described by the Teukolsky equation \cite{Teukolsky:1972my,1973ApJ...185..635T}. An early attempt towards constructing echo waveforms studied scalar perturbations around a Kerr-like wormhole \cite{Bueno:2017hyj}. Working on a sourceless system, Nakano \etal \cite{Nakano:2017fvh} imposed a complete reflecting boundary condition at a constant Boyer-Lindquist radius. %spinning ECO surface. 
Later, the effect of source terms was investigated \cite{Sago:2020avw,Maggio:2021uge,PhysRevD.101.084010,LongoMicchi:2020cwm,Xin:2021zir,Srivastava:2021uku}. Sago \etal \cite{Sago:2020avw} and Maggio \etal \cite{Maggio:2021uge} studied main GWs and echoes generated by  a particle that %is initially at rest at infinity and falls into a Kerr black hole. 
plunges into a Kerr black hole.  The case of a  particle (with scalar charge) sprialing into a Kerr black hole was studied in Ref.~\cite{PhysRevD.101.084010}. Refs.~\cite{LongoMicchi:2020cwm,LongoMicchi:2020cwm,Xin:2021zir,Srivastava:2021uku} further introduced the back-reaction of GW emissions on orbital motion.

%{[Do these papers above model the trajectory of a plunging particle? I don't know where to put them.  I think we have three themes: (i) boundary condition near horizon, (ii) plunging particle, (iii) comparable-mass BBH.}
%\B{No, they don't have plunging particles}
%

%The echo wave was then obtained by solving the inhomogenous Teukolsky equation. A similar study for scalar echo can be found in Ref.~\cite{PhysRevD.101.084010}.

%The boundary condition at a ECO surface, adopted by most of the above-mentioned studies, was imposed within the SN formalism \cite{Sasaki:2003xr}. 

Recently, Chen \etal \cite{Chen:2020htz} proposed a more physically-motivated boundary condition, by considering the tidal fields experienced by fiducial observers with zero angular momentum orbiting just above the ECO surface. This model established a relation between the ingoing component of the Weyl scalar $\psi_0$ and the outgoing piece of the Weyl scalar $\psi_4$. Using this new boundary condition, Xin \etal \cite{Xin:2021zir} calculated GW echoes by computing explicitly the $\psi_4$ falling down the ECO surface, and converting it to $\psi_0$ via the Teukolsky-Starobinsky (TS) identity
\cite{Starobinsky:1973aij,Teukolsky:1974yv}. They found weaker echoes than those obtained from other approaches \cite{Wang:2019rcf,Maggio:2019zyv}. A flaw in their calculation is that the TS identity is only applicable in the absence of source terms. A direct computation of $\psi_0$ propagating toward the ECO surface was later carried out by Srivastava \etal \cite{Srivastava:2021uku}.

% \begin{itemize}
% ==========take the mass ratio into account (BBH)==============%
% \item \cite{LongoMicchi:2020cwm} We include a more accurate prescription of the inspiral orbital motion incorporating the back reaction due to GW emission.  1) an adiabatic inspiral; 2) a transition phase; and 3) a geodesic plunge when the innermost stable circular orbit (ISCO) is crossed. similar to Shuo's paper \B{worth reading again}obtained echo waveforms from spinning ECOs by first deducing waves that propagate toward the horizon from waves that propagate toward infinity, and then imposing a reflectivity for Sasaki-Nakamura (SN) functions
% \item Shuo's paper; source 
% \item \cite{Annulli:2021dkw} CLA
% \item \cite{Wang:2020qcp} hybrid method
% \item \cite{Wang:2019rcf} qingwen's boltzmann reflectivity, Kerr. inside vs outside, no source; SN; to match with BBH
% \end{itemize}

\begin{figure}[htb]
    \includegraphics[width=0.48\textwidth]{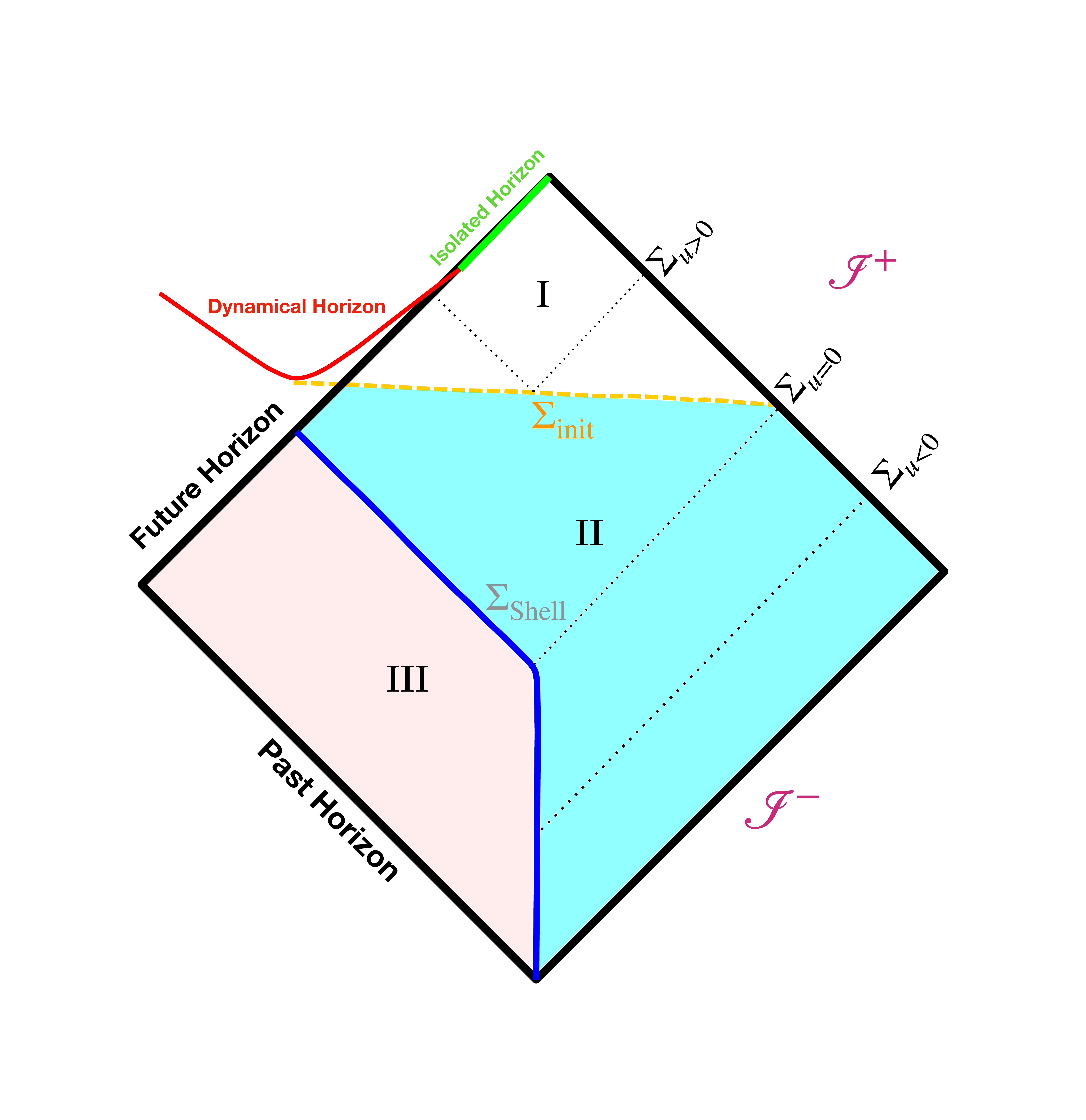}
    \caption{The space-time of a BBH merger event. The hybrid method divides the space-time into an inner PN region (III) and an outer BHP region (I+II). The two regions communicate via boundary conditions at the worldtube $\Sigma_{\rm Shell}$ (the blue curve), which was assumed to track the motion of the BH. The dynamical horizon (the red curve) lies inside the future horizon, and it eventually settles down to the isolated horizon. The common horizon forms at the time slice $\Sigma_{\rm init}$ (the horizontal dashed line). The time slice $\Sigma_{\rm init}$ is not unique and is determined by gauge conditions. The CLA focuses exclusively on the region I, where the system is treated as a Cauchy problem--- an initial data needs to be provided on $\Sigma_{\rm init}$, whereas the hybrid method gives attention to both region I and II and handles the system as a boundary value problem.  }
    \label{fig:my_label_new}
\end{figure}

%Although there have been extensive researches focusing on the echoes from extreme mass ratio inspirals (EMRIs), the study of echo for comparable-mass binaries is still at a preliminary stage. 
As we move away from extreme mass ratio inspirals, several approaches have been adopted to model echoes from comparable-mass binary black-hole (BBH) mergers. These include the {\it inside/outside} formulations, which do not involve modeling the merger dynamics; the adaptation of the Effective One-Body (EOB) \cite{Buonanno:1998gg,2014IJMPD..2350064H}; and the Close-Limit Approximation (CLA) approaches \cite{PhysRevLett.72.3297,PhysRevLett.77.4483,PhysRevD.56.6336,PhysRevLett.83.3581}, which have played important roles in modeling BBH ringdown waveforms in GR.
%The EOB formalism and the CLA has proven to be powerful tools to handle comparable-mass BBH systems, and they have been used to compute GW echoes. 

In the \textit{outside prescription}~\cite{Wang:2018gin,Conklin:2019fcs}, the main GR GW emitted by a BBH merger was modeled as having been generated by the reflection of an initial pulse originated from null infinity (see Fig.~1 in Ref.~\cite{Wang:2018gin}). The rest of this pulse travels through the light-ring potential, bounces back and forth between the surface of ECO and the peak of the potential. As a result, a sequence of echoes follows the main GR GW at null infinity. 
In the \textit{inside prescription}  \cite{Wang:2019rcf,Maggio:2019zyv}. 
%As
%opposed to the outside approach, this method models 
the main GR GW was modeled instead as the transmitted wave of an initial wave emerging from the past horizon (see Fig.~1 in Ref.~\cite{Wang:2019rcf}). Wang \etal \cite{Wang:2019rcf} computed this  initial wave by matching the main GW to that of a BBH merger event, whereas Maggio \etal \cite{Maggio:2019zyv} treated the main pulse as a superposition of QNMs, which led to analytical echo templates. Both the inside and outside prescriptions make direct connections between the main BBH GW and the ensuing echoes; they do not require detailed modeling of the merger dynamics. 
%, based on the assumption of the initial incident wave. 
%Therefore, these two methods do not need the detailed motion of the small perturber. 
%

%Wang \etal \cite{Wang:2019rcf} used the inside prescription and adjusted the incident pulse from the past horizon so that the main GW matches the LIGO template for GW150914 \cite{2016PhRvL.116f1102A}. 

In contrast, the approach based on the EOB formulation does rely on the orbital dynamics. Following the same spirit as the EOB method, Micchi \etal \cite{LongoMicchi:2020cwm} considered the back-reaction on the orbital evolution due to GW emissions. With a more accurate orbital dynamics, they were able to obtain a complete inspiral-merger-ringdown waveform and the subsequent echoes. Xin \etal \cite{Xin:2021zir} calibrated the dissipative force to a surrogate model \cite{Field:2013cfa,Varma:2019csw} so that the GW at infinity matches the prediction of numerical relativity (NR).

Recently, the CLA  approach was applied to computation of echoes from a head-on collision of two equal-mass ECOs \cite{Annulli:2021dkw}, where the Brill-Lindquist initial data \cite{PhysRev.131.471} for two BHs was ported into a linear perturbation of a single Schwarzschild space-time, with a modified boudary condition on a surface right above the horizon.

In addition to the EOB and CLA approaches, a so-called \textit{hybrid approach} \cite{Nichols:2010qi,Nichols:2011ih} has also been proposed to jointly use Post-Newtonian (PN) and Black-Hole Perturbation (BHP) theories to model comparable-mass BBH mergers. 
To illustrate this method, a 
%which was designed to study the coalescence of BBH, especially when the remnant object does not spin. 
%To exhibit the idea of the hybrid method, we provide a 
%
Penrose diagram of a BBH merger space-time is shown in Fig.~\ref{fig:my_label_new}. The space-time is split by a time-like world tube $\Sigma_{\rm Shell}$ (which asymptotes toward a null tube in its upper-left section) into an inner PN region III and an outer BHP region (I+II). The hybrid approach offers a way to construct space-time geometries in both regions --- including GWs at null infinity; it was able to accurately predict the GW waveform and kick velocity of a head-on collision \cite{Nichols:2010qi,Nichols:2011ih}. 

In this paper, we shall take a similar point of view as the hybrid approach --- by dividing the space-time into a linear BHP region (I and II in Fig.~\ref{fig:my_label_new}) and a region (III) in which the space-time is not a linear perturbation of the remnant BH.  We shall not attempt to {\it approximately solve for} the entire space-time geometry in all regions, but instead use gravitational waveform at the null infinity $\mathscr{I}^+$ already obtained from NR, and reconstruct the space-time geometry in the BHP region --- including GWs propagating toward the future horizon \Hp.  In particular, we find the location of the worldtube $\Sigma_{\rm Shell}$ at \Hp can be determined by looking for when the linearly quasi-normal ringing of horizon GW starts.
%we shall utilize this hybrid method to reconstruct the metric outside the shell $\Sigma_{\rm Shell}$ and to infer the GW that falls down into future horizon \Hp. 
Equipped with this information, together with the recent physically-motivated boundary condition near the would-be future horizon \cite{Chen:2020htz}, we can construct gravitational echoes at $\mathscr{I}^+$.

%Since the hybrid method has not been well justified when the remnant has a spin, here
As a first step toward demonstrating our space-time reconstruction approach, in this paper, we restrict ourselves to inspiraling BBHs whose remnants are non-rotating\footnote{The initial parameters of BBHs are fine-tuned so that the remnants are Schwarzschild BHs}\footnote{Our method will also be applicabl to head-on collisions.}. Specifically, we shall use a NR technique \textit{Cauchy-characteristic extraction} (CCE) \cite{PhysRevD.54.6153,PhysRevD.56.6298,Winicour:2008vpn,PhysRevLett.103.221101,Moxon:2020gha,Moxon:2021gbv} to extract the Weyl scalars $\psi_4$ and $\psi_0$ of the BBH events in question, and use them to reconstruct space-time geometry in the linear BH regions I and II. 

%they as inputs to construct our echo template. 

This paper is organized as follows. In Sec.~\ref{sec:qualitative_description_for_hybrid_method} we explain more details about space-time reconstruction using Fig.~\ref{fig:my_label_new} and outline the basic ideas of the hybrid method. We then describe our NR techniques and simulations in Sec.~\ref{sec:Numerical_Relativity_simulations}. Taking these NR simulations we explicitly carry out space-time reconstruction in Sec.~\ref{sec:kerr_spacetime}, in particular obtaining gravitational waves propagating toward the future horizon $\mathscr{H}^+$.  With these horizon waveforms, we construct gravitational-wave echoes at $\mathscr{I}^+$ in Sec.~\ref{sec:construction_echo}. Section~\ref{sec:Detectability_and_parameter_estimation} focuses on the detectability of GW echo and parameter estimation, using the Fisher information matrix formalism. Finally in Sec.~\ref{sec:conclusion} we summarize our
results.

Throughout this paper we use the geometric units with
$G=c=1$. Unless stated otherwise, we use the remnant mass $M_f$ to normalize all dimensional quantities\footnote{Namely $M_f=1$.} (e.g., time, length, and Weyl scalars). Note that this choice is different from the typical convention adopted by the NR community, where the initial total mass of the system $M_{\rm tot}$ is used.

\section{Space-time reconstruction from gravitational waves at future null infinity: theory}
\label{sec:qualitative_description_for_hybrid_method}

%In this section, we review some essentials of the hybrid method
%, and use it to give a quantitative description for the BBH spacetime of \ind~and \indnew. We 
In this section, we shall describe our theoretical strategy for space-time reconstruction based on BBH GWs at the future null infinity \scrip.  We shall divide the entire space-time into two regions, the black-hole perturbation region (I+II in Fig.~\ref{fig:my_label_new}), and the strong-field region (III in Fig.~\ref{fig:my_label_new}), as proposed during the construction of the hybrid model for BBH coalescence~\cite{Nichols:2010qi,Nichols:2011ih}.  In Sec.~\ref{sec:hybrid_method_intro}, we shall review the hybrid method, focusing on how space-time geometry in the bulk of the BHP region depends on boundary values.  In Sec.~\ref{sec:spacetime_reconstruction_from_homogeneous}, we discuss in particular how the bulk geometry can be expressed in terms of waves at \scrip. In Sec.~\ref{sec:spacetime_reconstruction_horizon_waveform}, we focus on GWs that propagate toward the future horizon \Hp, in particular propose a way to determine the boundary between the BHP region II and the strong field region III.  In Sec.~\ref{sec:spacetime_reconstruction_comparison_other_method}, we comment on how our approach is connected to previous works.

\subsection{From the hybrid method to space-time reconstruction}
\label{sec:hybrid_method_intro}
In the Penrose diagram of a coalescing BBH space-time (Fig.~\ref{fig:my_label_new}), the red curve represents the dynamical horizon, which is well-known to be inside the event horizon \cite{hawking1973large}.
Nichols and Chen \cite{Nichols:2010qi} proposed using a 3-dimensional time-like tube $\Sigma_{\rm Shell}$, shown as the blue curve, to divide the space-time into two regions.  The exterior regions (I+II) can be treated as a linearly perturbed Schwarzschild spacetime.  Interior to the tube $\Sigma_{\rm Shell}$, is a strong field region (III), which Nichols and Chen modeled using post-Newtonian theory; this PN metric is matched to the exterior perturbed Schwarzschild metric on the $\Sigma_{\rm Shell}$.  Note that the PN expansion for the interior space-time may break down toward the late stage of evolution, but the shell does fall rapidly to the horizon so the errors might stay within the BH potential and not propagate toward infinity.

For a head-on collision, the tube $\Sigma_{\rm Shell}$ passes through the centers of the two BHs, and follows plunge geodesic of the remnant BH (i.e., the BH on which regions I and II are based). 
%
%In their case \cite{Nichols:2010qi}, a head-on collision was studied and the matching tube passed the centers of two BHs --- it underwent a plunging geodesic motion in Schwarzschild space-time. Note that the PN prescription for the interior space-time may break down during the plunge, but the shell falls so rapidly to the horizon that the errors stay within the BH potential.
%
A more sophisticated framework was developed later \cite{Nichols:2011ih} to determine the motion of $\Sigma_{\rm Shell}$ for an inspiralling BBH system. This framework added a radiation-reaction force to account for the dissipative effect of GW emission. In the end, this PN-BHP system, accompanied by the no-incoming-wave condition at $\mathscr{I}^-$, forms a complete set of evolution equations, which leads to an approximated, {\it ab initio} waveform model. This method was able to predict a reasonable waveform for a BBH system merging in quasi-circular orbits.

In this paper, we focus mainly on the region I+II, where the space-time is treated as a linear perturbation to a Schwarzschild BH. Let us first examine this linear perturbation using the  Sasaki-Nakamura (SN) formalism \cite{Sasaki:2003xr}, in which the SN variable $\tensor[_{s}]{{\Psi_{lm}^{\rm SN}}}{}$ %and Zerilli functions $\Psi_{l,m}^{e}$ \cite{1969PhDT........13Z}, respectively. 
satisfies the Regge-Wheeler (RW) equation \cite{Regge:1957td}
\begin{align}
    \left(\frac{\partial^2 }{\partial u\partial v}+\frac{V_{\rm RW}^l}{4}\right)\tensor[_{s}]{{\Psi_{lm}^{\rm SN}}}{}=0, \label{RWZ}
\end{align}
where $u=t-r_*$ and $v=t+r_*$ are the retarded and advanced time, respectively, with the tortoise coordinate $r_*=r+2\ln \left(\frac{r}{2}-1\right)$. The RW potential reads \cite{Leaver:1985ax}
\begin{align}
	V_{\rm RW}^l=\frac{\Delta}{r^5}[(l^2+l)r-2(s^2-1)].
\end{align}
Here $s$ corresponds to the spin weight of $\tensor[_{s}]{{\Psi_{lm}^{\rm SN}}}{}$ and $\Delta=r^2-2r$. In the hybrid approach, no-incoming wave condition was imposed on \scrim, while PN data was imposed on $\Sigma_{\rm shell}$.  One way to obtain $\tensor[_{s}]{{\Psi_{lm}^{\rm SN}}}{}$ throughout regions I+II  from these boundary conditions is to use the characteristic method, as we discuss in Appendix~\ref{app:Characteristic_Approach}. 

In this paper, while keeping the no-incoming condition on \scrim,  we shall revert the rest of the  reconstruction process, by imposing outgoing waves obtained from NR on \scrip (e.g., with the CCE method).
%, we reconstruct $\tensor[_{s}]{{\Psi_{lm}^{\rm SN}}}{}$ in the entire region I+II. 
%
In particular,  we will obtain perturbative fields near \Hp, which will inform us the gravitational waveform going down the horizon, and serve as a foundation for obtaining GW echoes.

%Instead, we want to infer the amount of GW that falls down the future horizon, given a GW waveform at null infinity $\mathscr{I}^+$ (from CCE). Therefore, we need to reverse the process of the original hybrid method and evolve the sytem from $\mathscr{I}^+$ backward into the bulk. 

 %Boundary conditions are required to close the differential equation solution set. At $\mathscr{I}^-$, we use the no-incoming-wave condition, i.e., $\tensor[_{s}]{{\Psi_{lm}^{\rm SN}}}{}=0$, whereas at $\mathscr{I}^+$, we use the value of Weyl scalars $\psi_{4(lm)}^\infty(u)$ and $\psi_{0(lm)}^\infty(u)$ from CCE [Their definitions can be found in Eqs.~(\ref{psi0_psi4_inf_def})].  Note that here the retarded time $u$ is used as the temporal coordinate. To be concrete,  we impose
\mycomment{
\begin{subequations}
\begin{align}
    &\tensor[_{+2}]{{\Psi_{lm}^{\rm SN}}}{}=4\ddot{\psi}_{0(lm)}^\infty, \\
    &_{-2}\ddot{\Psi}_{lm}^{\rm SN}=\frac{1}{4}(l-1)l(l+1)(l+2)\psi_{4(lm)}^\infty+3\dot{\psi}_{4(lm)}^\infty,
\end{align}
\label{RW_boundary_condition}%
\end{subequations}
where the conditions are based on the generalized Chandrasekhar–Sasaki–Nakamura transformation \cite{1975RSPSA.343..289C,Sasaki:1981sx,Hughes:2000pf} (see Appendix \ref{app:CSN} for more details).}

\begin{figure}[htb]
        \includegraphics[width=\columnwidth,clip=true]{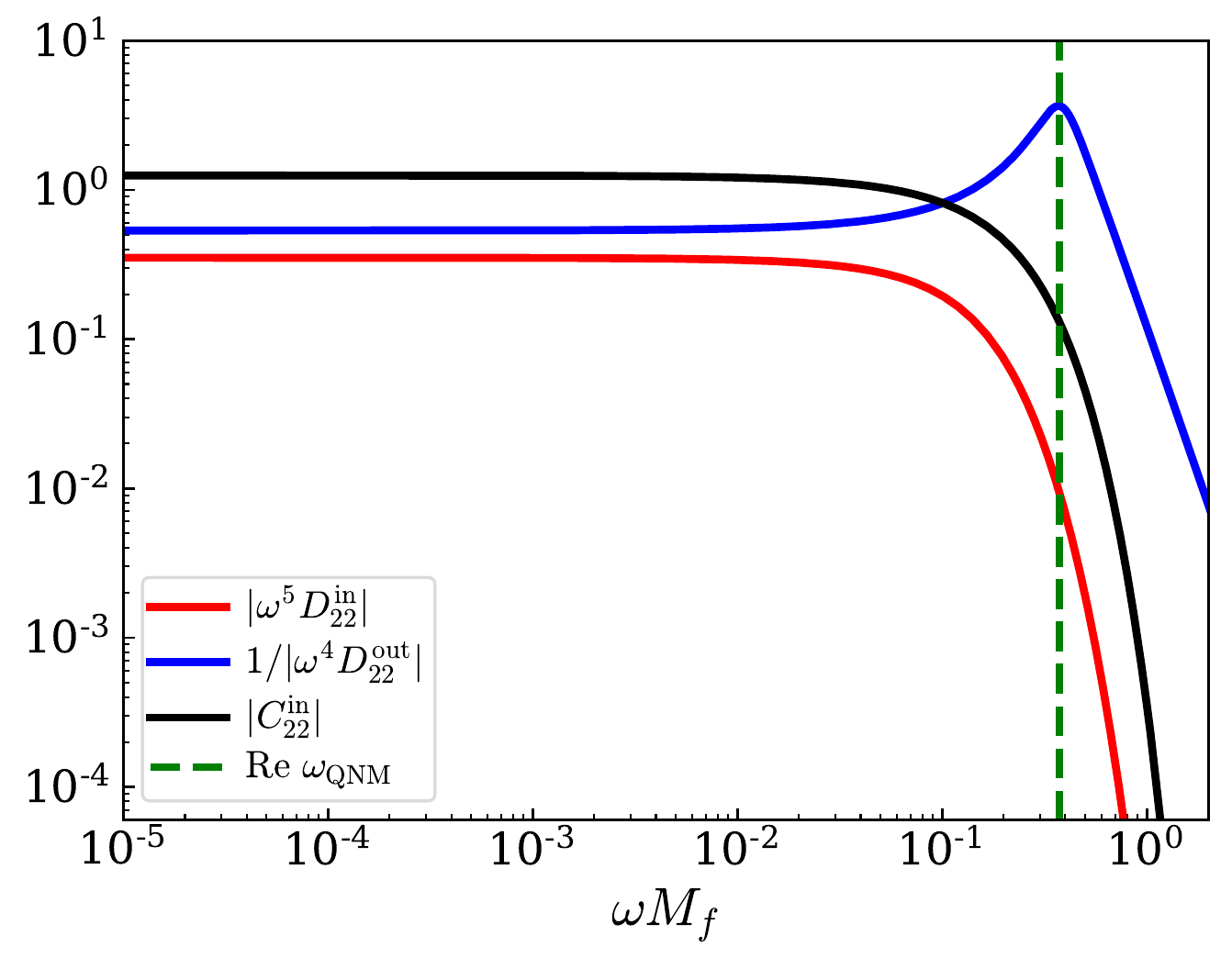}
  \caption{The coefficients $C^{\rm in}_{lm\omega}$ and $D^{\rm in/out}_{lm\omega}$ predicted by the Teukolsky equation, assuming a Schwarzschild BH. The vertical dashed line stands for the real part of the fundamental QNM $(0.374-0.0890i)$. Data are obtained from the Black Hole Perturbation Toolkit \cite{BHPToolkit}.}
 \label{fig:Din_Dout}
\end{figure}

\subsection{Space-time reconstruction using homogeneous Teukolsky solutions }
\label{sec:spacetime_reconstruction_from_homogeneous}

As we reconstruct space-time geometry, instead of SN variables, we will directly consider both 
%In this paper, we shall consider both 
$\psi_0$ and $\psi_4$, because they both have explicit physical meanings, as explained in Ref.~\cite{Chen:2020htz}.
%together play important roles in constructing gravitational-wave echoes 
  Since the new boundary \scrip $\cup$ \scrim for space-time reconstruction has a regular shape (unlike $\Sigma_{\rm shell}$), we can carry out space-time reconstruction by superimposing homogeneous solutions to the Teukolsky equation that already satisfy no-ingoing boundary condition --- traditionally referred to as the {\it up solutions}.

Let us first write general homogeneous solutions for $\psi_0$ and $\psi_4$ in mode expansions:
\begin{subequations}
\begin{align}
    \psi_4 (t,r,\theta,\phi) &= \frac{1}{r^4}\sum_{lm}\int d\omega\, _{-2}R_{lm\omega}(r) \, {}_{-2}Y_{lm}(\theta,\phi) e^{-i\omega t}, \\
    \psi_0 (t,r,\theta,\phi) &= \sum_{lm} \int d\omega\, _{+2}R_{lm\omega}(r) \, {}_{+2}Y_{lm}(\theta,\phi) e^{-i\omega t}.
\end{align}
\end{subequations}
Here $_{s}Y_{lm}$ are spin-weighted spherical harmonics. The radial functions $_{s}R_{lm\omega}(r)$ satisfy  the radial Teukolsky equation \cite{1973ApJ...185..635T}
\begin{align}
    \Delta^{-s}\frac{d}{dr}\left(\Delta^{s+1}\frac{d}{dr}\tensor[_{s}]{{R_{lm\omega}}}{}\right)+V\tensor[_{s}]{{R_{lm\omega}}}{}=0, \label{teukolsky-equation}
\end{align}
with
\begin{align}
    &V=4is\omega r-l(l+1)+\frac{r^4\omega^2-2is(r-M)r^2\omega}{\Delta}. \notag 
\end{align}
The up solutions, with their conventional normalization (with unity outgoing wave amplitude at infinity), have the following asymptotic forms near infinity and horizon
\begin{subequations}
\label{eq:upexp}
\begin{align}
&\tensor[_{-2}]{{R^{\rm up}_{lm\omega}}}{} \sim 
\begin{cases}
r^3 e^{i\omega r_*} ,\quad  & r_*\rightarrow +\infty, \\
\\
 D^{\rm out}_{lm\omega} e^{i \omega  r_*} + \Delta^2 D^{\rm in}_{lm\omega} e^{-i\omega r_*},   & r_*\rightarrow -\infty,
\end{cases} 
\\
&\tensor[_{+2}]{{R^{\rm up}_{lm\omega}}}{} 
\sim  
\begin{cases}
r^{-5} e^{i\omega r_*} \,,\quad  & r_*\rightarrow +\infty, \\
\\
 C^{\rm out}_{lm\omega} e^{i \omega  r_*} + \Delta^{-2} C^{\rm in}_{lm\omega} e^{-i\omega r_*} , \!\!  & r_*\rightarrow -\infty.
\end{cases}
\end{align}
\label{asymptotic-psi0-psi4_sourceless}%
\end{subequations}
Numerical values of the  coefficients $C^{\rm in/out}_{lm\omega}$ and $D^{\rm in/out}_{lm\omega}$ are available from the Black-Hole Perturbation Toolkit \cite{BHPToolkit}.

In a BBH coalescence space-time, the $\psi_0$ and $\psi_4$ in the I+II region have the following asymptotic forms:
\begin{subequations}
\label{eq:BBHexp}
\begin{align}
&\tensor[_{-2}]{{R^{\rm BBH}_{lm\omega}}}{} \sim 
\begin{cases}
r^3 Z^{\infty}_{lm\omega} e^{i\omega r_*} ,\quad  & r_*\rightarrow +\infty, \\
\\
 Z^{\rm H\, out}_{lm\omega} e^{i \omega  r_*} + \Delta^2 Z^{\rm H\,in}_{lm\omega} e^{-i\omega r_*},   & r_*\rightarrow -\infty,
\end{cases} 
\\
&\tensor[_{+2}]{{R^{\rm BBH}_{lm\omega}}}{} 
\sim  
\begin{cases}
r^{-5} Y^{\infty}_{lm\omega} e^{i\omega r_*} \,,\quad  & r_*\rightarrow +\infty, \\
\\
 Y^{\rm H\, out}_{lm\omega} e^{i \omega  r_*} + \Delta^{-2} Y^{\rm H\,in}_{lm\omega}  e^{-i\omega r_*} , \!\!  & r_*\rightarrow -\infty.
\end{cases}
\end{align}
\label{asymptotic-psi0-psi4}%
\end{subequations}
Here the amplitudes at infinity, $Z^{\infty}_{lm\omega}$ and $Y^{\infty}_{lm\omega}$ in Eq.~\eqref{eq:BBHexp}, can be directly obtained from NR simulations. For completeness, the strain $h_{lm}^{\infty}$ observed at $\mathscr{I}^+$ is related to $Z_{lm\omega}^{\infty}$ via
\begin{align}
    h_{lm}^{\infty}(\omega)=\frac{1}{\omega^2}Z_{lm\omega}^{\infty}.
\end{align}
Note that $h_{lm}^{\infty}$ is defined later in Eq.~(\ref{h_inf_def}). By comparing Eqs.~(\ref{asymptotic-psi0-psi4}) with the standard up solutions in Eqs.~\eqref{eq:upexp}, we can obtain amplitudes near the horizon: 
\begin{subequations}
\label{eq:recon}
\begin{align}
    \label{eq:recon:a}
    Z^{\rm H\, out}_{lm\omega} = D^{\rm out}_{lm\omega}  Z^{\infty}_{lm\omega} \,,\quad     Z^{\rm H\, in}_{lm\omega} = D^{\rm in}_{lm\omega}  Z^{\infty}_{lm\omega}, \\
    \label{eq:recon:b}
        Y^{\rm H\, out}_{lm\omega} = C^{\rm out}_{lm\omega}  Y^{\infty}_{lm\omega} \,,\quad     Y^{\rm H\, in}_{lm\omega} = C^{\rm in}_{lm\omega}  Y^{\infty}_{lm\omega}.
\end{align}
\label{solution_BBH_asym}%
\end{subequations}
%We have checked that Eqs.~(\ref{solution_BBH_asym}) are consistent with the numerical solutions to the RW equation in Eq.~(\ref{RWZ}).
In this way,  from waves escaping at infinity, $Z^{\infty}_{\lmw}$ and $Y^{\infty}_{\lmw}$, the coefficients  $D^{\rm in}_\lmw$ and $C^{\rm in}_\lmw$ will allow us to reconstruct ingoing waves $Z^{\rm in}_\lmw$ and $Y^{\rm in}_\lmw$ toward \Hp. We plot $D_{22\omega}^{\rm in}$ and $C_{22\omega}^{\rm in}$ in Fig.~\ref{fig:Din_Dout}.
%, assuming a Schwarzschild BH.

\begin{figure}
    \centering
    \includegraphics[width=0.475\textwidth]{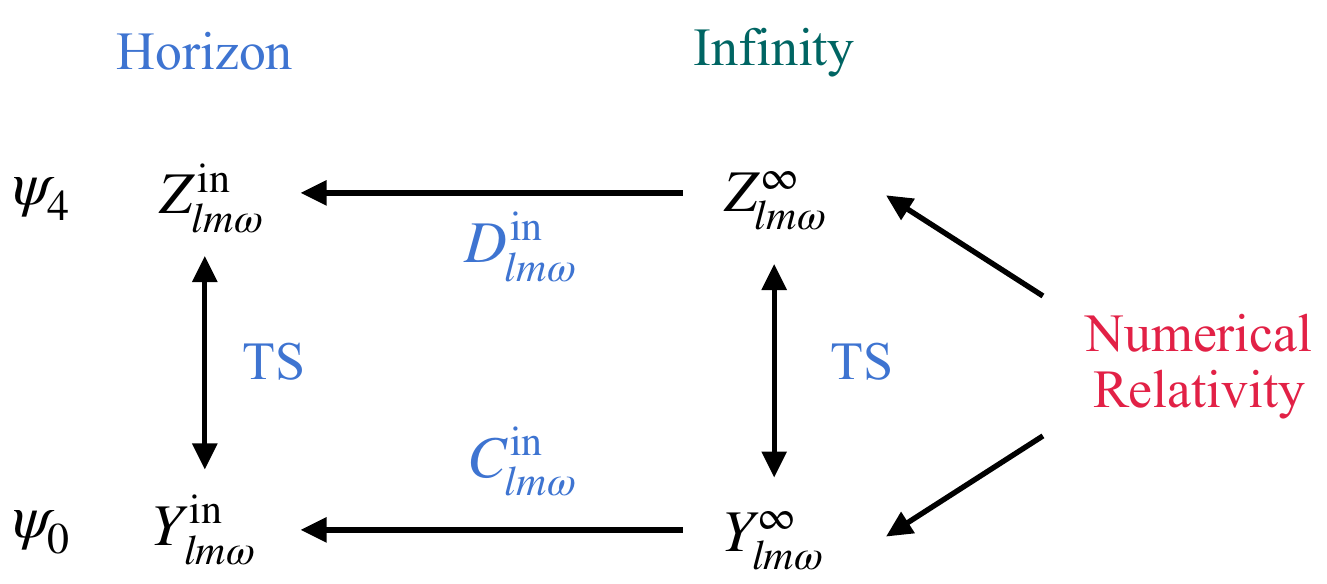}
    \caption{A diagram summarizing relations between BHP quantities on the horizon, $Z^{\rm in}_{lm\omega}$ and $Y^{\rm in}_{lm\omega}$, and those at infinity, $Z^{\infty}_{lm\omega}$ and $Y^{\infty}_{lm\omega}$. }
    \label{fig:relations}
\end{figure}

We note that for the same linear perturbative spacetime of Schwarzschild governed by the  the vacuum Teukolsky equation, the $\psi_0$ and $\psi_4$ can be related by the Teukolsky-Starobinsky (TS) relations, which state \cite{Starobinsky:1973aij,Teukolsky:1974yv}:
\begin{subequations}
\begin{align}
    &\frac{4\omega^4}{C^*}Y_\lmw^{\infty}= Z_\lmw^{\infty} \,,\quad \label{TS_inf}\\
    &Y_\lmw^{\rm H\,in} =\frac{D}{C}Z_\lmw^{\rm H\,in} \label{TS_hor}
\end{align}
    \label{TS}%
\end{subequations}
with
\begin{subequations}
\label{TS_CD}
\begin{align}
    C & =(l-1)l(l+1)(l+2)+12i\omega \label{TS_identity_C}  \\
D&=64i\omega\left(128\omega^2+8\right)\left(1-2i\omega\right).
\end{align}
\end{subequations}
These relations are consistent with coefficients in Eqs.~\eqref{eq:recon}. For example, because\footnote{We have checked that Eq.~(\ref{TS-infinity-hor}) holds up to numerical accuracy, which is at the order of  $10^{-13}$ for the Black Hole Perturbation Toolkit.}
\begin{align}
    \frac{|C|^2}{4\omega^4}C^{\rm in}_{lm}=DD^{\rm in}_{lm},
    \label{TS-infinity-hor}
\end{align}
one can obtain $Y^{\rm H\,in}$ from $Z^{\infty}$ either by: (i) using the TS relation at infinity to obtain $Y^{\infty}$, followed by Eq.~\eqref{eq:recon:b}, or (ii) using Eq.~\eqref{eq:recon:a} to obtain $Z^{\rm H\,in}$, and then use the TS relation near the horizon [i.e., Eq.~(\ref{TS_hor})]. Relations between the BHP quantities have been summarized in Fig.~\ref{fig:relations}.  We will check the TS relations directly in Sec.~\ref{sec:TS-identity}.

We would like to caution here that while it has been established~\cite{Starobinsky:1973aij,Teukolsky:1974yv} that the TS transformation maps between solutions of $\psi_0$ and $\psi_4$, these work alone did not explicitly establish the one-to-one relations in Eqs.~\eqref{TS} between $Z_{lm\omega}$ and $Y_{lm\omega}$ for the same GW. Further work by Wald \cite{PhysRevLett.41.203} explicitly related both $\psi_0$ and $\psi_4$ to the Hertz potential, while more recent work by Loutrel {\it et al.}~\cite{Loutrel:2020wbw} provided a new way to reconstruct metric (hence $\psi_0$)  from $\psi_4$.  From Ref.~\cite{Loutrel:2020wbw}, for the same, generic GW, the one-to-one relation is in between  $(Z_{l,m,\omega},Z_{l,-m,-\omega})$ and $(Y_{l,m,\omega},Y_{l,-m,-\omega})$, rather than simply between $Z_{lm\omega}$ and $Y_{lm\omega}$.  Nevertheless, as will be seen later in this paper (see Sec.~\ref{sec:TS-identity}), our numerical results for $\psi_0$ and $\psi_4$ do agree with Eqs.~\eqref{TS}.  This might be due to the fact that we have non-precessing systems which satisfy \cite{Boyle:2014ioa}
\begin{equation}
    Z_{l,m,\omega} =(-1)^lZ^*_{l,-m,-\omega}\,,\quad 
    Y_{l,m,\omega} =(-1)^lY^*_{l,-m,-\omega}\,.
\end{equation}
However, for more generic, e.g., precessing binaries, the naive TS relation Eq.~(\ref{TS}) may not hold.

\subsection{Connection to the inside prescription and determining the location of $\Sigma_{\rm Shell}$}
\label{sec:spacetime_reconstruction_horizon_waveform}

\begin{figure}[htb]
        \includegraphics[width=\columnwidth,clip=true]{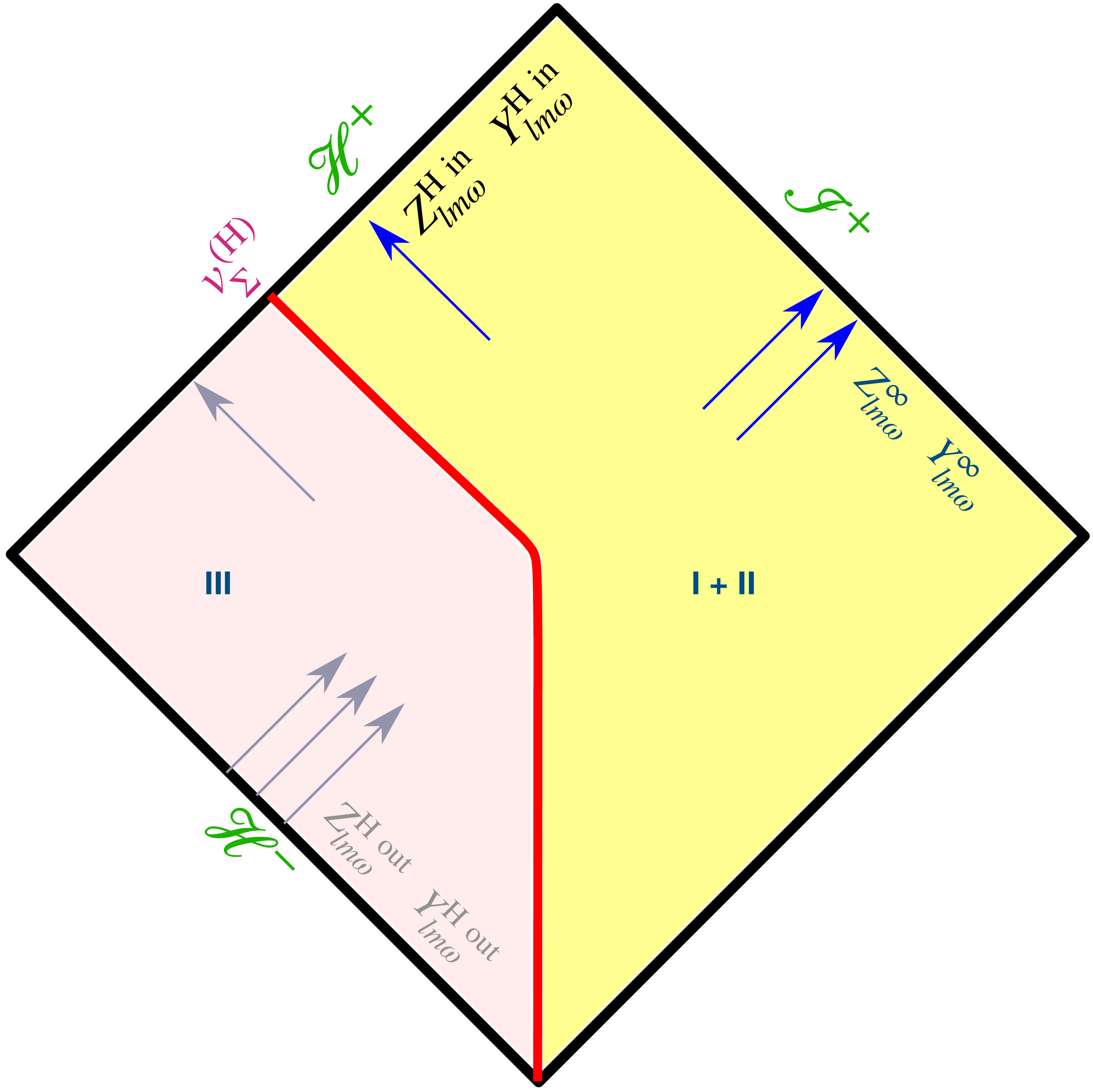}
  \caption{The space-time diagram illustrating the BHP region I+II and their linear extrapolation into region III. Outside the matching shell, curvature perturbations are linear combinations of the up-mode solutions to the homogeneous Teukolsky equation. At the infinity $\mathscr{I}^+$, the value of $Z_{lm\omega}^{\infty}$ and $Y_{lm\omega}^{\infty}$ are chosen to be consistent with the predictions of CCE. The past horizon exists in the strong gravity region III, where $Z_{lm\omega}^{\rm H~out}$ and $Y_{lm\omega}^{\rm H~out}$ represent the image wave that give rise to waves in the region I+II. They serve the same role as the initial wavepacket within the inside prescription \cite{Wang:2019rcf,Maggio:2019zyv}. The future horizon lies partially outside the matching shell, only the outside portion $(v>v_\Sigma^{(\rm H)})$ of $Z_{lm}^{\rm H~in}$ and $Y_{lm}^{\rm H~in}$ corresponds to the actual wave that falls down the horizon. One natural way to self-consistently determine the location of $\Sigma_{\rm shell}$ is to evaluate the starting time after which $Y^{\rm H~in}_{lm}(v)$ can be decomposed as a sum of QNM overtones. More details can be found in Sec.~\ref{sec:overtone_fit}.}
 \label{fig:image_wave}
\end{figure}

To understand the phyiscal meaning of  $Z^{\rm H~out}_\lmw$ and $Y^{\rm H~out}_\lmw$, which mathematically appears to be emitted from the past horizon \Hm, we have to go to Fig.~\ref{fig:image_wave} and remind ourselves that region I+II does not contain the past horizon of the background BH.  Anything below the red curve (the Shell) in Fig.~\ref{fig:image_wave} are {\it linear extrapolations}.  Nevertheless, this extrapolation asserts that waveforms at infinity can be thought of as generated by ``image waves'' with $Z^{\rm H\,out}$ and $Y^{\rm H\,out}$ that rise from the past horizon. This follows the same reasoning as the inside prescription \cite{Wang:2019rcf,Maggio:2019zyv}.

Since the image wave encounters the BH potential barrier (from the inside), it is partially transmitted toward \scrip, while partially reflected toward \Hp.  We can rewrite
\begin{subequations}
\begin{align}
    Z^{ \infty}_{lm\omega} = \frac{1}{D^{\rm out}_{lm\omega}}  Z^{\rm H~ out}_{lm\omega} \,,\quad     Z^{\rm H\, in}_{lm\omega} = \frac{D^{\rm in}_{lm\omega}}{D^{\rm out}_{lm\omega}} Z^{\rm H\, out}_{lm\omega} \\
            Y^{ \infty}_{lm\omega} = \frac{1}{C^{\rm out}_{lm\omega}}  Y^{\rm H~ out}_{lm\omega} \,,\quad     Y^{\rm H\, in}_{lm\omega} = \frac{C^{\rm in}_{lm\omega}}{C^{\rm out}_{lm\omega}} Y^{\rm H\, out}_{lm\omega}
\end{align}
\end{subequations}
Here $1/D^{\rm out}_{lm\omega}$ and $1/C^{\rm out}_{lm\omega}$ are the transmissivities from \Hm to \scrip, across the potential barrier, while $D^{\rm in}_{lm\omega}/D^{\rm out}_{lm\omega}$ and $C^{\rm in}_{lm\omega}/C^{\rm out}_{lm\omega}$ are reflectivities at the potential barrier that direct the wave toward \Hp.  (The dependence of  $1/D^{\rm out}_{22\omega}$  on $\omega$ is plotted in Fig.~\ref{fig:Din_Dout}.)

In this way, we have shown that the inside prescription and the hybrid method correspond to the same reconstruction of space-time geometry in the regime where the linear BHP applies. However, we want to emphasize that two methods adopt different ways when choosing the linear BHP region. In the hybrid method, it is given by the exterior region of $\Sigma_{\rm Shell}$. In particular, in order to compute echoes, we will need to terminate the linear perturbation region at the intersection of the shell $\Sigma_{\rm shell}$ and the future horizon, which is denoted by the advanced time $v=v_\Sigma^{({\rm H})}$ in Fig.~\ref{fig:image_wave}. One natural way to determine the intersection is to first evaluate the time-domain waveform
\begin{equation}
    Y^{\rm H~in}_{lm}(v) =\int  d\omega     Y^{\rm H~in}_\lmw e^{-i\omega v} \label{YHin-time-domain}
 \end{equation}
 and then define $v_\Sigma^{({\rm H})}$ as the starting time after which $Y^{\rm H~in}_{lm}(v)$ can be decomposed as a sum of QNM overtones. We shall provide more details when we carry out this decomposition in Sec.~\ref{sec:overtone_fit}. 

On the contrary, the inside prescription uses only the late-time evolution as the linear region. We shall give more discussions regarding this comparison in next subsection (Sec.~\ref{sec:spacetime_reconstruction_comparison_other_method}).

\subsection{Further comparisons with the inside prescription and the close limit approximation}
\label{sec:spacetime_reconstruction_comparison_other_method}

To fit the inside prescription into our framework, in Fig.~\ref{fig:my_label_new}, we choose a time slice $\Sigma_{\rm init}$ after which the space-time (i.e., the region I) is consistent with that of a single, perturbed BH. The time slice is usually not unique and is determined by a gauge condition. An appropriate choice is to let $\Sigma_{\rm init}$ represent a moment when the common horizon just forms, following the close limit approximation \cite{Price:1994pm,Abrahams:1995wd,Andrade:1996pc,Khanna:1999mh,Sopuerta:2006wj,Sopuerta:2006et,LeTiec:2009yf,Johnson-McDaniel:2009tvj}. Then the inside prescription corresponds to only taking the region I, and treating it as the linear BHP area. Consequently, one needs to take the ringdown of the main GWs at the null infinity as input, which is equivalent to  imposing a filter at $\mathscr{I}^+$ \cite{Maggio:2019zyv}, and use that information to calculate echoes. In fact, since the region II is not included, the indeterminate condition at past null infinity leaves a room for the outside prescription \cite{Wang:2018gin,Conklin:2019fcs}.

Similarly, the CLA corresponds to the region I as well. This is an approach to study the space-time based on the fact that the gravitational field in the region I can be modeled as the one of a single perturbed BH. The system in the region I is then treated as a Cauchy problem (i.e., an initial value problem) as long as an initial data is provided on $\Sigma_{\rm init}$. Previous studies have investigated the Misner initial data \cite{Misner:1960zz}, the Brill-Lindquist initial data \cite{PhysRev.131.471}, the Bowen-York initial data \cite{PhysRevD.21.2047} as well as numerically generated initial data \cite{Baker:2001sf,Campanelli:2005ia}. Once the gravitational field in region I is solved, one can read off the value of $Z^{\rm H\, in}_{lm\omega}$ and $Y^{\rm H\, in}_{lm\omega}$ at the future horizon and compute echo waveforms \cite{Annulli:2021dkw}. 

% The CLA \cite{Price:1994pm,Abrahams:1995wd,Andrade:1996pc,Khanna:1999mh,Sopuerta:2006wj,Sopuerta:2006et,LeTiec:2009yf,Johnson-McDaniel:2009tvj} is an approach to study the last stage of BBH coalescences,  when the two BHs are close enough that they are surrounded by a common horizon.  
% % The original formulation of CLA, proposed by Price and Pullin \cite{Price:1994pm}, was aimed for a head-on collision of a BBH. 
% In practice, a certain gauge condition is used to divide the entire spacetime into a bunch of time slices. In Fig.~\ref{fig:my_label_new}, for instance,  $\Sigma_{\rm init}$ represents the moment when the common horizon just forms. The CLA focuses exclusively on 

The hybrid method, however, is a boundary value problem. It divides the space-time into two regions via the time-like shell $\Sigma_{\rm Shell}$, as opposed to the space-like hypersurface $\Sigma_{\rm init}$ adopted by the CLA. In addition, both the region I and II are regarded as a BHP area.

%%%%%%%%%%%%%%%%%%%%%%%%%%%%%%%%%%%%%%%%%%%%%%%%%%%%%%%%%%%%%%%%%%%%%%%%%%%
\section{Numerical Relativity simulations}
\label{sec:Numerical_Relativity_simulations}
In this section, we adopt two BBH merger simulations performed using the Spectral Einstein Code (SpEC) \cite{spec}, developed by the Simulating eXtreme Spacetimes (SXS) collaboration \cite{Boyle:2019kee}. These binaries have their initial parameters fine-tuned, such that the remnant black holes are nearly non-spinning. 
%--- the accuracy of the hybrid method has only been well justified when the remnant does not rotate. 
%
%Below we shall neglect the final spin and treat the remnant as a Schwarzschild BH. 
Gravitational waveforms (at infinity) of these simulations are publicly available through the SXS catalog \cite{Boyle:2019kee}, with the identifier \ind~and \indnew. 

We summarize the properties of these binaries in Table \ref{table:SXS_runs}, where we adopt the standard convention in SpEC, namely labeling the heavier hole with `1' and the lighter one with `2', and assuming the $z-$axis to be aligned with the initial orbital angular momentum.  Our two systems have mass ratios $q=m_{2}/m_{1}=7$, 4, respectively; they undergo $N_{\rm cycle}=36$, 16.5 orbit cycles before the merger, with the initial orbital eccentricity already reduced to $\sim10^{-4}$. 
Both systems are non-precessing, with initial spins anti-aligned with the orbital angular momentum (or vanishing), as indicated by the negative signs of the dimensionless spin components, $\chi_1^z$ and $\chi_2^z$. 
%The negative sign of $\chi^z$ means that these initial spins are all antialigned with the orbital angular momentum. 
The remnant BHs have small spins at the $\chi_f\sim10^{-2}$ level, with the remnant mass $M_f$ slightly less than the initial total mass of the system $M_{\rm tot}=m_{1}+m_{2}$.

% \B{The other simulation we carry out is a head-on collision of an equal mass, nonspinning BBH system. The initial distance between two BHs is $150M_{\rm tot}$ --- large enough that the waveform can avoid the contamination of junk radiations. In the meantime, we also use the superposed harmonic initial data to reduce  junk radiations \cite{Varma:2018sqd}. The final BH is still non-rotating, with the mass $M_f=0.998M_{\rm tot}$.}

\begin{table}
    \centering
    \caption{A summary of NR simulations used in this paper. The first column is the identifier in the SXS catalog \cite{Boyle:2019kee}. The second column $q=m_2/m_1>1$ shows the mass ratio. The third column is the number of orbit cycles that a system undergoes before the merger. The fourth and fifth columns give the initial individual dimensionless spins. They have only the $z-$component, where the $z-$axis is chosen to be aligned with the orbital angular momentum. The sixth and seventh columns exhibit the remnant mass (in the unit of initial total mass $M_{\rm tot}$) and remnant spin. The final column corresponds to the radius of the extraction worldtube for CCE.}
    \begin{tabular}{c c c c c c c c} \hline\hline
ID& \multirow{2}{*}{$q$} & \multirow{2}{*}{$N_{\rm cycle}$} & \multirow{2}{*}{$\chi_{1}^z$} & \multirow{2}{*}{$\chi_{2}^z$} & \multirow{2}{*}{$\frac{M_f}{M_{\rm tot}}$} & \multirow{2}{*}{$\chi_f$} & Extraction \\ 
SXS:BBH: & & & & & & & Radius$(M_{\rm tot})$ \\ \hline
0207 & 7.0 & 36 & $-0.6$ & $10^{-6}$ & $0.991$ & $-0.077$ & 300 \\ \hline
1936 & 4.0 & 16.5 & $-0.8$ & $-0.8$ & 0.985 & $0.022$ & 273 \\ \hline\hline
     \end{tabular}
     \label{table:SXS_runs}
\end{table}

We extract gravitational waveforms at the null infinity $\mathscr{I}^+$ using the Cauchy Characteristic Extraction (CCE) method~\cite{Moxon:2020gha,Moxon:2021gbv}, implemented in the new NR code SpECTRE \cite{Kidder:2016hev,spectrecode}. The CCE system evolves the Einstein field equations on a foliation of null hypersurfaces, where the metric is written in the Bondi-Sachs coordinates \cite{Madler:2016xju}. This method is most efficient in evolving the space-time far from the BBH system, and is reliable enough to produce all Weyl scalars $\psi_{0,1,2,3,4}$ with high accuracy \cite{Moxon:2020gha,Moxon:2021gbv}. In practice, CCE first reads off boundary data on a worldtube covered by the inner Cauchy evolution, and then evolves a hierarchical system from the worldtube towards future null infinity. The radii of the extraction worldtubes for \ind~and \indnew~are summarized in Table \ref{table:SXS_runs}. Same as the standard treatment in NR, CCE decomposes each of the Weyl scalars $\psi_{0,1,2,3,4}$, and the strain $h$, into sums over a set of spin-weighted spherical harmonics $\tensor[_{s}]{{Y_{lm}}}{}(\theta,\phi)$. Using the notation defined in Eqs.~(\ref{asymptotic-psi0-psi4}), the decomposition reads
\begin{subequations}
\begin{align}
&[rM_{f}\psi_{4}]_{\mathscr{I}^+} =\sum_{l,m}\tensor[_{-2}]{{Y_{lm}}}{}(\theta,\phi)Z^{\infty}_{lm},\\ 
&[rh/M_{f}]_{\mathscr{I}^+} =\sum_{l,m}\tensor[_{-2}]{{Y_{lm}}}{}(\theta,\phi)h_{lm}^\infty, \label{h_inf_def}\\
&[r^5M_f^{-3}\psi_{0}]_{\mathscr{I}^+}=\sum_{l,m}\tensor[_{+2}]{{Y_{lm}}}{}(\theta,\phi)Y^{\infty}_{lm},  \label{psi0_inf_def}
\end{align}
\label{psi0_psi4_inf_def}%
\end{subequations}
where $\theta$ and $\phi$ are the polar and azimuthal angles, respectively, on the sky in the source frame. Note that in Eqs.~(\ref{psi0_psi4_inf_def}) the asymptotic $r$-dependences of $\psi_4$, $h$ and $\psi_0$, as $r\to\infty$,  are consistent with the peeling theorem \cite{penrose1984spinors}. 
%to  account for the fall-off rate in the limit of $r\to\infty$. 
%
Furthermore, these fields are normalized by the appropriate powers of $M_f$ so that $Z^{\infty}_{lm}$, $Y^{\infty}_{lm}$ and $h_{lm}^\infty$ are dimensionless. We want to emphasize again that as opposed to the usual NR convention, where the initial total mass of the system $M_{\rm tot}$ is used as the unit for time and length, in this paper, we use the remnant mass $M_f$ to normalize all dimensional quantities, because we mainly deal with perturbations of the remnant (approximately) Schwarzschild BH. 

\begin{figure}[htb]
        \includegraphics[height=11cm,width=\columnwidth,clip=true]{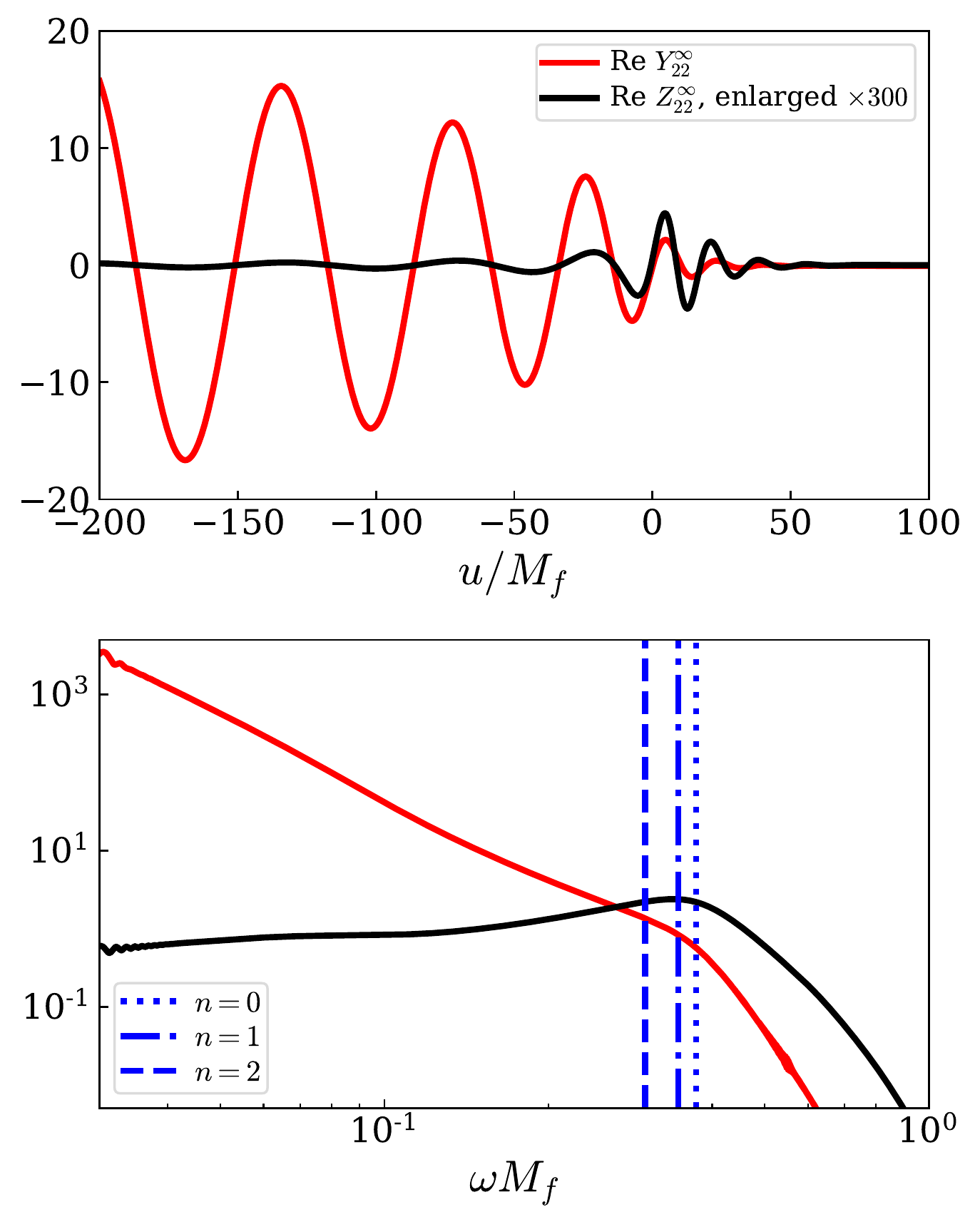}
  \caption{The spherical modes $Y_{22}^\infty$ and $Z_{22}^\infty$ of \ind, in the time domain (the upper panel), and in the frequency domain (the lower panel). The vertical lines in the lower panel stand for QNM frequencies of a Schwarzschild BH, labeled by the overtone index $n$. The absolute value of $Z^{\infty}_{22}$ is amplified by a factor of 300 for ease of read.}
 \label{fig:psi4_psi0_scri}
\end{figure}

\begin{figure*}[htb]
   \centering
\includegraphics[width=\textwidth,clip=true]{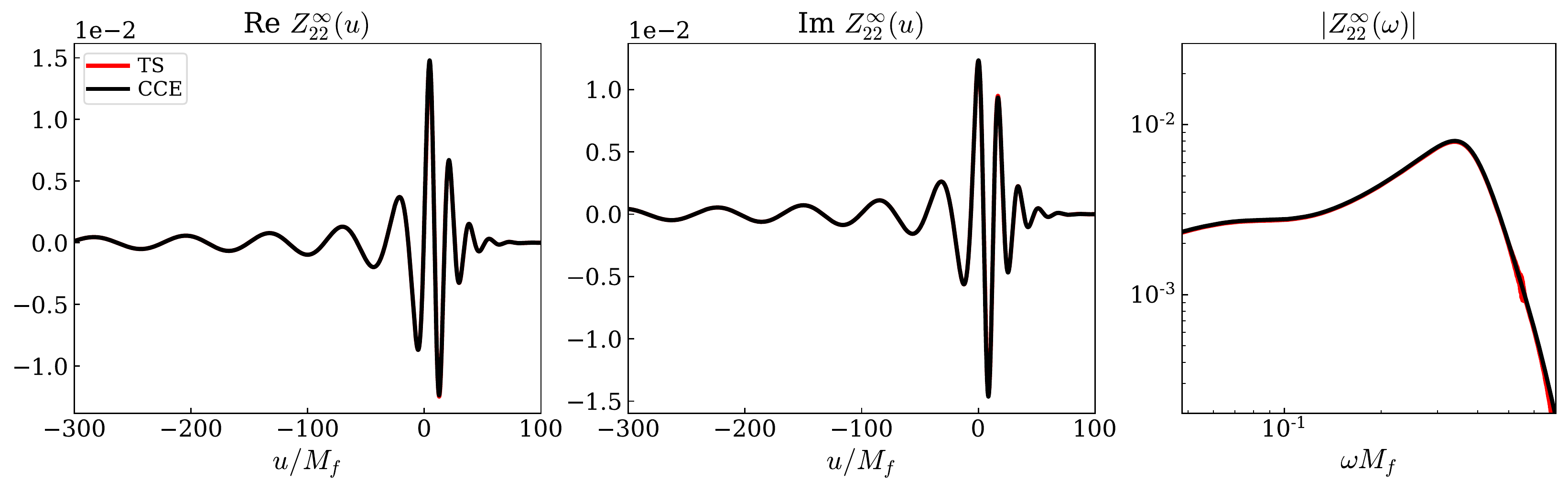}
  \caption{The validity of the TS identity at infinity [Eq.~(\ref{TS_inf})], using \ind. The predicted form $\frac{4\omega^4}{C^*}Y^{\infty}_{22}$ (in red) is compared to the actual $Z^{\infty}_{22}$ (in black), in the time domain (the left two panels), and in the frequency domain (the right panel). The comparison for \indnew~is in Fig.~\ref{fig:TS_identity_1936}.}
 \label{fig:TS_identity}
\end{figure*}

Furthermore, we shift all temporal coordinates such that $u=0$ corresponds to the peak of total rms strain amplitude:
\begin{align}
\left.\sqrt{\sum_{lm}|h_{lm}(u)|^2}\right|_{u=0}={\rm peak}. \label{peak_time}
\end{align}

\section{Numerical implementations of the hybrid method}
\label{sec:kerr_spacetime}

In this section, we apply the space-time reconstruction procedure of Sec.~\ref{sec:qualitative_description_for_hybrid_method} to  \ind~and \indnew. In Sec.~\ref{sec:TS-identity}, we first investigate the validity of TS identities at future null infinity $\mathscr{I}^+$ [see Eq.~(\ref{TS_inf})], given that the future null infinity lies completely in the BHP region. We also provide the horizon-$\psi_0$ at future horizon \Hp. Then in Sec.~\ref{sec:overtone_fit}, we use the  horizon-$\psi_0$ to determine the location of the matching tube $\Sigma_{\rm Shell}$ by looking for when its linearly quasi-normal ringing starts.

\subsection{At null infinity and future horizon: The Weyl scalars and the Teukolsky-Starobinsky identities}
\label{sec:TS-identity}
For \ind, we plot its $Z^{\infty}_{l=2,m=2}$ and $Y^{\infty}_{l=2,m=2}$ in Fig.~\ref{fig:psi4_psi0_scri}, in both time domain (upper panel) and frequency domain (lower panel). In the frequency domain, $Z^{\infty}_{22}$ (black curve) peaks at the fundamental (2,2) quasi-normal mode frequency (the vertical dotted line). On the other hand,   $Y^{\infty}_{22}$ rises up sharply in low frequencies, where its magnitude is much greater than that of $Z^{\infty}_{22}$. This feature in the frequency domain  is consistent with the TS identity at infinity [see Eq.~(\ref{TS_inf})]. To be concrete, we test the validity of Eq.~(\ref{TS_inf}) in Fig.~\ref{fig:TS_identity}. The actual $Z^{\infty}_{22}$ (in black) is compared to  $\frac{4\omega^4}{C^*}Y^{\infty}_{22}$ (in red), in the time domain (the left two panels) and frequency domain (the right panel). We see the TS identity holds throughout the entire region. The comparison for \indnew~is similar and can be found in Appendix \ref{app:1936}.

\begin{figure}[htb]
        \includegraphics[width=\columnwidth,clip=true]{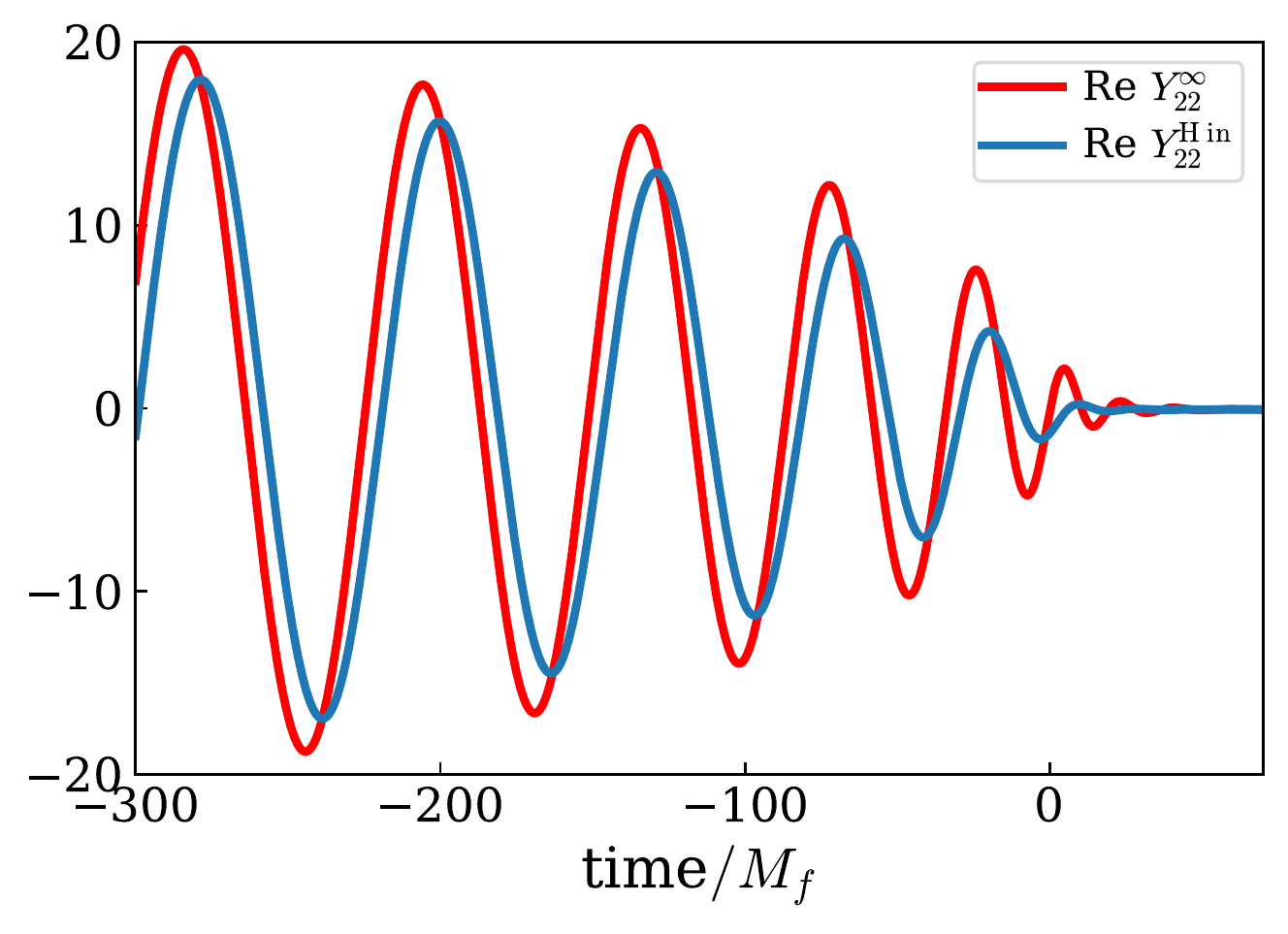}
  \caption{The real part of $Y_{22}^{\rm H\,in}$ [Eq.~(\ref{asymptotic-psi0-psi4})] and $Y^{\infty}_{22}$ in the time domain, using \ind. The temporal coordinate for $Y_{22}^{\rm H\,in}$ is $v$, while is $u$ for $Y^{\infty}_{22}$. Both coordinates are in the unit of final mass.}
 \label{fig:psi0_inf_hor}
\end{figure}

At the future horizon, $Y_{lm}^{\rm H\,in}$~[Eq.~(\ref{asymptotic-psi0-psi4})] is essential for us to compute echoes (see Sec.~\ref{sec:baoyi_reflection_condition} for more details).  In Fig.~\ref{fig:psi0_inf_hor}, we plot $Y_{22}^{\rm H\,in}$ of \ind~in the time domain (blue curve), where the advanced time $v$ is used as the time coordinate. Similar to $Y^{\infty}_{22}$ [see Fig.~\ref{fig:psi4_psi0_scri}], $Y_{22}^{\rm H\,in}$ has a dominated low-frequency content. At early stage, $Y_{22}^{\rm H\,in}$ is inside the strong gravity region III and  should be excised --- as we shall discuss in Secs.~\ref{sec:overtone_fit} and \ref{sec:echo_inside}.  For comparison, we also plot $Y^{\infty}_{22}$ in the same figure (red curve) --- using  $u$  as the time coordinate.  We caution that this comparison only has a qualitative meaning, because the two waveforms are emitted toward different directions. Showing the $v$ dependence of $Y_{22}^{\rm H\,in}$ and the $u$ dependence of $Y^{\infty}_{22}$ in the same plot effectively traces both of these waves back to the same time $t$ at $r_*=0$.  This is qualitatively meaningful because the ringdown  wave can be thought of as having originated from the light ring at $r=3M$, where $r_* \approx 0$.   From this comparison, we can see $Y_{22}^{\rm H\,in}$ decreases faster and undergoes fewer cycles of oscillation at the late phase than $Y^{\infty}_{22}$. 

%It is worth to emphasize that this comparison may not be directly meaningful because the two waveforms are emitted toward different direction, and that advanced time $v$ at future horizon is not necessarily synchronized with the retarded time $u$ at null infinity.

%For completeness, we close this subsection by the TS identity at future horizon \cite{Starobinsky:1973aij,Teukolsky:1974yv}. The identity writes
%\begin{align}
%\psi_{0(lm)}^{\rm H} =\frac{D}{C}\psi_{4(lm)}^{\rm H}, \label{TS-horizon}
%\end{align}
%where $C$ is defined in Eq.~(\ref{TS_C}).
%Plugging Eqs.~(\ref{psi0_future_horizon}) and (\ref{psi4_future_horizon}) into Eq.~(\ref{TS-horizon}), we obtain
%\begin{align}
%    \frac{|C|^2}{4\omega^4}C^{\rm in}_{lm}=DD^{\rm in}_{lm}.
%    \label{TS-infinity-hor}
%\end{align}

\subsection{Determining the location of \texorpdfstring{$\Sigma_{\rm Shell}$}{Lg}}
\label{sec:overtone_fit}
As mentioned in Sec.~\ref{sec:qualitative_description_for_hybrid_method},  the region outside the matching tube $\Sigma_{\rm Shell}$ is consistent with a sourceless, linearly perturbed Schwarzschild space-time.  Accordingly, the part of $Y_{lm}^{\rm H\,in}$ that is in region I+II can be decomposed into a sum of QNMs (in the time domain).  Conversely,  we can use this fact to determine the location of $\Sigma_{\rm Shell}$. Indeed,  this method has been used not only to determine the start time of a BBH ringdown at the future infinity\footnote{The linear perturbation regime was found to be valid as early as the peak of strain if seven overtones are included.} \cite{Giesler:2019uxc}, but also to investigate the dynamics of a final apparent horizon in a BBH system approaching to equilibrium \cite{Mourier:2020mwa}. More specifically, we write \cite{Lim:2019xrb},
\begin{subequations}
\begin{align}
&h_{22}^\infty (u>u^{(h)})=\sum_{n=0}^{n_{\rm max}} [\mathcal{A}_{n}^{(h)}e^{-i\omega_{n}u}+\mathcal{B}_{n}^{(h)}e^{i\omega_{n}^*u}], \label{hinf_fit_over} \\
&Y^{\infty}_{22} (u>u^{(\infty)})=\sum_{n=0}^{n_{\rm max}} [\mathcal{A}_{n}^{(\infty)}e^{-i\omega_{n}u}+\mathcal{B}_{n}^{(\infty)}e^{i\omega_{n}^*u}],  \\
&Y_{22}^{\rm H\,in} (v>v_\Sigma^{({\rm H})})=\sum_{n=0}^{n_{\rm max}} [\mathcal{A}_{n}^{({\rm H})}e^{-i\omega_{n}v}+\mathcal{B}_{n}^{({\rm H})}e^{i\omega_{n}^*v}], 
\end{align}
\label{overtone-fit}%
\end{subequations}
where $\omega_{n}$ is the QNM frequency of a Schwarzchild BH, and $n$ refers to the overtone index (we have restricted to $l=2$). Note that for a Schwarzchild BH, the QNM frequency is independent of its spin weight and azimuthal quantum number. Unlike Ref.~\cite{Giesler:2019uxc},  we include both prograde modes $\mathcal{A}_n$ and retrograde modes $\mathcal{B}_n$ for generality \cite{Dhani:2020nik}. In Eq.~(\ref{overtone-fit}) we use $u^{(\infty /h)}$ and $v_\Sigma^{(\rm H)}$ to indicate the time at which ringdown begins, and we emphasize again that the retarded time $u$ is used for $h_{22}^{\infty}$ and $Y^{\infty}_{22}$  at the null infinity, whereas the advanced time $v$ is used for $Y_{22}^{\rm H\,in}$ at the future horizon .

\begin{figure*}[htb]
   \centering
    \subfloat[\ind]{\includegraphics[width=\textwidth,clip=true]{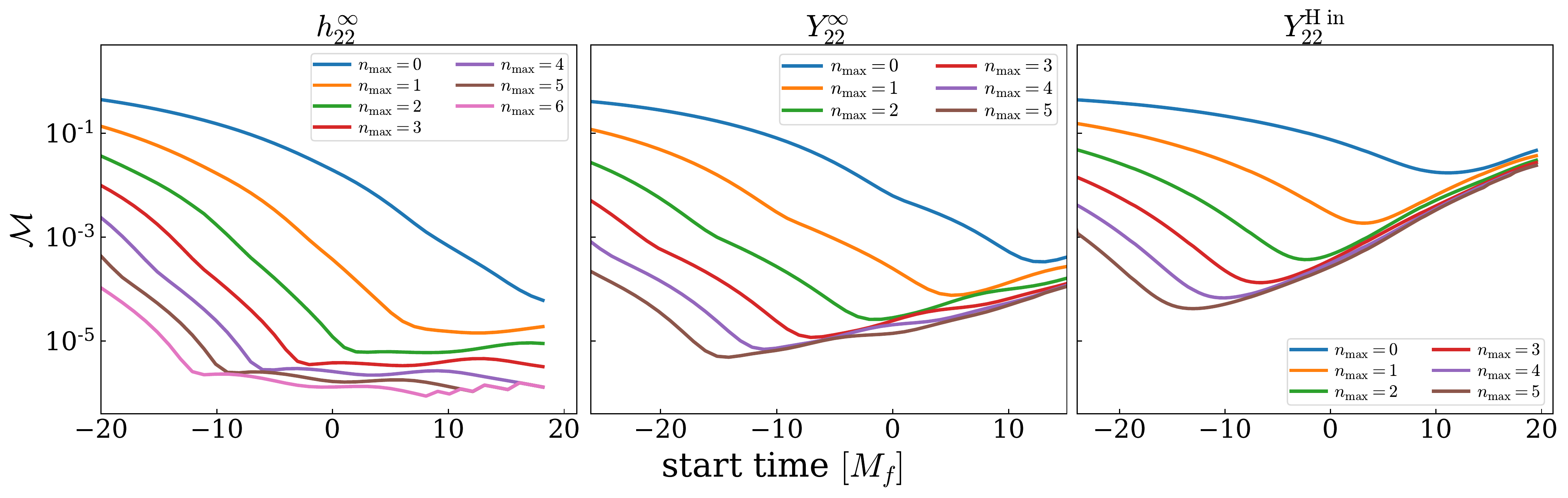}} \\
    \subfloat[\indnew]{\includegraphics[width=\textwidth,clip=true]{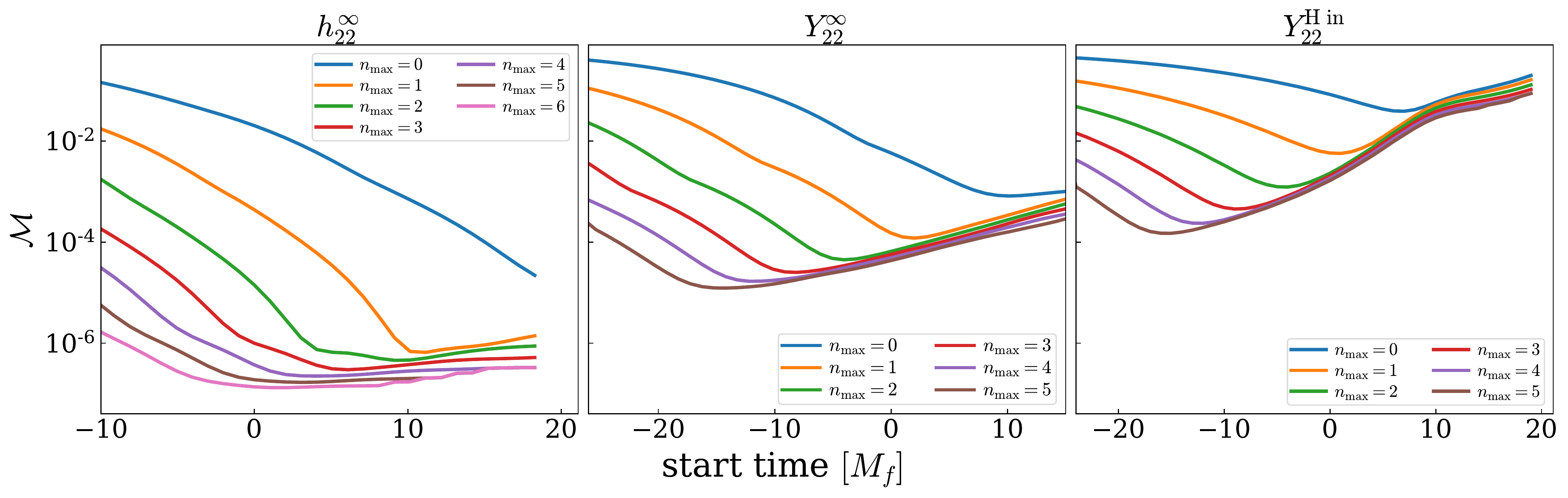}}
  \caption{Mismatch as a function of start time (in the unit of remnant mass) for different models [Eq.~(\ref{overtone-fit})]. Each model includes up to $n_{\rm max}$ overtones. The left panel corresponds to the strain $h_{22}^\infty$ at infinity, the middle one $Y_{22}^\infty$, and the right panel $Y_{22}^{\rm H\,in}$ [see Eqs.~(\ref{asymptotic-psi0-psi4}) and (\ref{h_inf_def})]. The upper row refers to \ind, whereas the lower one \indnew. All waveforms are aligned such that $t=0$ occurs at the peak of $\sqrt{\sum_{lm}|h_{lm}(t)|^2}$.}
 \label{fig:overtone_mismatch}
\end{figure*}

In making the decomposition, we follow the procedure of Ref.~\cite{Giesler:2019uxc}, namely we use the mismatch $\mathcal{M}$  between the quasi-normal mode ringdown waveform model (e.g., $h_{22}^{\rm Ringdown}$) and the NR result (e.g., $h_{22}^{\rm NR}$) as a loss function
\begin{align}
\mathcal{M}=1-\frac{(h_{22}^{\rm NR},h_{22}^{\rm Ringdown})}{\sqrt{(h_{22}^{\rm Ringdown},h_{22}^{\rm Ringdown})(h_{22}^{\rm NR},h_{22}^{\rm NR})}},
\end{align}
with
\begin{align}
(h_{22}^{\rm NR},h_{22}^{\rm Ringdown})=\text{Re} \int_{u_\Sigma^{(h)}}^{T}h_{22}^{\rm NR}h_{22}^{{\rm Ringdown}~*}dt,
\end{align}
where the upper limit of the integral $T$ is taken to be $90M_f$ after the peak of total rms strain amplitude. In addition, we use unweighted linear least squares to fit the mode amplitudes and use nonlinear least squares to fit the final spin and mass. The mode frequency $\omega_n$ is obtained from a Python package $\textsf{qnm}$ \cite{Stein:2019mop}. During the fit, we find that the numerical accuracy of $Y^{\infty}_{22}$ and $Y_{22}^{\rm H\,in}$ is much worse than that of $h^{\infty}_{22}$, which makes the remnant mass and spin more difficult to recover.

In Fig.~\ref{fig:overtone_mismatch}, we plot the mismatch $\mathcal{M}$ for $h_{22}^\infty$ (the left panel), $Y_{22}^\infty$ (the middle panel), and $Y_{22}^{\rm H\,in}$ (the right panel), for \ind~(the upper panel) and \indnew~(the lower panel). We see the strain $h_{22}^\infty$ can be decomposed into a sum of the fundamental mode and 6 overtones\footnote{Including more overtones no longer improves the match.}. For \ind, the linear regime can be extended to $16M_f$ before the peak of $h_{22}^\infty$, whereas for \indnew, the linear quasinormal ringing regime starts from $2.0M_f$, similar to the case of GW150914 \cite{Giesler:2019uxc} and superkick systems \cite{Ma:2021znq}.

On the other hand, since the numerical accuracy of $Y_{22}^\infty$ and $Y_{22}^{\rm H\,in}$ from CCE is not as high as $h_{22}^\infty$, only 5 overtones can be resolved. In particular, the late-time portion is dominated by numerical noise, therefore the mismatch $\mathcal{M}$ tends to increase significantly. The start times of the linear regime for $h_{22}^\infty$, $Y_{22}^\infty$, and $Y_{22}^{\rm H\,in}$ 
%is around $\sim -10M_f$, and we summarize the 
are summarized  in Table~\ref{table:overtone_fit}. Below, we will use the start time of $Y_{22}^{\rm H\,in}$, denoted by $v_\Sigma^{\rm (H)}$, as the advanced time of
%these numbers to represent the location of 
the matching tube $\Sigma_{\rm Shell}$ (Figs.~\ref{fig:my_label_new} and \ref{fig:image_wave}), and utilize the exterior portion of the GW to approximate the actual wave falling down the future horizon.

Apart from searching for the start time of quasi-normal ringing regime of $Y_{22}^{\rm H\,in}$, it is also interesting to investigate their QNM amplitudes \cite{Oshita:2021iyn,Ma:2021znq}. This topic is beyond the scope of our study and we only provide a brief discussion in Appendix \ref{app:qnm_spectrum}.

\begin{table}
    \centering
    \caption{A summary for the QNM decomposition of $h_{22}^\infty$, $Y_{22}^\infty$ and $Y_{22}^{\rm H\,in}$. The second row refers to the maximum number of overtones that we include into Eq.~(\ref{overtone-fit}). The third and fourth rows correspond to the time from which the waveform is consistent with a linear quasinormal ringing. The values are from the minimum of the corresponding curves in Fig.~\ref{fig:overtone_mismatch}.}
    \begin{tabular}{c c c c c} \hline\hline
    &     &   $h_{22}^\infty$ &$Y_{22}^\infty$ & $Y_{22}^{\rm H\,in}$  \\ \hline
 $n_{\rm max}$ & &   6 & 5 & 5 \\ \hline
\multirow{2}{*}{$u^{(\infty/h)}$ or $v_\Sigma^{(\rm H)}$}  & \ind & $-11.1$ & $-14.1$ & $-13$ \\ \cline{2-5}
   & \indnew & $2.0$ & $-14.2$ & $-15$ \\\hline\hline
     \end{tabular}
     \label{table:overtone_fit}
\end{table}

\section{Constructing Echoes}
\label{sec:construction_echo}
Now we utilize the horizon-going GW obtained above  to construct GW echoes at infinity. 
In Sec.~\ref{sec:baoyi_reflection_condition}, we first introduce physical boundary conditions near an ECO surface \cite{Chen:2020htz}, and obtain formulas that relate horizon waves to echoes at infinity. 
Then in Sec.~\ref{sec:num_ref_bol}, we focus on the Boltzmann reflectivity and discuss QNM structures of the ECO. 
%features of the reflectivity adopted in this work --- the Boltzmann reflectivity. In particular, we investigate the QNMs of the ECO.
Next in Sec.~\ref{sec:echo_inside}, we compute echo waveforms numerically and investigate the impact of prescriptions made at the matching shell $\Sigma_{\rm Shell}$ (see Fig.~\ref{fig:my_label_new}), taking \ind~for example. Finally, we compare the hybrid method with the inside prescription in Sec.~\ref{sec:num_compare_to_inside}.

\subsection{Constructing echoes using the physical boundary condition near an ECO surface}
\label{sec:baoyi_reflection_condition}
%As proposed recently by 
Chen \etal \cite{Chen:2020htz} recently proposed imposing boundary conditions near the ECO surface using the 
%the boundary condition is based on the 
Membrane Paradigm, in which a family of zero-angular-momentum fiducial observers (FIDOs) are considered. Within their own rest frame, the FIDOs experience a tidal tensor field \cite{Zhang:2012jj}
\begin{align}
    \mathcal{E}_{ij}=h^a_ih^c_j C_{abcd}U^bU^d,
\end{align}
where $C_{abcd}$ is the Weyl tensor, $U^b$ is the four-velocity of the FIDOs, and $h_i^a=\delta^a_i+U^aU_i$ is the projection operator. The transverse component of $\mathcal{E}_{ij}$ is of particular interest \cite{Chen:2020htz}
\begin{align}
    \mathcal{E}_{\rm transverse}\sim-\frac{\Delta}{4r^2}\psi_0-\frac{r^2}{\Delta}\psi_4^*, \label{tidal_tensor}
\end{align}
since it represents the stretching and squeezing effect due to GW. In analogous to the tidal response of a neutron star, the response of the ECO was proposed to be linear in $\mathcal{E}_{\rm transverse}$, namely \cite{Chen:2020htz}
\begin{align}
    \left[-\frac{r^2}{\Delta}\psi_4^*\right]_{\rm surface}=\left[\frac{\mathscr{R}^{\rm ECO}}{\mathscr{R}^{\rm ECO}-1}\mathcal{E}_{\rm transverse}\right]_{\rm surface}. \label{reflection_model}
\end{align}
The reflectivity $\mathscr{R}^{\rm ECO}$ depends on the (non-GR) property of ECO  as we shall discuss in Sec.~\ref{sec:num_ref_bol}.

Near the ECO surface, $\psi_0$ is dominated by the incident wave (toward the horizon), whereas $\psi_4$ by the reflected wave (by the ECO), i.e.,
\begin{subequations}
\begin{align}
    &\tensor[_{+2}]{{R^{\rm ECO}_{lm}}}{}(u,v)\sim\int\frac{d\omega}{\Delta^2}Y_{lm\omega}^{\rm H~in~ECO}e^{-i\omega v},\\
    &\tensor[_{-2}]{{R^{\rm ECO}_{lm\omega}}}{}(u,v)\sim \int d\omega Z_{lm\omega}^{\rm H~out~ECO}e^{-i\omega u}, 
\end{align}
\label{psi4_expansion_horizon}%
\end{subequations}
with $\tensor[_{\pm2}]{{R^{\rm ECO}_{lm}}}{}(u,v)$ the radial Teukolsky function for the ECO. Here we use the same notation as Eq.~(\ref{asymptotic-psi0-psi4}), and we emphasize that $Y_{lm\omega}^{\rm H~in~ ECO}$ stands for the \textit{actual} $\psi_0$-wave that falls down the future horizon. 
%In Sec.~\ref{sec:echo_inside}, we shall discuss how to use the hybrid method to compute $Y_{lm\omega}^{\rm H~in~ ECO}$.

After simplification, the boundary condition in Eq.~(\ref{reflection_model}) becomes
\begin{align}
    Z_{lm\omega}^{\rm H~out~ECO}=\frac{(-1)^{l+m+1}}{4}\mathscr{R}^{\rm ECO}Y_{lm\omega}^{\rm H~in~ECO}, \label{ref_echo}
\end{align}
where we have used the symmetry of a nonprecessing BBH system under reflection across the orbital plane \cite{Boyle:2014ioa}
\begin{align}
    [Y_{l,-m,-\omega}^{\rm H~in~ECO}]^*=(-1)^lY_{lm\omega}^{\rm H~in~ECO}.
\end{align}
Subsequently, the echo waveform at null infinity reads \cite{Xin:2021zir}
\begin{align}
    Z_{lm\omega}^{\infty~{\rm echo}}=\mathcal{K}(\omega)Y_{lm\omega}^{\rm H~in~ECO},
    \label{hyrbid_echo1}
\end{align}
with the transfer function $\mathcal{K}(\omega)$
\begin{align}
&\mathcal{K}(\omega)=\frac{(-1)^{l+m+1}\mathscr{R}^{\rm ECO}}{1-\mathscr{R}^{\rm ECO}\mathcal{R}^{\rm BH~T}} \frac{1}{4D^{\rm out}_{lm}}=\frac{C}{DD^{\rm in}_{lm}}\sum_{n=1}\left(\mathscr{R}^{\rm ECO}\mathcal{R}^{\rm BH~T}\right)^n,  \label{transfer_function}
\end{align}
and
\begin{align}
\mathcal{R}^{\rm BH~T} =(-1)^{l+m+1}\frac{D^{\rm in}_{lm}}{D^{\rm out}_{lm}}\frac{D}{4C}. \label{BH_T_R}
\end{align}
In Eq.~(\ref{transfer_function}), we have written the total echo signal as a sum of individual 
%$n$-th 
echoes.

\begin{figure}[htb]
        \includegraphics[width=\columnwidth,clip=true]{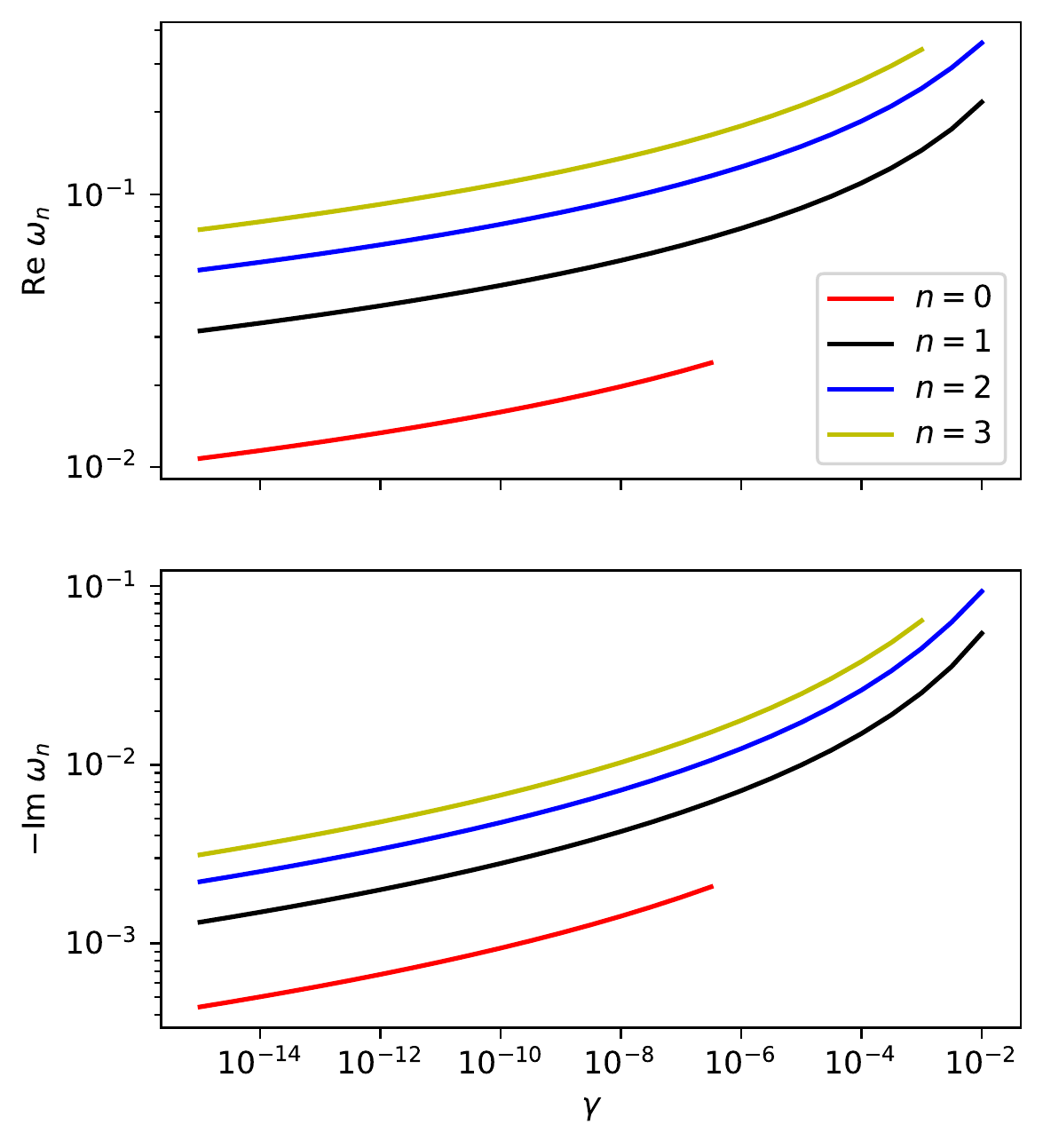}
  \caption{ The real and imaginary parts of QNMs for an irrotational ECO, as functions of $\gamma$. They are the solutions to Eq.~(\ref{ECO_QNM_def}). The Boltzmann reflectivity is used, assuming $T_{\rm QH}=T_H$. Each mode is labeled by the overtone index $n$. The imaginary part of QNMs is negative, meaning that the mode is stable.}
 \label{fig:ECO_QNM}
\end{figure}

\begin{figure}[htb]
        \includegraphics[width=\columnwidth,clip=true]{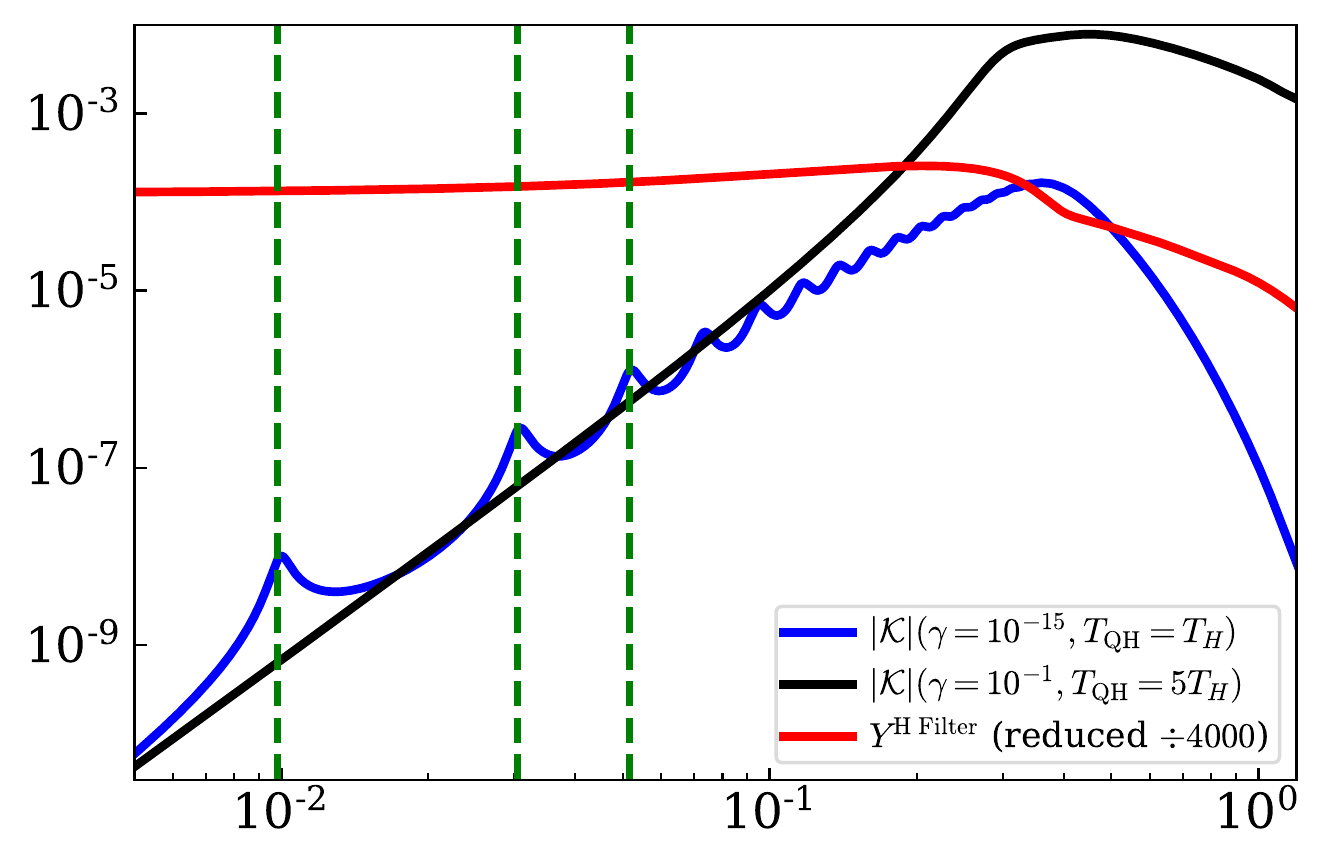}
  \caption{The transfer function $\mathcal{K}$ of the ECO using $(\gamma=10^{-15},T_{\rm QH}=T_H)$ (the blue curve), and $(\gamma=10^{-1},T_{\rm QH}=5T_H)$ (the black curve). The QNM resonances are visible in the former case, where the location of first three resonances are labeled by the dashed vertical lines, based on the estimation in Eq.~(\ref{omega_peak_overtone}). By comparison, the red curve corresponds to the absolute value of the filtered horizon wave $Y^{\rm H~Filter}$ for \ind, assuming $v_\Sigma^{\rm H}=-13$ and $\Delta v=2/\kappa$ [see Eq.~\eqref{Y_in_filter}]. Its value is decreased by a factor of 4000 for ease of read.}
 \label{fig:transfer_function}
\end{figure}

\begin{figure}[htb]
        \includegraphics[width=\columnwidth,clip=true]{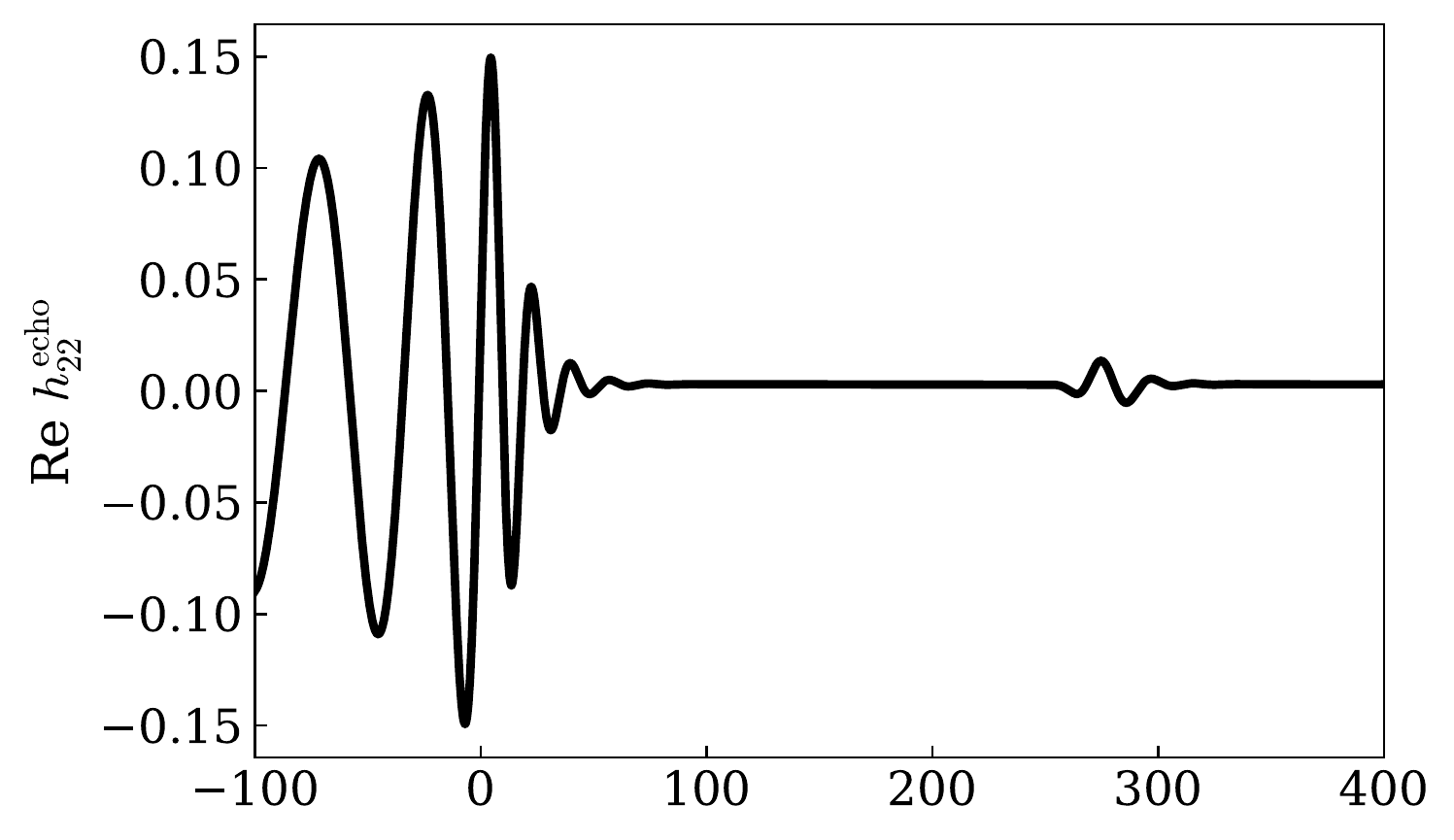}
  \caption{The echo emitted by \ind, following the main GW. Here we set $v_\Sigma^{(\rm H)}=-13,\Delta v=2/\kappa=8,\gamma=10^{-15}$, and $T_{\rm QH}=T_H$. }
 \label{fig:mainWF_echo}
\end{figure}

\begin{figure*}[htb]
        \includegraphics[width=\textwidth,clip=true]{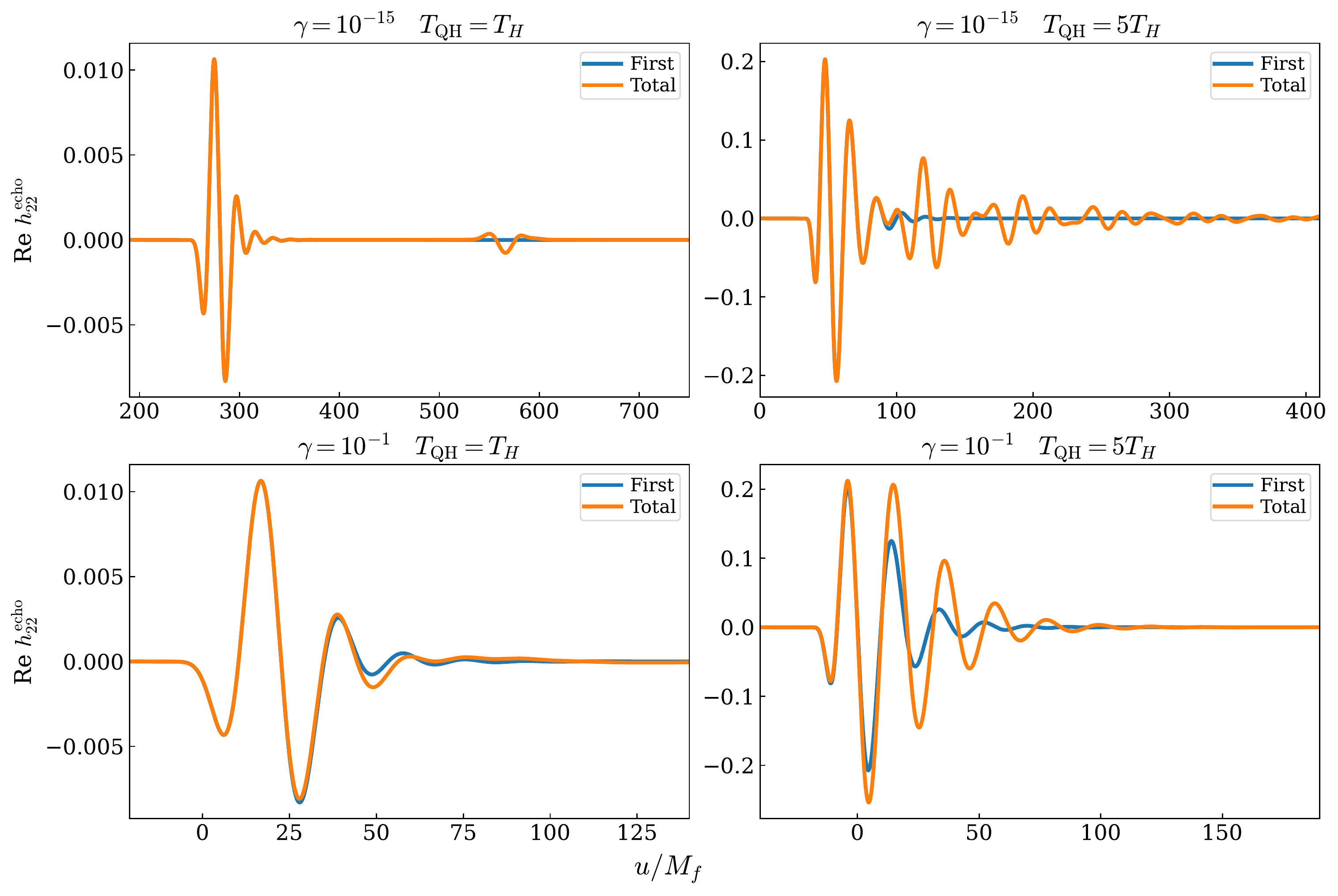}
  \caption{The echoes emitted by \ind, with a variety of  $T_{\rm QH}$ and $\gamma$. The width of filer $\Delta v$ is equal to $2/\kappa$. The total echoes (orange curves) are compared with the first echoes (blue curves). In the upper left panel, the values of $T_{\rm QH}$ and $\gamma$ are small enough that the spacing between echoes is greater than the echo duration, hence the individual pulses are well separated, whereas in the other three panels, different pulses overlap and interfere with each other.}
 \label{fig:echo_interferece}
\end{figure*}

\subsection{The Boltzmann reflectivity}
\label{sec:num_ref_bol}
%The quantum effects gives rise to thermal features near BH horizon. 
To model quantum effects around the horizon, Oshita \etal \cite{Oshita:2019sat} and Wang \etal \cite{Wang:2019rcf} proposed that GWs around the horizon interact with a quantum thermal bath.  Specifically, these waves are subject to a position-dependent dissipation $\Omega(r_*)/E_{\rm Pl}$, and driven by a position-dependent stochastic source $\xi(r_*)$ --- levels of the driving and the dissipation are related by the fluctuation-dissipation theorem \cite{Kubo_1966}. Then the BHP equation is modified to \cite{Oshita:2019sat,Wang:2019rcf}
\begin{align}
    \left[-i\gamma\frac{\Omega(r_*)}{E_{\rm Pl}}\frac{d^2}{dr_*^2}+\frac{d^2}{dr_*^2}+\omega^2-V_{\rm RWZ}^l\right]\tensor[_{s}]{{\Psi_{lm}^{\rm SN}}}{}(r_*)=\xi(r_*), \label{RWZ_quantum_modification}
\end{align}
where $\Omega(r_*)=|\omega|/\sqrt{|g_{00}(r_*)|}$ is the proper frequency measured in the frame of the Schwarzschild observers, $E_{\rm Pl}$ is the Planck energy, and $\gamma$ is a dimensionless dissipation parameter that controls how the damping ramps up as the wave gets close to the horizon. Note that Eq.~\eqref{RWZ_quantum_modification} reduces to the classical Zerilli-RW equation in the limit of $\gamma\to 0$ (vanishing of the dissipative effect) and $\xi\to0$ (vanishing of the fluctuation source). Consequently, the modified equation leads to the Boltzmann reflectivity \cite{Oshita:2019sat,Wang:2019rcf}:
\begin{align}
    \mathscr{R}^{\rm ECO}=\exp\left[-i\frac{\omega}{\pi T_{\rm{QH}}}\ln(\gamma|\omega|)\right]\exp\left(-\frac{|\omega|}{2T_{\rm QH}}\right)\,, \label{bol_refl}
\end{align}
where the quantity $T_{\rm QH}$ is the effective horizon temperature. 
The first term on the right hand side of Eq.~(\ref{bol_refl}) implies that as $\gamma\ll 1$, the region between $r_*\sim\frac{\ln\gamma}{2\pi T_{\rm QH}}$ and the peak of the BH potential forms a cavity. In this way, the 
%This results in the existence of 
ECO's QNM frequencies, $\omega_n$, are determined as poles of the transfer function $\mathcal{K}(\omega)$ [see Eq.~(\ref{transfer_function})]
\begin{align}
\mathscr{R}^{\rm ECO}(\omega_n)\mathcal{R}^{\rm BH~T}(\omega_n)=1.\label{ECO_QNM_def}
\end{align}
%with $n$ the index of overtones.
We solve Eq.~(\ref{ECO_QNM_def}) numerically and plot the value of $\omega_n$ as a function of $\gamma$ in Fig.~\ref{fig:ECO_QNM}, where the quantum horizon temperature $T_{\rm QH}$ is set to be the Hawking temperature $T_H$:
\begin{align}
    T_H\coloneqq\frac{\kappa}{2\pi} =\frac{1}{8\pi}, \label{hawking-temp}
\end{align}
with $\kappa=1/4$ the surface gravity. We can see that the absolute value of the real and imaginary parts of $\omega_n$ increases with $\gamma$ and $n$. In particular, the negative sign of ${\rm Im}~\omega_n$ ensures the stability of the QNMs. For the fundamental mode $n=0$, its decay rate is less than $10^{-3}$, hence it is long-lived.

The feature of ECO's QNMs is also visible in the transfer function $\mathcal{K}$, as shown in Fig.~\ref{fig:transfer_function}. The blue curve corresponds to the case with $(\gamma=10^{-15},T_{\rm QH}=T_H)$. There are a number of local maxima (resonances) whose locations are close to the real part of the corresponding QNMs.  In the limit of $\gamma\ll 1$, the peak frequency $\omega^{(n)}_{\rm peak}$ is given by  
\begin{align}
&\omega^{(n)}_{\rm peak}=\omega_{\rm FSR}^{(n)}-\frac{\omega_{\rm FSR}^{(n)}}{(2n+1)\pi}{\rm Im}\ln\left[ \mathcal{R}^{\rm BH~T}(\omega_{\rm FSR}^{(n)}) \right], \label{omega_peak_overtone}
\end{align}
where the free spectral range (SFR) of the cavity writes
\begin{align}
&\omega_{\rm FSR}^{(n)}=(2n+1)\frac{T_{\rm QH}\pi^2}{|\ln \gamma|}\left\{1-\frac{1}{\ln\gamma}\ln\left[(2n+1)\frac{T_{\rm QH}\pi^2}{|\ln \gamma|}\right]\right\} \notag \\
&+\mathcal{O}\left[(\ln \gamma)^{-2}\right], \quad n=0,1\ldots
\end{align}
In Fig.~\ref{fig:transfer_function} we label the location of $\omega^{(n)}_{\rm peak}$ for $n=0,1,2$ using the dashed vertical lines. Additionally, $\mathcal{K}$ has a global maximum at the fundamental QNM of a Schwarzschild BH $(0.374-0.0890i)$, contributed by the factor $1/D_{22}^{\rm out}$ (see the blue curve in Fig.~\ref{fig:Din_Dout}). Within the frequency band $\omega<0.374$, $\mathcal{K}$ is dominated by $1/D_{22}^{\rm out}$, hence its asymptotic behavior is $\sim\omega^4$ as $\omega\to0$. Whereas for the band $\omega>0.374$, $\mathcal{K}$ decays exponentially due to the second term on the right hand side of Eq.~(\ref{bol_refl}).

%Oppositely,
On the other hand,  
when $\gamma$ is comparable to 1, GWs cannot be effectively trapped near the ECO surface, and  the ECO QNMs do not exist. This fact is 
%presented 
clearly manifested 
in the transfer function of the case with $(\gamma=10^{-1},T_{\rm QH}=5T_H)$, as shown in the black curve in Fig.~\ref{fig:transfer_function}. Moreover, since the value of $T_{\rm QH}$ is greater than the previous one, more high-frequency contents can be reflected by the ECO surface hence emerge at infinity.

\begin{figure}[htb]
        \includegraphics[width=\columnwidth,clip=true]{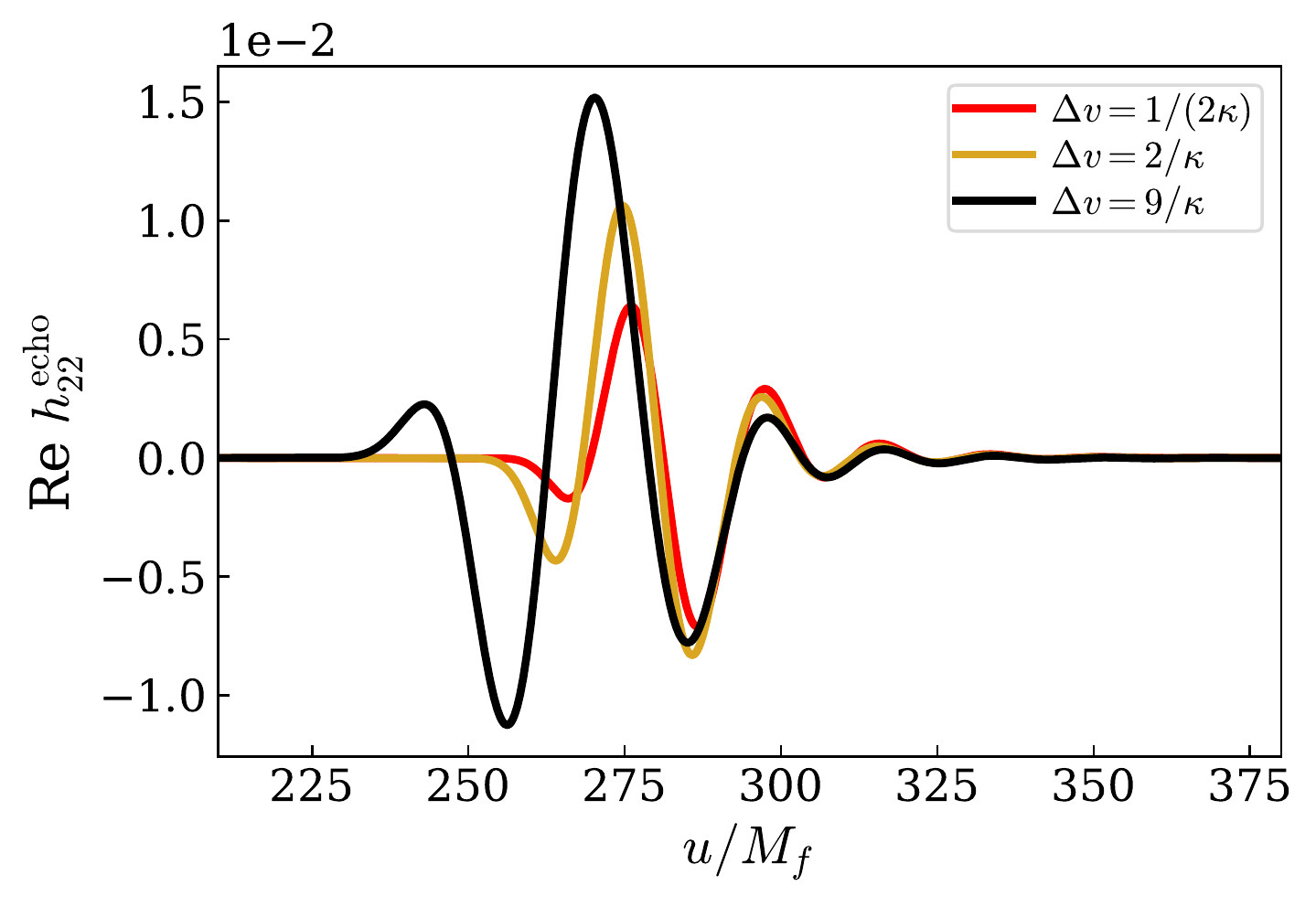}
  \caption{The influence of the filter parameter $\Delta v$ on echo waveforms. Each curve corresponds to the real part of the first echo (with different $\Delta v$), using \ind~and the Boltzmann reflectivity ($\gamma=10^{-15}$ and $T_{\rm QH}=T_H$) . The filter is applied at the future horizon with $v_\Sigma^{(\rm H)}=-13$.}
 \label{fig:filter_echo_lorentz_deltav}
\end{figure}

\subsection{Numerical computation of echo waveforms}
\label{sec:echo_inside}
In order to use Eq.~(\ref{hyrbid_echo1}) to compute echo waveforms, we first need to estimate the actual wave $Y_{lm\omega}^{\rm H~in~ ECO}$ [see Eq.~(\ref{psi4_expansion_horizon})] that falls down the future horizon. In the context of hybrid method, the future horizon exists partially in region I+II, only the late-time portion of $Y_{lm}^{\rm H~in}$ [see Eq.~(\ref{YHin-time-domain})] can represent $Y_{lm}^{\rm H~in~ECO}$, namely
\begin{align}
    &Y_{lm}^{\rm H~in~ECO}(v)=Y_{lm}^{\rm H~in}(v), & {\rm when}~~v>v_\Sigma^{({\rm H})}.
    \label{Y_Psi_filter}
\end{align}
Note again that the condition is in the time domain. The value of $v_\Sigma^{({\rm H})}$ was determined by searching for the starting time after which $Y_{lm}^{\rm H~in}(v)$ can be decomposed as a sum of QNM overtones, as discussed in Sec.~\ref{sec:overtone_fit}. In practice, we impose the condition in Eq.~(\ref{Y_Psi_filter}) via a filter:
\begin{align}
    Y_{lm}^{\rm H~ in~ECO}(v)&\to Y^{\rm H~Filter}_{lm}(v), \notag \\
    &
    =Y_{lm}^{\rm H~in}(v)\mathcal{F}(v)+{\rm Const.}\times[1-\mathcal{F}(v)],
    \label{Y_in_filter}
\end{align}
where the Planck-taper filter $\mathcal{F}(v)$ is given by \cite{McKechan:2010kp}
\begin{align}
    \mathcal{F}(v;v_\Sigma^{(\rm H)},\Delta v)=
    \begin{cases}
    0, &  v<v_\Sigma^{({\rm H})}-\Delta v, \\
    \frac{1}{\exp{z}+1},& v_\Sigma^{({\rm H})}-\Delta v<v<v_\Sigma^{({\rm H})}, \\
    1, & v>v_\Sigma^{({\rm H})}.
    \end{cases}
    \label{F_planck_filter}
\end{align}
and $z=\frac{\Delta v}{v-v_\Sigma^{({\rm H})}}+\frac{\Delta v}{v-v_\Sigma^{({\rm H})}+\Delta v}$. The Planck-taper filter $\mathcal{F}(v)$ is a function that gradually ramps up from 0 to 1 within the time interval $[v_\Sigma^{({\rm H})}-\Delta v,v_\Sigma^{({\rm H})}]$. Therefore, $Y^{\rm H~Filter}_{lm}(v)$ in Eq.~(\ref{Y_in_filter}) represents a quantity that switches from a constant value to $Y_{lm}^{\rm H~in}(v)$ that is predicted by the hybrid method. The value of the constant does not affect the echo waveform since this zero-frequency content cannot penetrate the BH potential (see the value of $D_{22}^{\rm out}$ in Fig.~\ref{fig:Din_Dout}). In our case, we set the constant to 0.

With the transfer function at hand, we are able to compute echo waveforms. Figure \ref{fig:mainWF_echo} shows an echo signal following the main GW, emitted by the system \ind, assuming $v_\Sigma^{(H)}=-13$, as summarized in Table \ref{table:overtone_fit}, and $(\Delta v=2/\kappa=8,\gamma=10^{-15},T_{\rm QH}=T_H)$. To further investigate how the echo signal is impacted by the parameters $(\gamma,T_{\rm QH})$, we vary their values and exhibit the results in Fig.~\ref{fig:echo_interferece}. The echo waveform of \indnew~looks similar to that of \ind, and it can be found in Appendix \ref{app:1936}. The total echo waveform is compared with the first echo. In the case of $(\gamma=10^{-15},T_{\rm QH}=T_H)$ (shown in the upper left panel), distinct echo pulses are separated by an equal time interval of 
\begin{align}
    \Delta u^{\rm echo}\sim|\ln\gamma|/(\pi T_{\rm QH}), \label{echo-time-interval}
\end{align}
which is long compared with the duration of BBH ringdown.  These well-separated echoes do result mathematically  from a collective excitation of ECO's multiple QNMs displayed in Fig.~\ref{fig:transfer_function} --- even though each individual QNM bears little resemblance to the echo pulse.   On the other hand, for greater values of $T_{\rm QH}$ and $\gamma$ ($\gamma=10^{-1},T_{\rm QH}=5T_H$, shown in the lower right panel), the spacing between nearby pulses becomes comparable to the pulse duration, distinct echo pulses interfere with each other, and we cannot resolve any single pulse. In addition, since the ECO with greater $T_{\rm QH}$ reflects a broader frequency band, the final echo is stronger.

We then investigate the impact of the filter parameter $\Delta v$ in  Eq.~(\ref{F_planck_filter}). As shown in Fig.~\ref{fig:filter_echo_lorentz_deltav}, we compute the first echo emitted by \ind, using $(\gamma=10^{-15},T_{\rm QH}=T_H)$ and $v_\Sigma^{(\rm H)}=-13$ --- for a variety of $\Delta v$. We can see that the waveforms have different amplitude evolution within the first two cycles, but the distinction is suppressed shortly afterwards.

\begin{figure}[htb]
        \includegraphics[width=\columnwidth,clip=true]{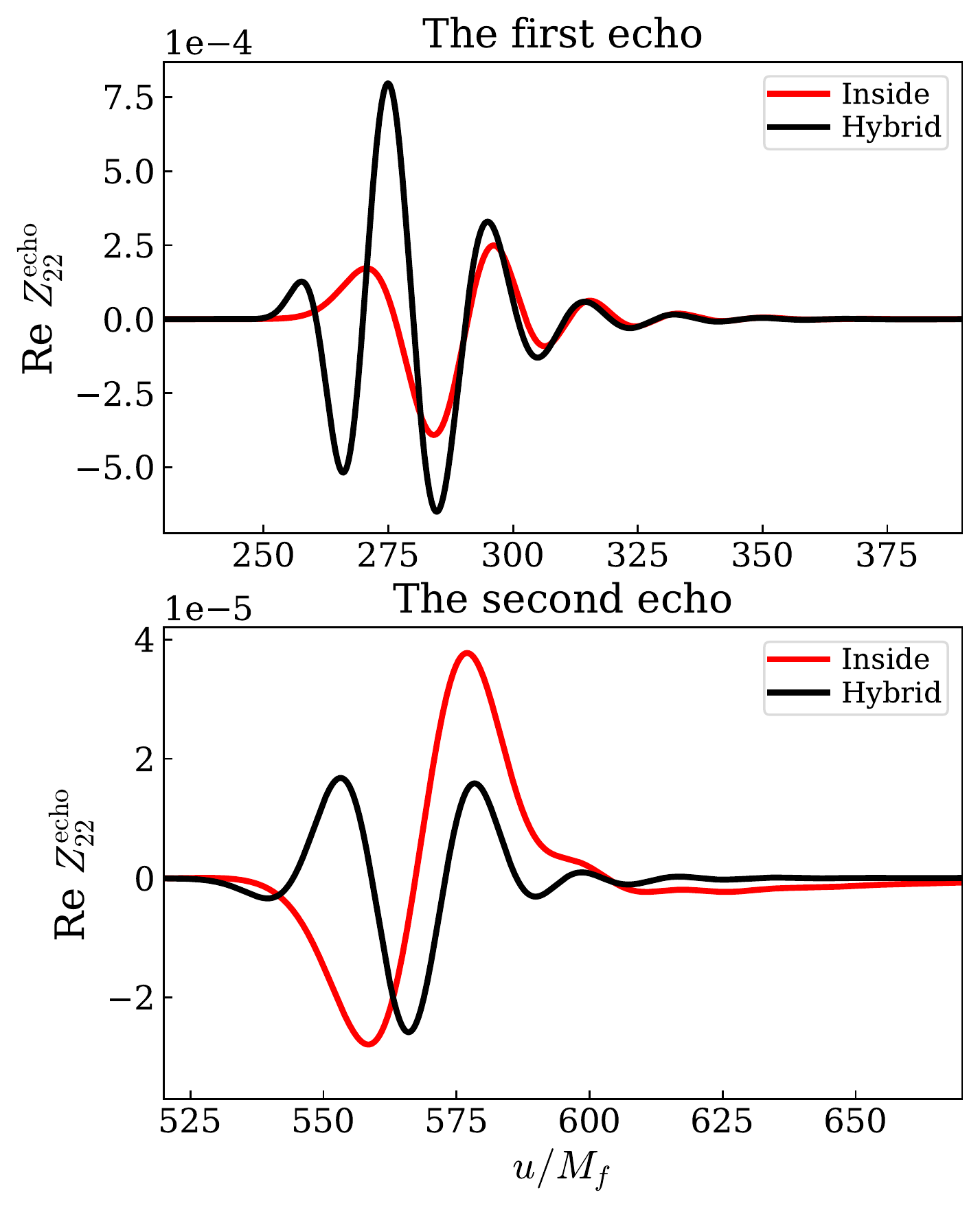}
  \caption{A comparison between the hybrid approach and the inside prescription, using \ind. We choose the Boltzmann reflectivity with $(\gamma=10^{-15},T_{\rm QH}=T_H)$. The upper panel shows the first echo, whereas the bottom panel is the second echo. The filter is applied at null infinity (labeled by "Inside", in red), and at future horizon (labeled by "Hybrid", in black). The width of both filter $\Delta v$ is $2/\kappa$.}
 \label{fig:inside_vs_hybrid}
\end{figure}

\begin{figure*}
    \centering
    \subfloat[aLIGO, \ind]{\includegraphics[width=\columnwidth]{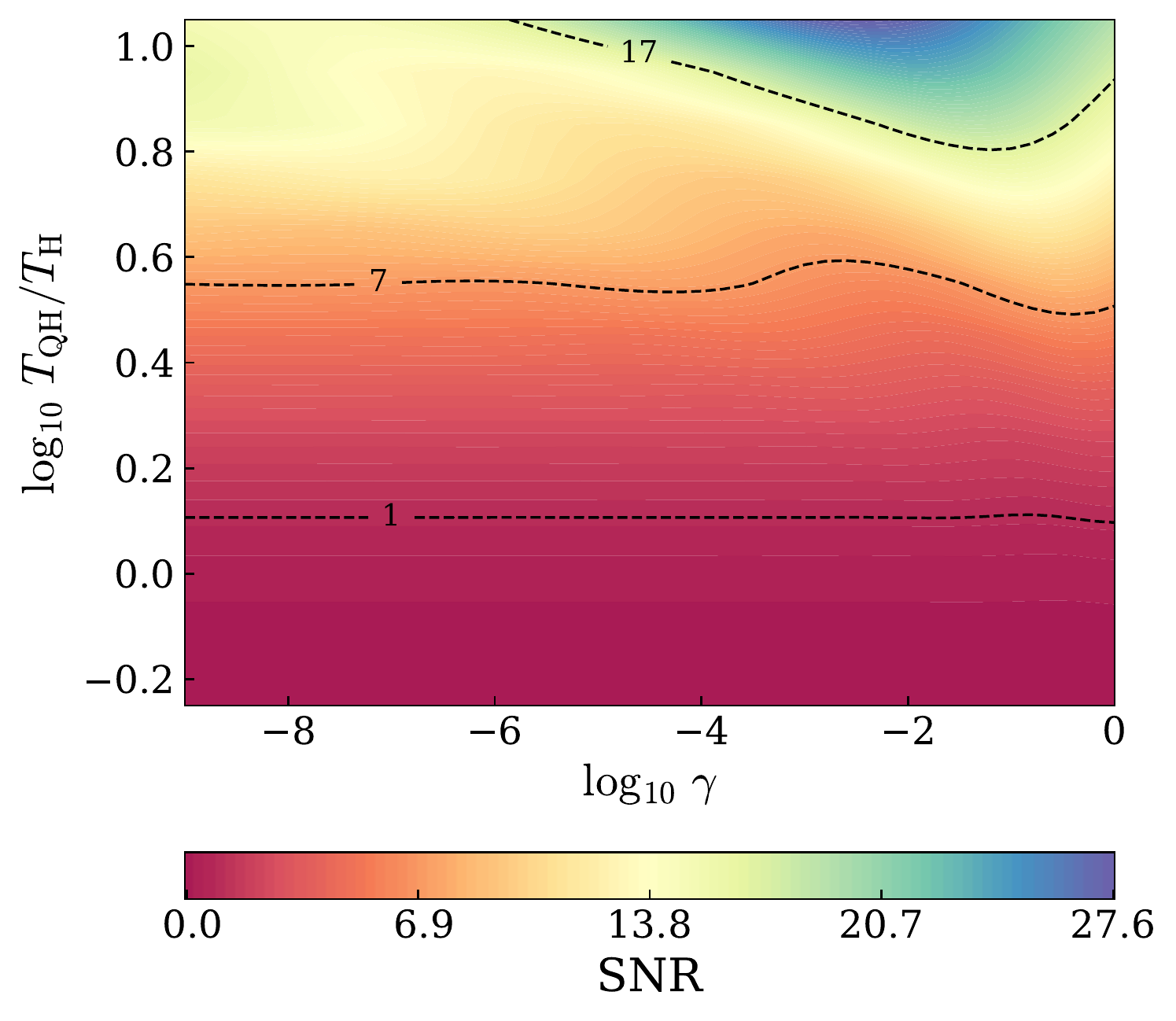}}
    \subfloat[CE, \ind]{\includegraphics[width=\columnwidth]{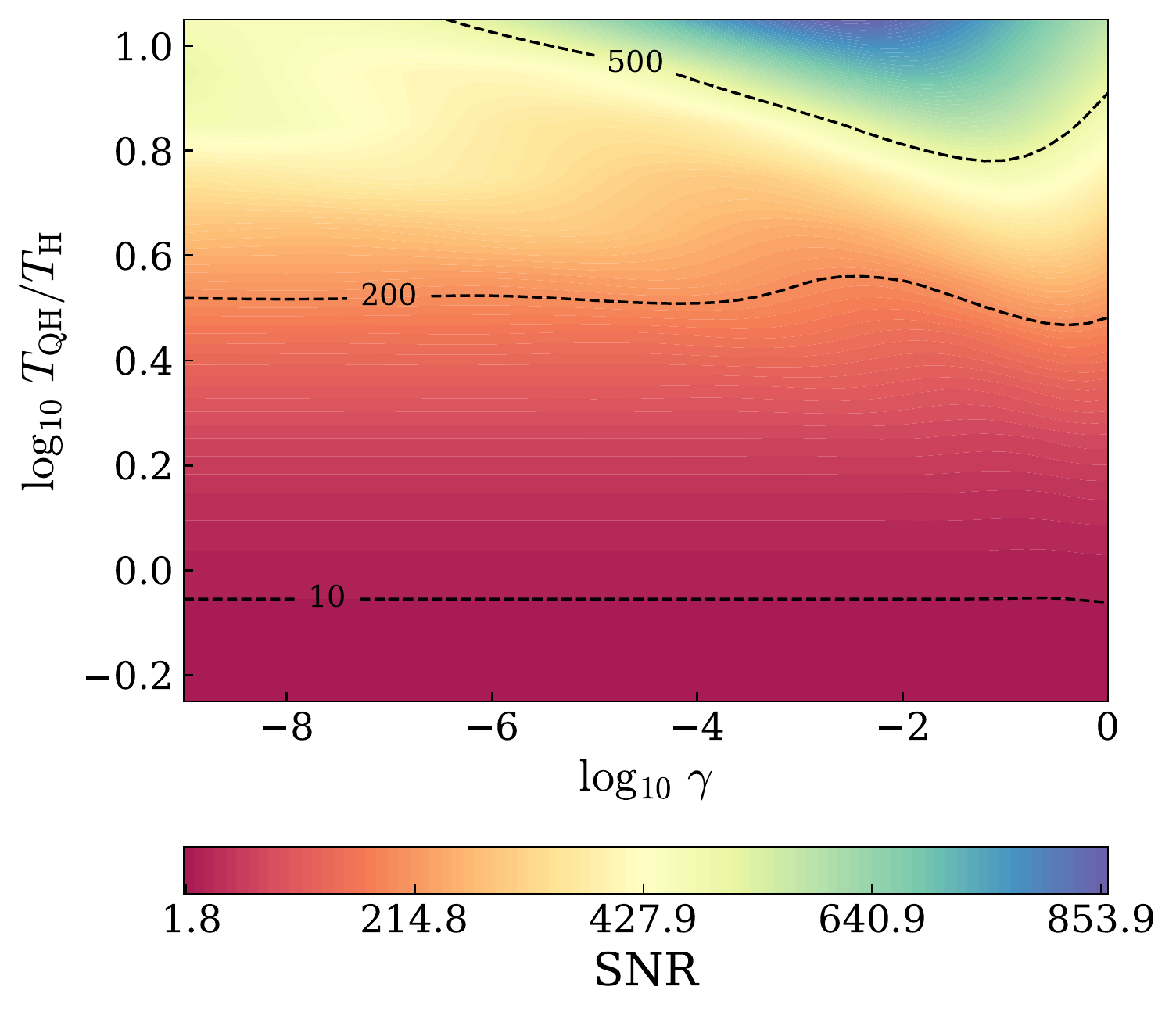}} \\
      \subfloat[aLIGO, \indnew]{\includegraphics[width=\columnwidth]{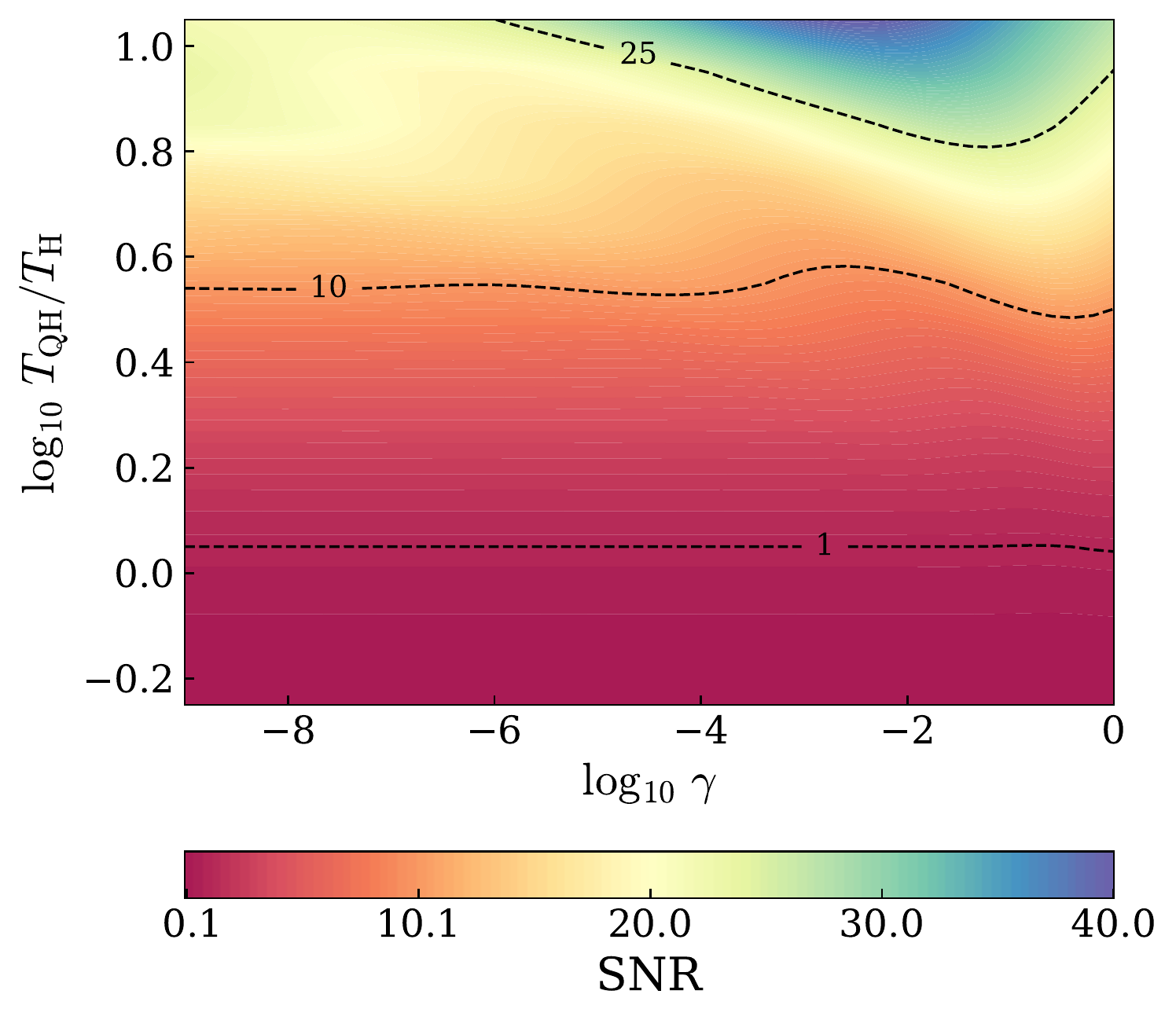}}
    \subfloat[CE, \indnew]{\includegraphics[width=\columnwidth]{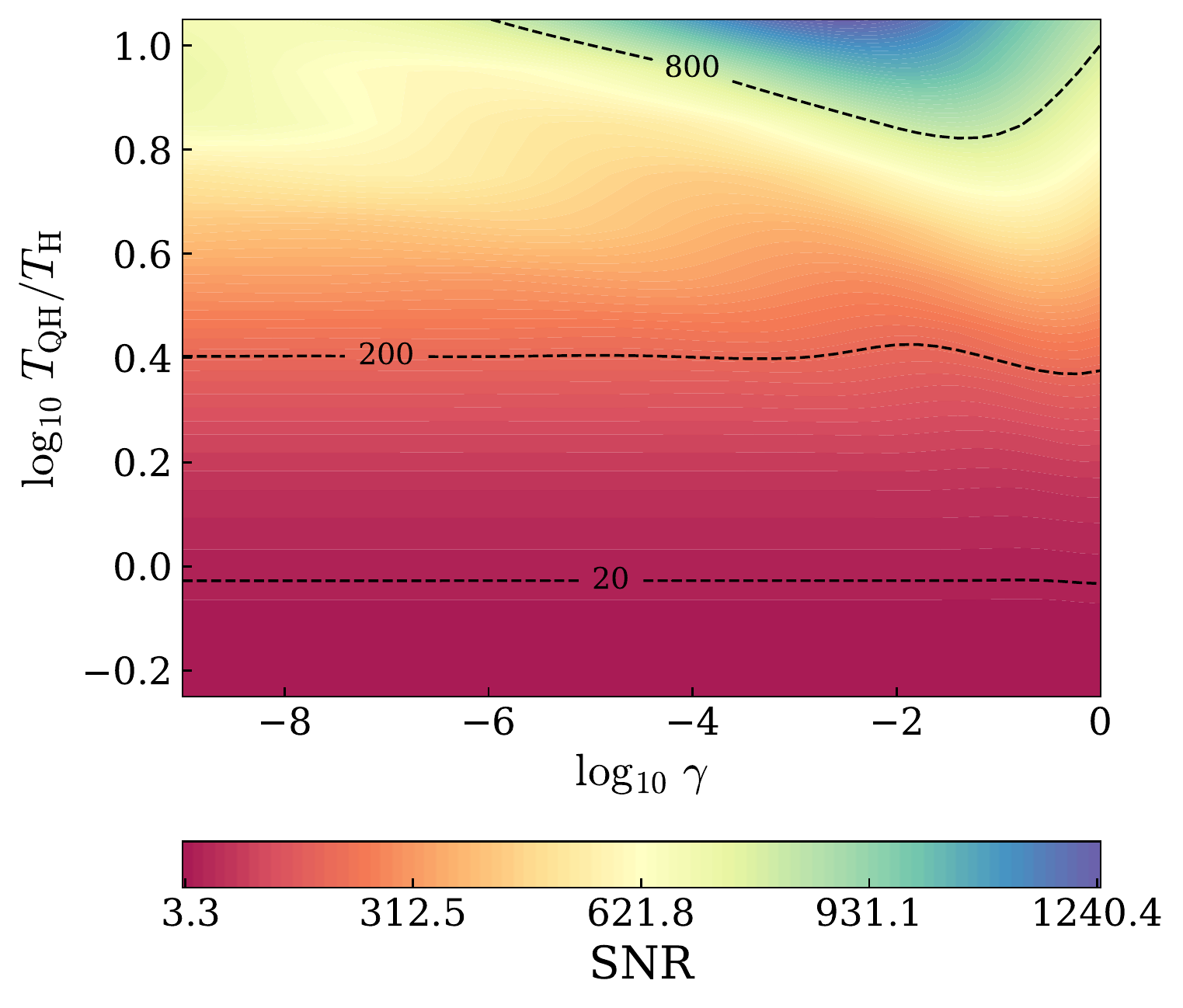}}
  \caption{The sky-averaged echo SNR across the $T_{\rm QH}-\gamma$ space, using \ind~(the upper panel) and \indnew~(the lower panel), as well as aLIGO (the left column) and CE (the right column). The binary system is 100Mpc away from the detector, with a total mass of $60M_\odot$. We set $\Delta v$ to $2/\kappa$ and the values of $v_\Sigma^{(\rm H)}$ are listed in Table \ref{table:overtone_fit}.}
 \label{fig:snr_bolt}
\end{figure*}

\subsection{Comparison with the inside prescription}
\label{sec:num_compare_to_inside}

The horizon filter is absent in the framework of inside prescription \cite{Maggio:2019zyv,Wang:2019rcf}. Taking $v_\Sigma^{(\rm H)}\to -\infty$, Eq.~(\ref{Y_Psi_filter}) reduces to
\begin{align}
    Y_{lm\omega}^{\rm H~ in~ECO}=C^{\rm in}_{lm}(\omega)Y^{\infty}_{lm\omega}, \label{Yin_Psi_inf}
\end{align}
and Eq.~(\ref{hyrbid_echo1}) becomes 
\begin{align}
    Z_{lm\omega}^{\infty~\rm echo}=\sum_{n=1}^\infty\left(\mathscr{R}^{\rm ECO}\mathcal{R}^{\rm BH~T}\right)^nZ_{lm\omega}^{\infty},
    \label{inside_formula}
\end{align}
where we have used the TS identities in Eqs.~(\ref{TS}). A direct usage of Eq.~(\ref{inside_formula}) will lead to undesired low-frequency contents, contributed by the inspiral stage. A workaround would be taking only the ringdown portion of $Z_{lm}^{\infty}(u)$, following Ref.~\cite{Maggio:2019zyv}. We compare the hybrid method [Eq.~(\ref{hyrbid_echo1})] with the inside formula [Eq.~(\ref{inside_formula})] in Fig.~\ref{fig:inside_vs_hybrid}, assuming \ind. Here we choose $\Delta v=2/\kappa$ and $(\gamma=10^{-15},T_{\rm QH}=T_H)$. We see for the first echo, the hybrid method leads to a stronger signal, but the inside prescription has a stronger second echo. Meanwhile, for the initial part of the first echo, the hybrid method gives rise to one more cycle, but the evolution is almost identical afterwards.

\section{Detectability and parameter estimation}
\label{sec:Detectability_and_parameter_estimation}
In this section, we focus on the detectability of the echoes computed in this paper by current and future detectors. We first give a brief summary of detector response, signal-to-noise ratio (SNR) and Fisher matrix calculations in  Sec.~\ref{sec:fisher_intro}. 
%we give a brief introduction to signal-to-noise ratio and the Fisher matrix.  
%first give a brief introduction to our computation method. 
Then we study the detectability of echoes by calculating SNR in Sec.~\ref{sec:Detectability}, and  discuss parameter estimation by adopting the Fisher matrix in Sec.~\ref{sec:fisher}.

\subsection{The signal-to-noise ratio and Fisher-matrix formalism}
\label{sec:fisher_intro}
We first construct two polarizations of an echo $h_{+,\times}^{\rm echo}$ by assembling $h^{\rm echo}_{lm}$:
\begin{align}
    h_+^{\rm echo}-ih_{\times}^{\rm echo}=\sum_{m=\pm2} \tensor[_{-2}]{Y}{_{l=2,m}}(\theta,\phi)h^{\rm echo}_{l=2,m},
\end{align}
where we are using the leading contributions $h^{\rm echo}_{2,\pm2}$, who satisfy the condition $h^{\rm echo}_{2,-2}=(h^{{\rm echo}}_{2,2})^*$. The echo strain $h^{\rm echo}$ detected by a detector is given by 
\begin{align}
    h^{\rm echo}=F_+(\theta_S,\phi_L,\psi_L)h_+^{\rm echo}+F_\times(\theta_S,\phi_L,\psi_L)h_\times^{\rm echo}, \label{h_echo_SNR}
\end{align}
with $(\theta_S,\phi_L)$ the sky location of a source with respect to the detector, and $\psi_L$ the polarization angle. The SNR of a given GW signal $h$ is written as $\sqrt{(h|h)}$, where the inner product between two waveforms $(h|g)$ reads
\begin{align}
(h|g)=4{\rm Re} \int\frac{h^*(f)g(f)}{S_n(f)}df.
\end{align}
Here $S_n(f)$ is the spectral density of the noise when detecting GWs. 
% In this paper, we mainly focus on AdvLIGO and the Cosmic Explorer (CE), whose $S_n(f)$ are shown in Fig.~\ref{fig:psd}. 
The averaged SNR over angular parameters $(\theta_S,\phi_L,\psi_L,\theta,\phi)$ is given by \cite{Finn:1992xs}
\begin{align}
    \left<\rho^2\right>=\frac{16}{25}\int \frac{|h_+|^2(\theta=0)}{S_{n}(f)}df. \label{snr_average}
\end{align}
We shall adopt the sky-averaged SNR all through this paper. 

% \begin{figure}[htb]
%         \includegraphics[width=\columnwidth,clip=true]{psd.pdf}
%   \caption{}
%  \label{fig:psd}
% \end{figure}

On the other hand, the Fisher matrix for a given gravitational waveform $h(\lambda^i)$ can be written as
\begin{align}
    \Gamma_{ij}=\left(\left.\frac{\partial h}{\partial \lambda^i}\right|\frac{\partial h}{\partial \lambda^j}\right),
\end{align}
where $\lambda^i$ are parameters to be estimated. In this paper, we restrict ourselves to  $(\gamma,T_{\rm QH})$ that determine the Boltzmann reflectivity [Eq.~\eqref{bol_refl}]. By inverting $\Gamma_{ij}$, we obtain parameter estimation accuracies for $\lambda^i$ as 
\begin{align}
    \Delta\lambda^i=\sqrt{(\Gamma^{-1})_{ii}}.
\end{align}
%\begin{figure*}
%    \centering
%    \subfloat[aLIGO]{\includegraphics[width=\columnwidth]{snr_quantum_v1_deltav_ligo.pdf}}
%    \subfloat[CE]{\includegraphics[width=\columnwidth]{snr_quantum_v1_deltav_ce.pdf}}
%  \caption{The sky-averaged echo SNR as a function of filter parameters $v_1$ and $\Delta v$ [see Eq.~(\ref{F_planck_filter})], using aLIGO (a) and CE (b). The binary system is chosen to be \ind, with a total mass $60M_\odot$. The distance between the binary system and the detector is 100Mpc. We use the Boltzmann reflectivity with $T_{\rm QH}=T_H$ and $\gamma=10^{-2}$.}
% \label{fig:snr_filter}
%\end{figure*}

\subsection{Detectability of echoes}
\label{sec:Detectability}
To study how the SNR is impacted by the reflectivity parameters $(\gamma,T_{\rm QH})$, we adopt a aLIGO-like detector \cite{LIGOScientific:2014qfs} and a Cosmic Explorer (CE)-like detector \cite{LIGOScientific:2016wof}, for both \ind~and \indnew. We assume the binaries to have a total mass of $60M_\odot$, and to be located  100Mpc from the detector. 

In the baseline case with $T_{\rm QH}=T_H$, $\gamma=10^{-1}$,  $\Delta v=2/\kappa$ and using values of $v_\Sigma^{(\rm H)}$ in Table \ref{table:overtone_fit}, we obtain (sky-averaged) echo SNR of $\sim 0.45$ for aLIGO, and $\sim 15$ for CE.  Echo SNRs of \indnew\, are greater than \ind\, by a factor of $\sim 1.5$ in both detectors. In order to compare with Ref.~\cite{LongoMicchi:2020cwm}, we also estimate the ratios between echo SNR and ringdown SNR.  To first obtain the ringdown SNR, we choose the lower limit of integration in Eq.~\eqref{snr_average} to be the frequency of $h_{22}^{\infty}$ evaluated at $u^{(h)}$ [see Eq.~\eqref{hinf_fit_over} and Table~\ref{table:overtone_fit}]. For aLIGO, the ringdown SNR for \ind\, is around 7.0, and the ratio ${\rm SNR}_{\rm echo}/{\rm SNR}_{\rm ringdown}=6.5\%$, close to the blue curve in the bottom left panel of Fig.~9 in Ref.~\cite{LongoMicchi:2020cwm}.

%Using aLIGO, the sky-averaged echo SNR, ${\rm SNR}_{\rm echo}$, is around $0.45$ when $T_{\rm QH}=T_H$. 

%On the other hand, for CE, the echo SNR is $\sim15$ as $T_{\rm QH}= T_H$. Given the same detector, the SNR of \indnew~is greater than that of \ind~by a factor of $\sim1.5$.

In Figure~\ref{fig:snr_bolt}, we explore how the echo SNR depends on values of $\gamma$ and $T_{\rm QH}$, for both detectors and both binaries, respectively, assuming $\Delta v=2/\kappa$ and the values of $v_\Sigma^{(\rm H)}$ being listed in Table \ref{table:overtone_fit}. The SNR increases with $T_{\rm QH}$ since a larger $T_{\rm QH}$ corresponds to a broader reflection frequency band, and more incident waves are reflected. The $\gamma$ dependence of SNR is more complex. For small values of $T_{\rm QH}$ (i.e., around unity, as originally proposed by Ref.~\cite{Wang:2019rcf}), the SNR barely depends on $\gamma$, because in this case the echoes are weak and mainly dominated by the first pulse, where $\gamma$ only controls the separation between the echoes in time, then it does not affect the SNR.  By contrast, for $T_{\rm QH}\gtrsim 5T_{H}$, the echoes may overlap with each other, and (constructively) interfere, elevating the SNR.

%For the $\gamma-$dependence, we find that the SNR depends barely on $\gamma$ when $T_{\rm QH}$ is small. This is because the total echo signal is dominated by the first pulse (e.g., the left column in Fig.~\ref{fig:echo_interferece}), and $\gamma$ controls only its phase. In contrast, as $T_{\rm QH}\gtrsim 5T_{H}$, the subsequent echo pulses play a non-negligible role. Therefore, $\gamma$ determines their time interval and how they interfere, see the right column of Fig.~\ref{fig:echo_interferece} for instance. 

\begin{figure}
    \centering
    \includegraphics[width=\columnwidth]{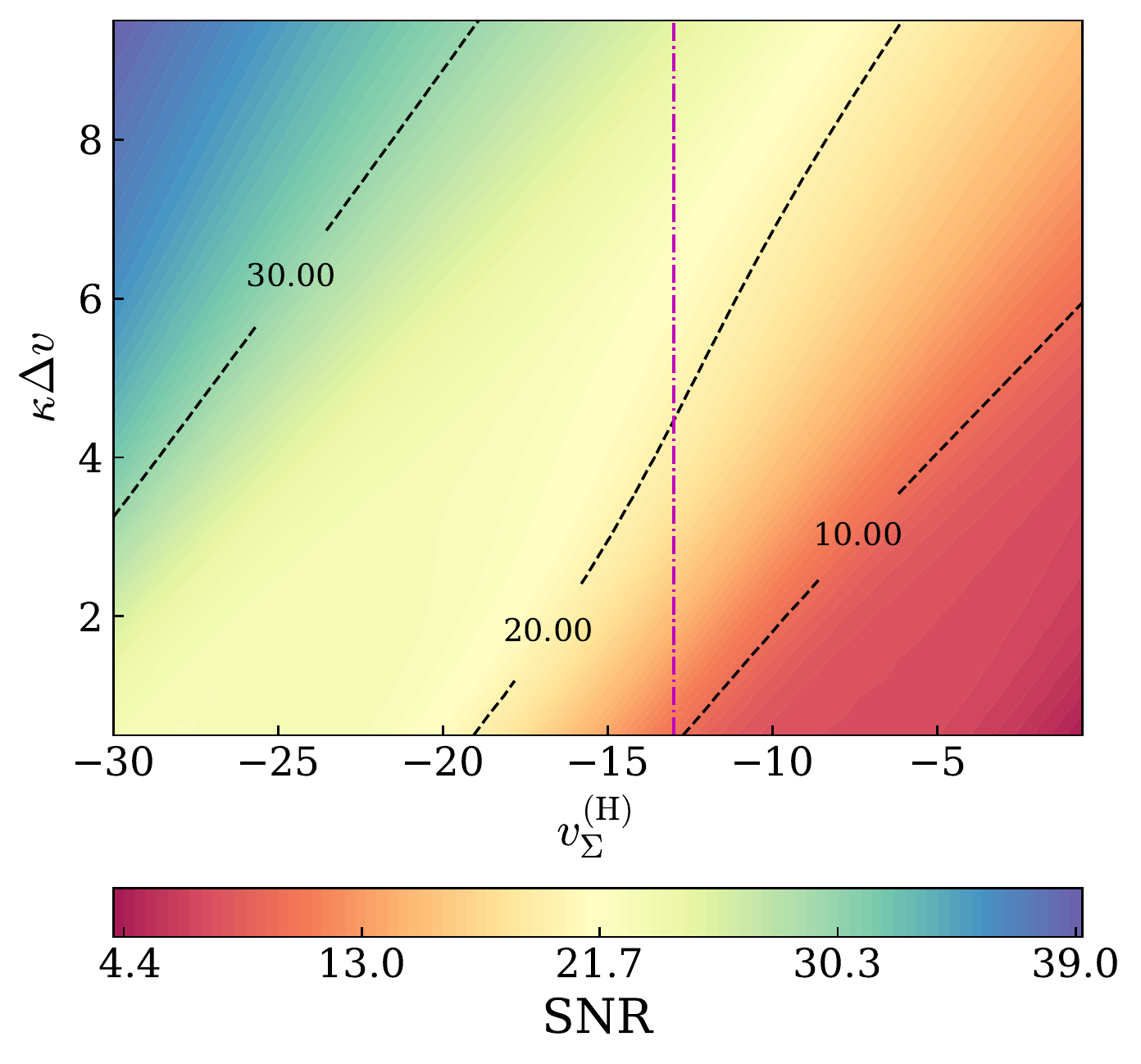}
  \caption{The sky-averaged echo SNR as a function of filter parameters $v_\Sigma^{(\rm H)}$ and $\Delta v$ [see Eq.~(\ref{F_planck_filter})], using CE. The binary system is \ind~and has the same total mass and distance as Fig.~\ref{fig:snr_bolt}. We use the Boltzmann reflectivity with $\gamma=10^{-15}$ and $T_{\rm QH}=T_H$. The vertical dot-dashed line stands for the value of $v_\Sigma^{(\rm H)}$ in Table \ref{table:overtone_fit}.}
 \label{fig:snr_filter}
\end{figure}

Next we investigate the impact of filters on the horizon, namely the advanced time $v_\Sigma^{(\rm H)}$  at which the shell $\Sigma$ crosses the horizon, and the thickness $\Delta v$ of the transition region in which we cut off reflection.  Taking \ind~and CE for example, we plot, in Fig.~\ref{fig:snr_filter}, the sky-averaged echo SNR as a function of two filter parameters $v_\Sigma^{(\rm H)}$ and $\Delta v$ [see Eq.~(\ref{F_planck_filter})], where we choose $\gamma=10^{-15}$ and $T_{\rm QH}=T_H$. As expected, the SNR decreases as either $v_\Sigma^{(\rm H)}$ increases or $\Delta v$ decreases. The global pattern suggests that the dependence on $v_\Sigma^{(\rm H)}$ and $\Delta v$ is linearly correlated.

\begin{figure}[htb]
%\subfloat[ \ind]{\includegraphics[width=\columnwidth,clip=true]{bolt_fisher.pdf}}\\
%\subfloat[ \indnew]{\includegraphics[width=\columnwidth,clip=true]{1936_bolt_fisher.pdf}}
\includegraphics[width=\columnwidth,clip=true]{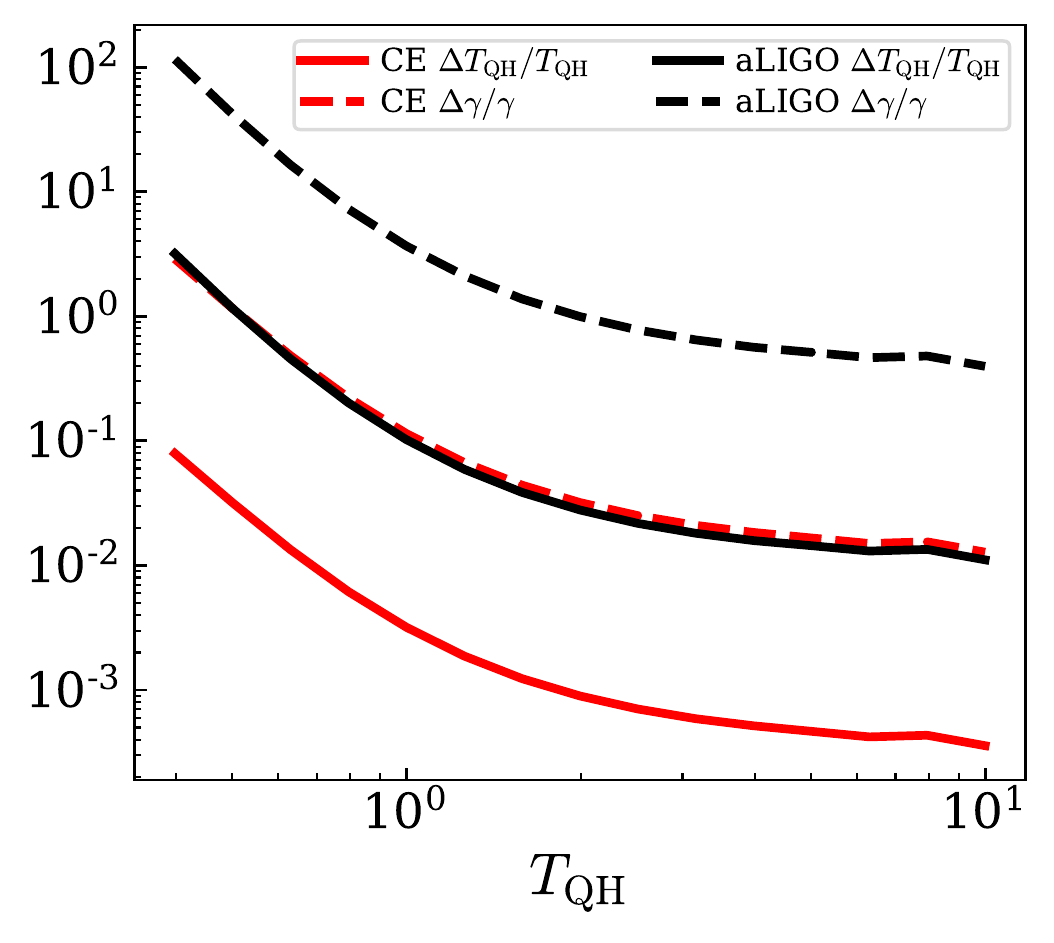}
  \caption{The fractional error of $T_{\rm QH}$ (solid curves) and $\gamma$ (dashed curves) as functions of $T_{\rm QH}$, using aLIGO (in black) and CE (in red). The binary system is \ind, who has a total mass of $60M_\odot$, and is located 100Mpc from the detector. Two filter parameters $v_\Sigma^{(\rm H)}$ and $\Delta v$ are still set to $-13$ and $2/\kappa$, respectively. We vary the value of $T_{\rm QH}$ from 0.4 to 10 while fixing the value of $\gamma$ to $10^{-15}$.}
 \label{fig:fisher_boltz}
\end{figure}

\subsection{Parameter estimation}
\label{sec:fisher}
We now use the Fisher-matrix formalism to study parameter estimation. Here we restrict ourselves to reflectivity parameters $(\gamma,T_{\rm QH})$, resulting in 2-D Fisher Matrices. 
%
%focus on the echo sector alone, meaning that the Fisher matrix is 2 dimensional, corresponding to $[\gamma,T_{\rm QH}]$. 
%
%Other parameters (e.g., chirp mass and mass ratio) have been inferred from the main GW. 
This will result in an under-estimate of measurement errors. 
As shown in Fig.~\ref{fig:fisher_boltz}, we compute the fractional errors of $T_{\rm QH}$ and $\gamma$, using \ind. We still assume that the system has a total mass of $60M_\odot$, and is located 100Mpc from the detector. Two filter parameters $v_\Sigma^{(\rm H)}$ and $\Delta v$ are still set to $-13$ and $2/\kappa$, respectively. We vary the value of $T_{\rm QH}$ from 0.4 to 10 while fixing the value of $\gamma$ to $10^{-15}$. We see the fractional error decreases as $T_{\rm QH}$ increases, since the echo signal is stronger. The constraint on $T_{\rm QH}$ is greater than $\gamma$ since it has bigger impact on the echo's profile and SNR. Choosing $T_{\rm QH}=T_{H}$, the aLIGO can constrain $\gamma$ and $T_{\rm QH}$ to the level of  366.7\% and 10.2\%, respectively. These two constraints lead to 20.9\% measurement uncertainty in the time interval $\Delta u^{\rm echo}$ between individual echoes, based on Eq.~(\ref{echo-time-interval}). For CE, the fractional errors of $\gamma$, $T_{\rm QH}$, and $\Delta u^{\rm echo}$ are 11.4\% and 0.3\%, and 0.65\%, respectively.

%In this subsection, the Fisher formalism is used to study the constraints that we could put on the reflectivity. Here we focus on the reflectivity alone rather than the information of main wave (e.g., chirp mass, and mass ratio). Therefore, the Fisher matrix for the Lorentzian reflectivity is 3-dimensional, corresponds to $[\epsilon,b,\Gamma]$. Figure \ref{fig:fisher_lorent} shows the fractional error of $b$, $\epsilon$, $\Gamma$ across the $\epsilon-\Gamma$ space, where we adopt the "hybrid"$-\psi_0$ prescription with $v_1=5M_f$ and $\Delta v_1=5M_f$. We can see the pattern of contours is consistent with that of SNR (see Fig.~\ref{fig:snr_lorentz}). The most constrained parameter is $b$, since it controls the time lag between each echo. For the parameter space we explore, the estimation error of $b$ is less than $10\%$, whereas $\Gamma$ is constrained to the level of $1\%-200\%$, and $\epsilon$ is $0.1\%-147\%$. As for the Boltzmann reflectivity, the Fisher matrix is 2-dimensional, corresponding to $[\gamma,T_{\rm QH}]$, respectively. Their fractional errors are shown in Fig.~\ref{fig:fisher_boltz}, where $\Delta T_{\rm QH}$ can be constrained to less than $1\%$ whereas $\gamma$ to $\sim 10\%$. We have checked that $\gamma$ has little impact on those estimations.

\section{CONCLUSION}
\label{sec:conclusion}
In this paper, we made use of the hybrid method \cite{Nichols:2010qi,Nichols:2011ih} to 
establish an echo waveform model for comparable-mass merging binaries whose remnants do not rotate. The hybrid method was proposed originally to predict GWs emitted by BBH coalescences --- it separates the space-time of a BBH event into an inner PN region and an outer BHP region (see Fig.~\ref{fig:my_label_new}). The two regions communicate via boundary conditions on a worldtube $\Sigma$. To build the echo model, we first took the Weyl scalars of the BBH systems from CCE \cite{Moxon:2020gha} at the future null infinity. Then we {\it reversed} the process of the hybrid method by evolving Weyl scalars back into the bulk, and 
%it turned out that 
the solution in the BHP region is proportional to the up-mode solution to the homogeneous Teukolsky equation, as required by the uniqueness of solutions. With the solution at hand, we were able to compute the GW that falls down the future horizon. 

Since the BHP theory is not valid inside the matching shell $\Sigma$, only the portion of GW that lies outside the worldtube $\Sigma_{\rm Shell}$ is physical.  Consequently, the usefulness of our method is limited to the ringdown phase.  We determined the location of $\Sigma$, namely the advanced time $v_\Sigma^{(\rm H)}$ at which it crosses the future horizon, by  looking for the quasi-normal ringing regime of the horizon$-\psi_0$ --- we fitted $Y^{\rm H~in}_{lm}$ to a superposition of five overtones [Eq.~(\ref{overtone-fit})]. 
We then removed the earlier piece of $\psi_0$ (with $v <v_\Sigma^{(H)}$) by applying a Planck-taper filter, whose width $\Delta v$ (a free parameter in our model) can be viewed as the effective thickness of the matching shell. 
% On the other hand, the width of the worldtube is a free parameter in our model; 

Next, by utilizing the physical boundary condition near ECO surfaces \cite{Chen:2020htz} and the Boltzmann reflectivity \cite{Wang:2019rcf}, we computed the QNMs of irrotational ECOs, as well as echo signals of two systems: \ind~and \indnew. We picked these two runs because their remnant spins vanish, in which the prediction of the hybrid method for ringdown signals has proved to be accurate \cite{Nichols:2010qi}. Finally, we studied the detectability and parameter estimation of echoes.

We summarize our main conclusions as follows: 

(i) The hybrid method is similar to the inside prescription of Refs.~\cite{Wang:2019rcf,Maggio:2019zyv} in the sense that both of them treat the main GW as a transmitted wave of an initial pulse emerging from the past horizon (see Fig.~\ref{fig:image_wave}). Furthermore, filters are involved in both treatments, which, however, have different physical interpretations. The inside prescription (also the CLA) handles the system as an initial value problem (the Cauchy problem), where the whole process is split into two stages. Only the late time portion lies in the BHP region. Therefore, the filter needs to be applied at the future null infinity. Oppositely, in our case, the exterior system is described by a boundary value problem --- a spatial volume is separated at every moment. Accordingly, the filter is imposed at the future horizon to remove the unrealistic portion of the incoming GW. We took \ind~as an example and compared the hybrid method with the inside prescription. We found that the inside prescription leads to fewer cycles than the hybrid method for the initial part of the echo. Meanwhile, the first echo predicted by the inside prescription is weaker than the result by the hybrid method.

\begin{figure*}[htb]      
   \centering
    \subfloat[\ind]{\includegraphics[width=\columnwidth,clip=true]{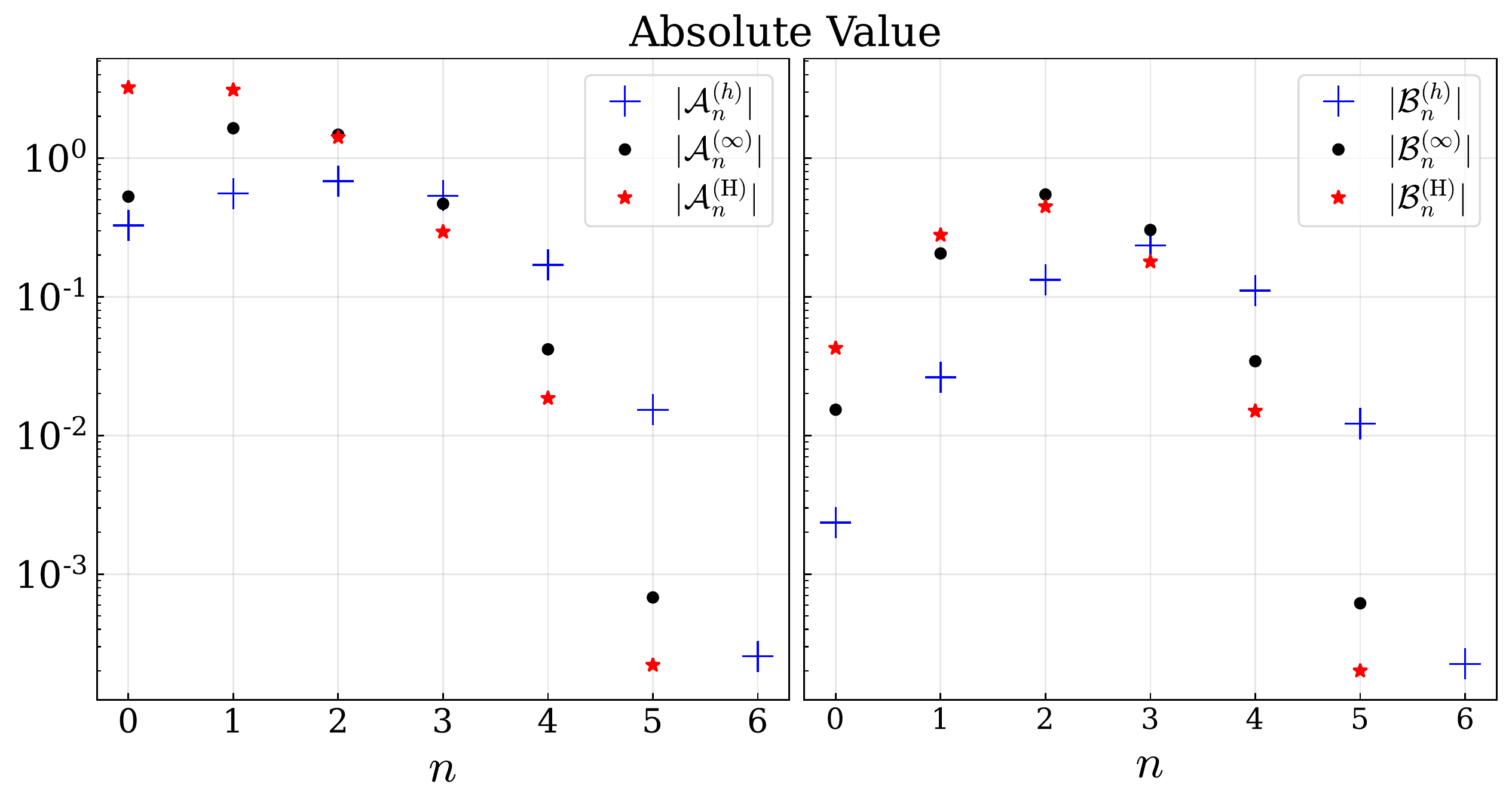}
    \includegraphics[width=\columnwidth,clip=true]{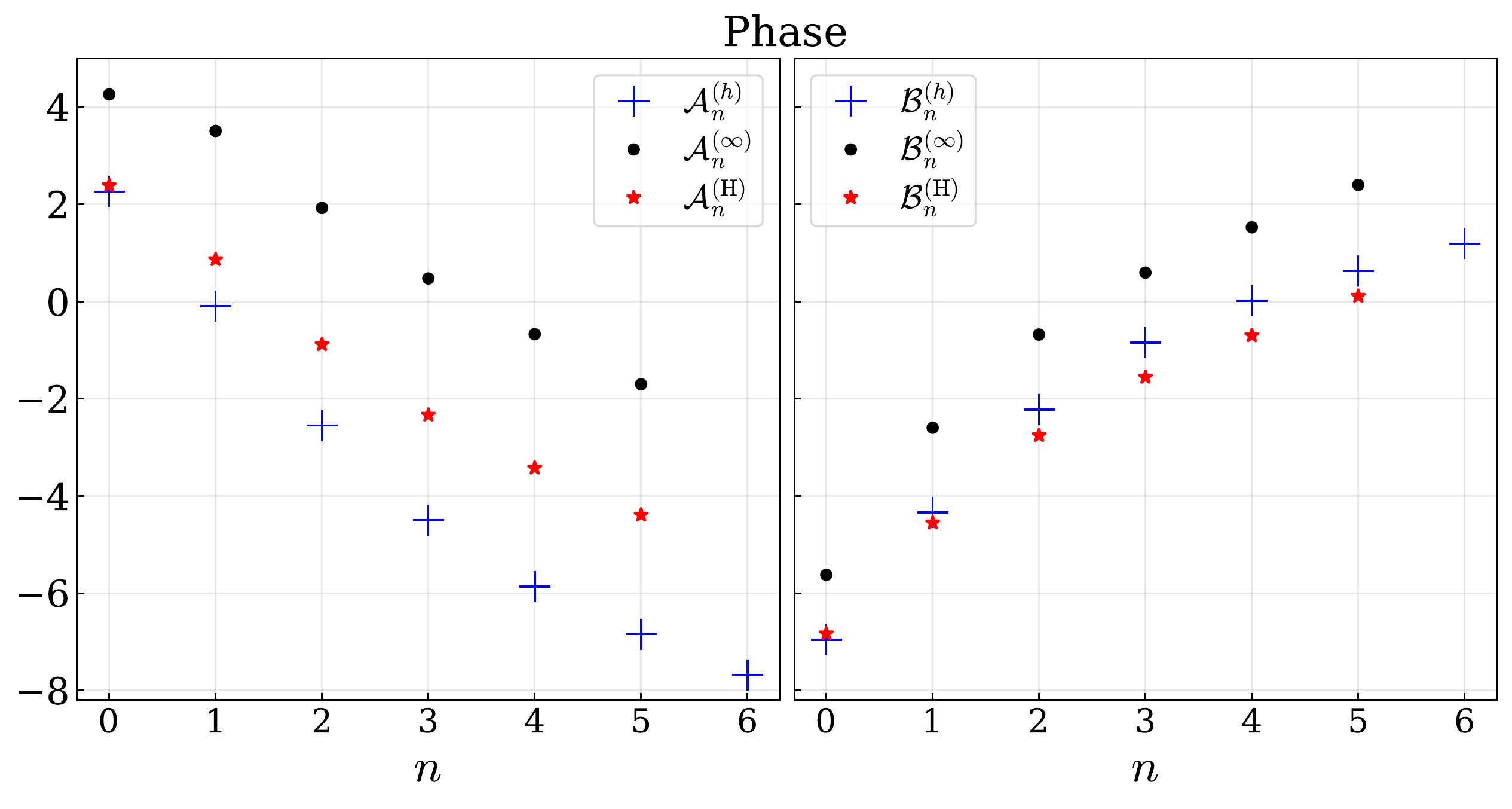}} \\
    \subfloat[\indnew]{\includegraphics[width=\columnwidth,clip=true]{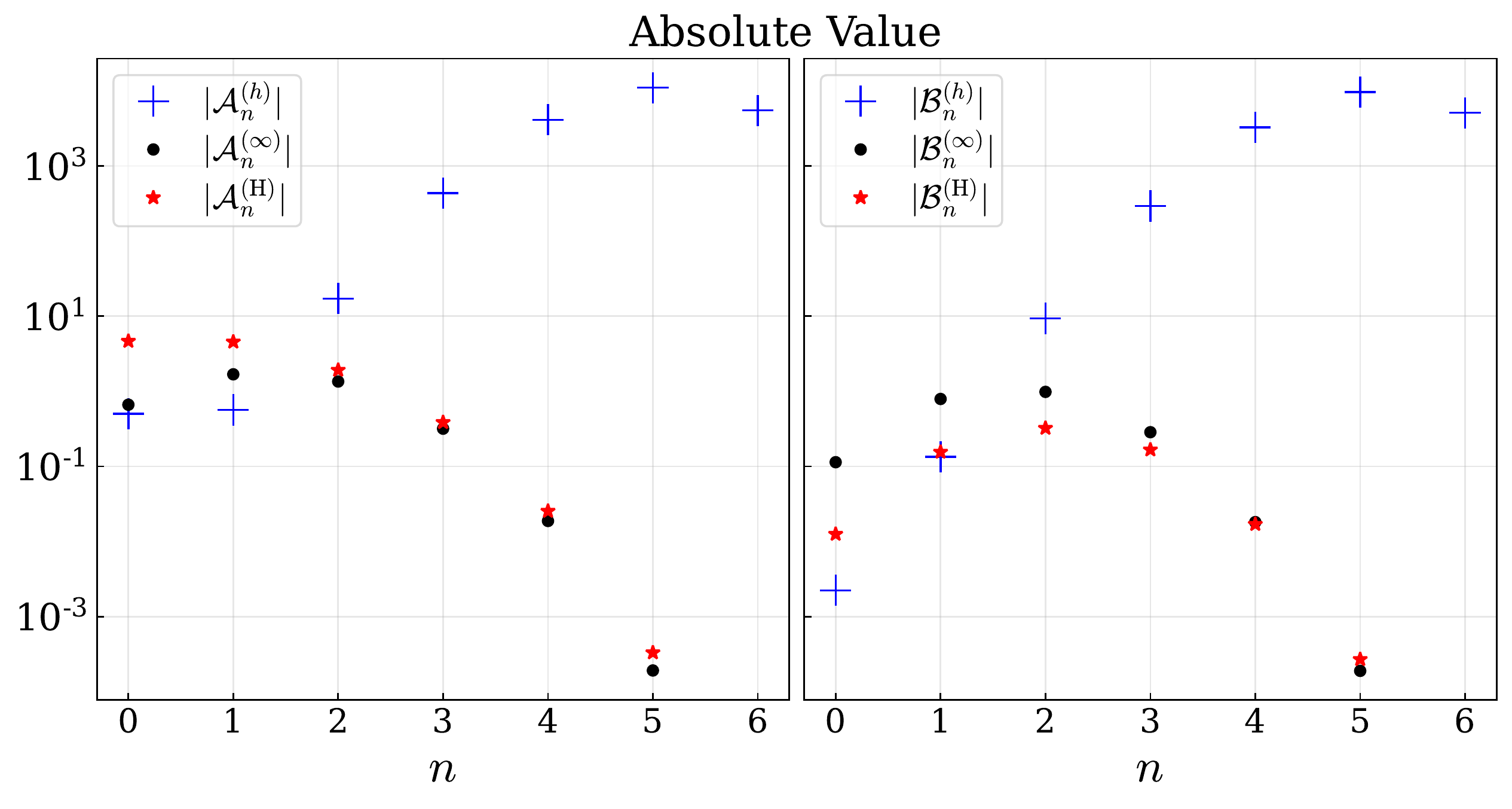}
    \includegraphics[width=\columnwidth,clip=true]{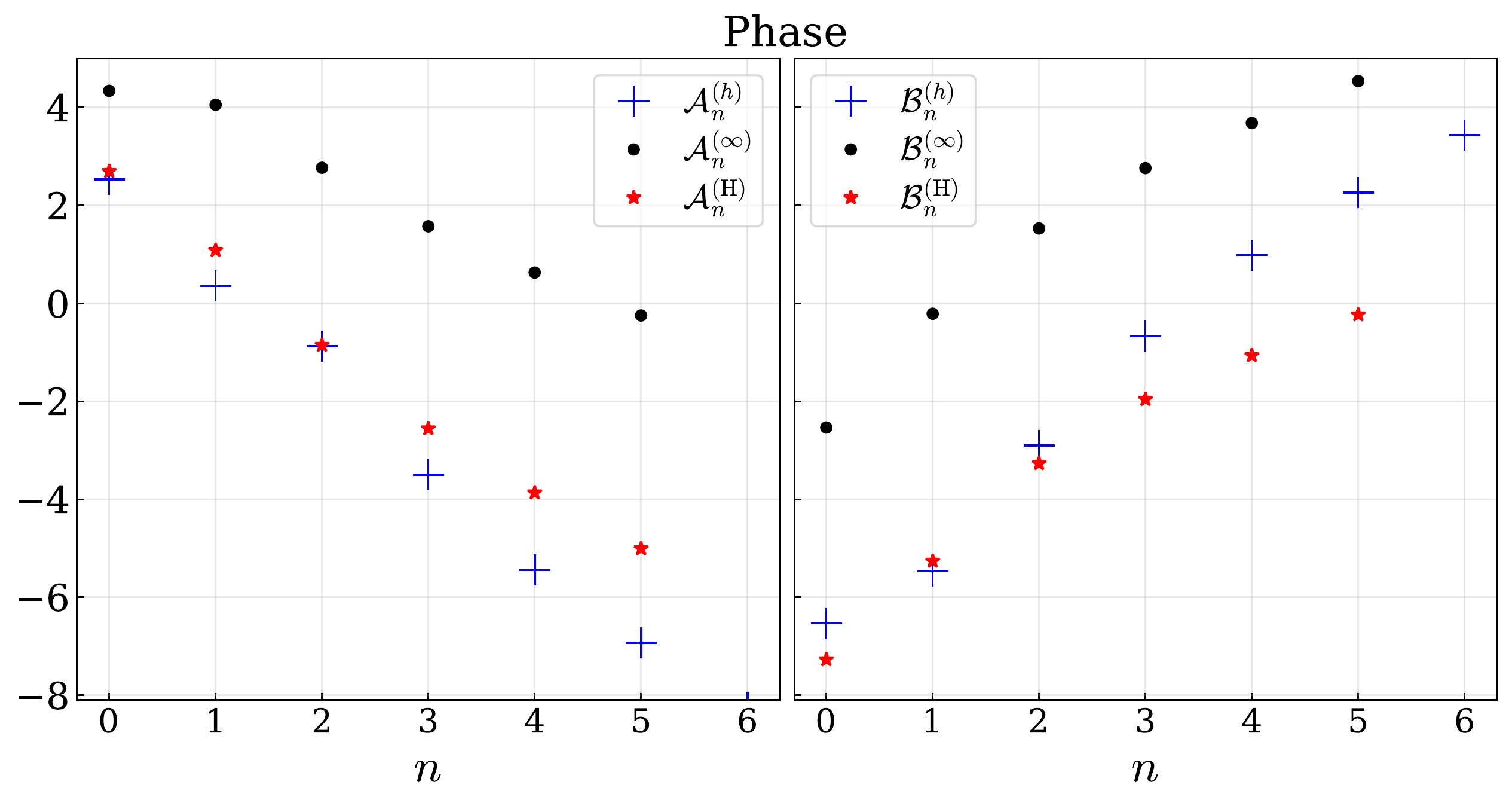}} 
  \caption{The absolute value (the left two panels) and phase (the right two panels) of the prograde mode $\mathcal{A}_n$ and the retrograde mode $\mathcal{B}_n$, assuming \ind~(the upper row) and \indnew~(the lower row). We fit Eqs.~(\ref{overtone-fit}) to the data of $h_{22}^\infty$ (blue), $Y_{22}^\infty$ (black) as well as $Y_{22}^{\rm H~in}$ (red) obtained from CCE.}
 \label{fig:overtone_mag}
\end{figure*}

(ii) The Weyl scalars $\psi_{0,4}$ from CCE are consistent with the TS identities throughout the entire frequency band in question. This supports the treatment of the hybrid method that uses the BHP theory to describe the exterior region, at least when the remnant object does not rotate.

(iii) Similar to the studies of Refs.~\cite{Giesler:2019uxc,Ma:2021znq}, using six overtones, the ringdown of the strain for \indnew~starts at $2M_f$ after the peak. However, the time for \ind~can be extended to $\sim11M_f$ before the peak. For the horizon and infinity $\psi_0$: $Y_{22}^{\rm H/\infty}$, the prediction of CCE is less accurate, and we were only able to resolve five overtones. The linearly quasi-normal ringing regime of $Y_{22}^{\rm H~in}$ for \ind~and \indnew~are similar and they start at $\sim 13-15M_f$ before the peak.

% (vi) The \indnew-like system leads to greater SNR than the \ind-like system by a factor of 1.5. Using aLIGO, the echo SNR is around 0.45 as $T_{\rm QH}=T_{H}$, while it is 15 when using CE. The measurement uncertainty of $\gamma$ is greater than that of $T_{\rm QH}$ since the latter one has bigger impact on echo's profile and SNR. Choosing $T_{\rm QH}=T_H$, aLIGO can constrain $\Delta\gamma$ and $\Delta T_{\rm QH}$ to the level of 39.6\% and 10.2\%, respectively. For CE, the fractional errors are $1.2\%$ and $0.3\%$.

We have restricted ourselves to  inspiralling compact binaries whose remnants are Schwarzschild-like ECOs. Future work could extend the hybrid method to Kerr-like ECOs and utilize it to compute echoes emitted by more general comparable-mass coalescence systems. It is worth pointing out that throughout the process, the Kerr-like background should have an adiabatically evolving mass and angular momentum due to GW emission. It will be a limitation for the hybrid method if one fails to capture this feature. Another possible avenue for future work is to apply our calculations to head-on collisions and compare the echo waveform with the results in Ref.~\cite{Annulli:2021dkw}.

%We want to remark that, during the early inspiral stage, the Kerr-like background should have an adiabatically evolving mass and angular momentum. But within such a low frequency regime, the properties of the Kerr BH are expected to have a weak impact on the GW \sma{cite?}. Indeed, Nichols and Chen \cite{Nichols:2011ih} treated the exterior spacetime as Schwarzschild BH, and the phasing already matched NR simulations well. Near the merger, the effect of mass and angular momentum emission cannot be neglected. But since the ringdown was shown to start from the peak of the strain \cite{Giesler:2019uxc}, and we are mostly interested in the post-merger echoes, we shall just ignore the mass and angular momentum changes. This will be a limitation for this approach. 

%==========================================================================
\begin{acknowledgments}
%We want to thank Jordan Moxon for useful discussions.
We thank Manu Srivastava, Shuo Xin, Rico K.L.\ Lo, Ling Sun for discussions. 
This work makes use of the Black Hole Perturbation Toolkit. The computations presented here were conducted on the
Caltech High Performance Cluster, partially supported by a
grant from the Gordon and Betty Moore Foundation.  This work was supported by the  Simons Foundation (Award Number 568762), the Brinson Foundation, Sherman
 Fairchild Foundation, and by NSF Grants No.~PHY-2011961, No.~PHY-2011968, PHY--1836809, and
 No.~OAC-1931266 at Caltech, and NSF Grants No.~PHY- 1912081 and No.~OAC-1931280
 at Cornell.
\end{acknowledgments}  

\appendix
\section{The QNM amplitudes of  \ind~and \indnew}
\label{app:qnm_spectrum}
Figure~\ref{fig:overtone_mag} shows the absolute value and phase of $\mathcal{A}^{(h/\infty/\rm H)}_n$ and $\mathcal{B}^{(h/\infty/\rm H)}_n$ [see Eq.~\eqref{overtone-fit}]. For \indnew, $\mathcal{A}_n^{(h)}$ peaks at $n=5$, consistent with previous studies \cite{Giesler:2019uxc,Ma:2021znq,Oshita:2021iyn}. However, in this case the absolute value of the retrograde mode $\mathcal{B}^{(h)}_n$ is comparable with that of $\mathcal{A}_n^{(h)}$, thus it is not negligible. For \ind, the contribution of the retrograde mode $\mathcal{B}_n^{(h)}$ is considerable as well, and $\mathcal{A}_n^{(h)}$ peaks at $n=2$ and $\mathcal{B}_n^{(h)}$ at $n=3$. 

\begin{figure}[htb]
        \includegraphics[width=\columnwidth,clip=true]{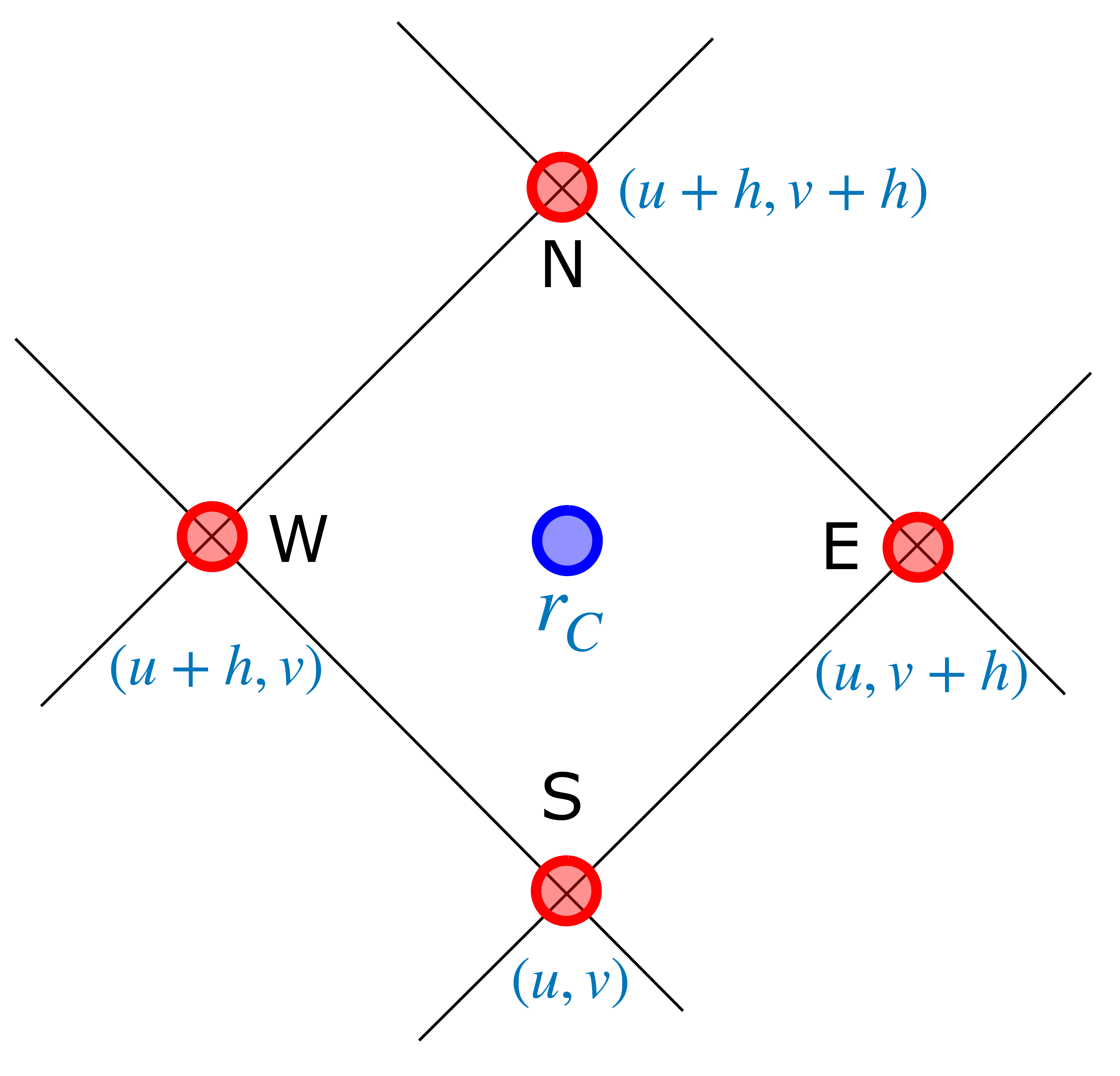}
  \caption{The $(u,v)$ grid cell in characteristic evolution scheme for the RW equation.}
 \label{fig:teukolsky_equation_characterisitic}
\end{figure}

\section{The characteristic approach for solving the RW equation}
\label{app:Characteristic_Approach}
 Eq.~(\ref{RWZ}) can be solved numerically via a second-order-accurate, characteristic method, proposed by Gundlach \etal \cite{Gundlach:1993tp}. As shown in Fig.~\ref{fig:teukolsky_equation_characterisitic}, Gundlach \etal \cite{Gundlach:1993tp} picked four points on a discretized $(u,v)$ grid:
\begin{align}
    &\tensor[_{s}]{{\Psi_{lm}^{\rm N}}}{}=\tensor[_{s}]{{\Psi_{lm}^{\rm SN}}}{}(u+h,v+h), && 
    \tensor[_{s}]{{\Psi_{lm}^{\rm E}}}{}=\tensor[_{s}]{{\Psi_{lm}^{\rm SN}}}{}(u,v+h), \notag \\
    &\tensor[_{s}]{{\Psi_{lm}^{\rm W}}}{}=\tensor[_{s}]{{\Psi_{lm}^{\rm SN}}}{}(u+h,v), && 
    \tensor[_{s}]{{\Psi_{lm}^{\rm S}}}{}=\tensor[_{s}]{{\Psi_{lm}^{\rm SN}}}{}(u,v),
\end{align}
with $h$ the step size.
The value on left corner $\tensor[_{s}]{{\Psi_{lm}^{\rm W}}}{}$ can be obtained through
\begin{align}
    \tensor[_{s}]{{\Psi_{lm}^{\rm W}}}{} = &\tensor[_{s}]{{\Psi_{lm}^{\rm N}}}{}+\tensor[_{s}]{{\Psi_{lm}^{\rm S}}}{}-\tensor[_{s}]{{\Psi_{lm}^{\rm E}}}{} \notag \\
    &+\frac{h^2}{8}V^l_{\rm RW}(r_c)(\tensor[_{s}]{{\Psi_{lm}^{\rm N}}}{}+\tensor[_{s}]{{\Psi_{lm}^{\rm S}}}{})+\mathcal{O}(h^3),\label{RWZ-characteristic}
\end{align}
where $V^l_{\rm RW}(r_c)$ is the value of the RW potential at the center $r_c=(u+h/2,v+h/2)$. We note that Eq.~(\ref{RWZ-characteristic}) is different from the one used in Refs.~\cite{Nichols:2010qi,Nichols:2011ih}, where $\tensor[_{s}]{{\Psi_{lm}^{\rm N}}}{}$ was calculated based on the other three. This is because we evolve the system backward into the bulk (from $\mathscr{I}^+$ to past horizon).

% On the other hand, for $\psi_{0(22)}^{\rm H}$ of \ind~and \indnew, the absolute value of $\mathcal{A}_{n}^{\rm (H)}$ decreases with increasing $n$ while $\mathcal{B}_{n}^{(\rm H)}$ has a peak at $n=2$. 

% Using the the physical boundary condition near an ECO surface \cite{Chen:2020htz}, the QNM of the ECO is determined by the condition in  Eq.~(\ref{ECO_QNM_def})
% \begin{align}
% \frac{D^{\rm in}_{lm}}{D^{\rm out}_{lm}}\frac{D}{4C}=\frac{B^{\rm in}_{lm}}{B^{\rm out}_{lm}}\frac{2-i\omega}{2+i\omega}\frac{4\omega-i}{4\omega+i}.
% \end{align}

% The value of $B^{\infty}_{lm}$ and $A^{\infty}_{lm}$ in Eq.~(\ref{RW-AB-Teukolsky}) leads to the boundary condition for $\tensor[_{s}]{{\Psi_{lm}^{\rm SN}}}{}$ in Eq.~(\ref{RW_boundary_condition}).

\section{\indnew}
\label{app:1936}
\begin{figure*}[htb]
   \centering
\includegraphics[width=\textwidth,clip=true]{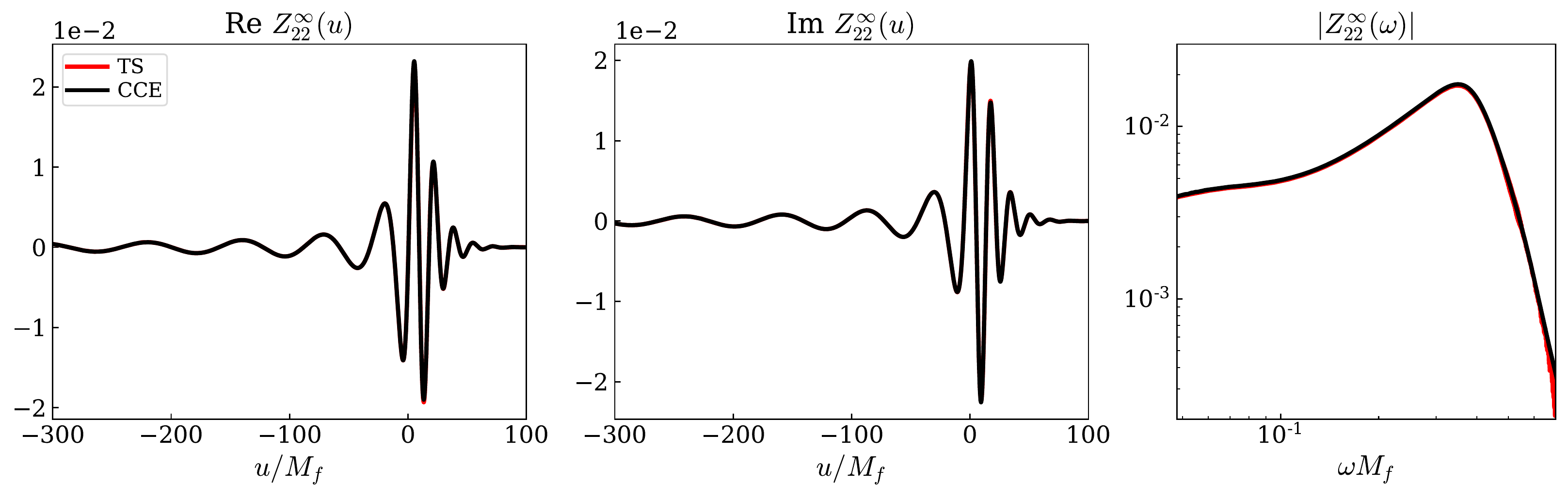}
  \caption{Same as Fig.~\ref{fig:TS_identity}, using \indnew.}
 \label{fig:TS_identity_1936}
\end{figure*}

Using \indnew, we test the validity of the TS identity at the null infinity [see Eq.~\eqref{TS_inf}] in Fig.~\ref{fig:TS_identity_1936}. Conventions are the same as Fig.~\ref{fig:TS_identity}.

\begin{figure*}[htb]
        \includegraphics[width=\textwidth,clip=true]{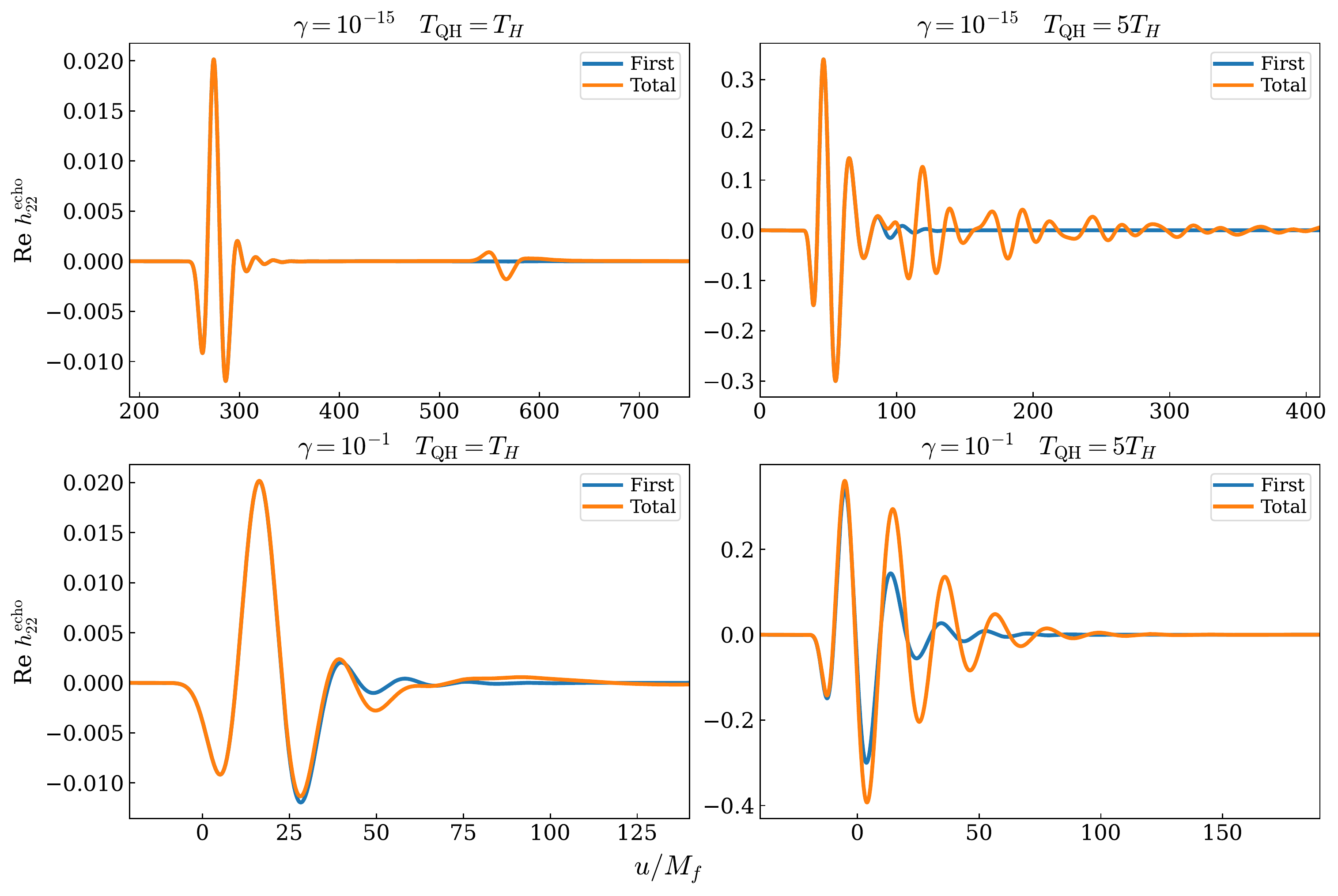}
  \caption{Same as Fig.~\ref{fig:echo_interferece}, using \indnew.}
 \label{fig:1936_echo_interferece}
\end{figure*}

In Fig.~\ref{fig:1936_echo_interferece}, we present the total echo and the first echo with a variety of $(\gamma,T_{\rm QH})$. The location of the filter is listed in Table \ref{table:overtone_fit}, and the width of the filter is set to $2/\kappa$.

\section{Chandrasekhar–Sasaki–Nakamura transformation}
\label{app:CSN}
The generalized Chandrasekhar–Sasaki–Nakamura transformation reads \cite{Hughes:2000pf}
\begin{subequations}
\begin{align}
    &\tensor[_{s}]{{\Psi_{lm}^{\rm SN}}}{}=
    \begin{cases}
    r^{|s|+1} D_-^{|s|}\left(\frac{1}{r^{|s|}}\tensor[_{s}]{{R^{\rm BH }_{lm}}}{}\right) & s<0, \\
    r^{s+1}D_+^s\left[\left(\frac{\Delta}{r}\right)^s\tensor[_{s}]{{R^{\rm BH }_{lm}}}{}\right] & s\geq 0,
    \end{cases} \\
&\tensor[_{s}]{{R^{\rm BH }_{lm}}}{}=
    \begin{cases}
    \frac{1}{c_0}\left(\frac{\Delta}{r}\right)^{|s|}D_+^{|s|}\left(r^{|s|-1}\tensor[_{s}]{{\Psi_{lm}^{\rm SN}}}{}\right) & s\leq 0, \\
    \frac{1}{c_0}\frac{1}{r^s}D_-^s\left(r^{s-1}\tensor[_{s}]{{\Psi_{lm}^{\rm SN}}}{}\right) & s>0,
    \end{cases}
\end{align}
\label{CSN_transformation}%
\end{subequations}
with $D_{\pm}=\frac{d}{dr}\pm \frac{i\omega r^2}{\Delta}$ and the constant $c_0$ given by
\begin{align}
    c_0=\begin{cases}
    C^* & s=-2, \notag \\
    l(l+1) & s=\pm1, \notag \\
    C & s=2,
    \end{cases}
\end{align}
where $C$ is defined in Eq.~(\ref{TS_identity_C}).

The up-mode solution, $\tensor[_{s}]{{\Psi_{lm\omega}^{\rm up}}}{}$, to the RW equation [Eq.~(\ref{RWZ})] takes an asymptotic expansion:
\begin{subequations}
\begin{align}
&\tensor[_{-2}]{{\Psi_{lm\omega}^{\rm up}}}{} \sim  
\begin{cases}
B^{\infty}_{lm\omega} e^{i\omega r_*} \,,\quad  & r_*\rightarrow +\infty, \\
\\
 B^{\rm out}_{lm\omega} e^{i \omega  r_*} +  B^{\rm in}_{lm\omega} e^{-i\omega r_*} \,,\quad   & r_*\rightarrow -\infty,
\end{cases} 
\\
&\tensor[_{+2}]{{\Psi_{lm\omega}^{\rm up}}}{}
\sim
\begin{cases}
A^{\infty}_{lm\omega} e^{i\omega r_*} \,,\quad  & r_*\rightarrow +\infty, \\
\\
 A^{\rm out}_{lm\omega} e^{i \omega  r_*} + A^{\rm in}_{lm\omega} e^{-i\omega r_*} \,,\quad   & r_*\rightarrow -\infty.
\end{cases}
\end{align}
\label{asymptotic-RW}%
\end{subequations}
Plugging Eqs.~(\ref{asymptotic-psi0-psi4_sourceless}) and (\ref{asymptotic-RW}) into Eq.~(\ref{CSN_transformation}), we obtain
\begin{subequations}
\begin{align}
&B^{\infty}_{lm\omega} =-\frac{C^*}{4\omega^2}, &&B^{\rm out}_{lm\omega}=-\frac{C^*D^{\rm out}_{lm}}{8\omega(i+4\omega)}, \notag \\
&B^{\rm in}_{lm\omega}=16(1-6i\omega-8\omega^2) D^{\rm in}_{lm}, \notag \\
&A^{\infty}_{lm\omega} =-4\omega^2, && A^{\rm in}_{lm\omega}=\frac{C}{8\omega (i-4\omega)} C^{\rm in}_{lm}, \notag \\
&A^{\rm out}_{lm\omega}=16(1+6i\omega-8\omega^2)C^{\rm out}_{lm},
\end{align}
\label{RW-AB-Teukolsky}%
\end{subequations}
and the TS identity in Eq.~(\ref{TS-infinity-hor}) implies
\begin{align}
\frac{B^{\rm in}_{lm\omega}}{B^{\infty}_{lm\omega}}=\frac{A^{\rm in}_{lm\omega}}{A^{\infty}_{lm\omega}}.
\end{align}

\clearpage
\bibliography{reference}
\end{document}